\documentclass[preprint2,twocolappendix]{aastex701}
\usepackage{amsmath}
\DeclareMathOperator\arctanh{arctanh}

\begin{document}

\newcommand{\kms}{\ensuremath{\mathrm{km}\,\mathrm{s}^{-1}}}
\newcommand{\Hunits}{\ensuremath{\mathrm{km}\,\mathrm{s}^{-1}\,\mathrm{Mpc}^{-1}}}
\newcommand{\galunits}{\ensuremath{\mathrm{km}\,\mathrm{s}^{-1}\,\mathrm{kpc}^{-1}}}
\newcommand{\galacc}{\ensuremath{\mathrm{km}^2\,\mathrm{s}^{-2}\,\mathrm{kpc}^{-1}}}
\newcommand{\MLsun}{\ensuremath{\mathrm{M}_{\sun}/\mathrm{L}_{\sun}}}
\newcommand{\Lsun}{\ensuremath{\mathrm{L}_{\sun}}}
\newcommand{\Msun}{\ensuremath{\mathrm{M}_{\sun}}}
\newcommand{\Ha}{\ensuremath{\mathrm{H}\alpha}}
\newcommand{\SFR}{\ensuremath{\mathit{SFR}}}
\newcommand{\aveSFR}{\ensuremath{\langle \mathit{SFR} \rangle}}
\newcommand{\sfrate}{\ensuremath{\mathrm{M}_{\sun}\,\mathrm{yr}^{-1}}}
\newcommand{\Aunits}{\ensuremath{\mathrm{M}_{\sun}\,\mathrm{km}^{-4}\,\mathrm{s}^{4}}}
\newcommand{\surfdens}{\ensuremath{\mathrm{M}_{\sun}\,\mathrm{pc}^{-2}}}
\newcommand{\voldens}{\ensuremath{\mathrm{M}_{\sun}\,\mathrm{pc}^{-3}}}
\newcommand{\gevcc}{\ensuremath{\mathrm{GeV}\,\mathrm{cm}^{-3}}}
\newcommand{\etal}{et al.}
\newcommand{\LCDM}{$\Lambda$CDM}
\newcommand{\ML}{\ensuremath{\Upsilon_*}}
\newcommand{\Mst}{\ensuremath{M_*}}
\newcommand{\Mg}{\ensuremath{M_g}}
\newcommand{\Mb}{\ensuremath{M_b}}
\newcommand{\Mmiss}{\ensuremath{M_X}}
\newcommand{\MCGM}{\ensuremath{M_{\mathrm{CGM}}}}
\newcommand{\Mhalo}{\ensuremath{M_{\mathrm{200}}}}
\newcommand{\Vhalo}{\ensuremath{V_{\mathrm{200}}}}
\newcommand{\Vf}{\ensuremath{V_f}}
\newcommand{\sigstar}{\ensuremath{\sigma_*}}
\newcommand{\gobs}{\ensuremath{\mathrm{g}_{\mathrm{obs}}}}
\newcommand{\gtot}{\ensuremath{\mathrm{g}_{\mathrm{tot}}}}
\newcommand{\gbar}{\ensuremath{\mathrm{g}_{\mathrm{bar}}}}
\newcommand{\azero}{\ensuremath{\mathrm{g}_{\dagger}}}

\title{The Baryonic Mass--Halo Mass Relation of Extragalactic Systems}

\author[0000-0002-9762-0980]{Stacy S. McGaugh} 
\affiliation{Department of Astronomy, Case Western Reserve University, 10900 Euclid Avenue, Cleveland, OH 44106, USA}
\email{stacy.mcgaugh@case.edu}

\author[0000-0001-7048-3173]{Tobias Mistele}
\affiliation{Department of Astronomy, Case Western Reserve University, 10900 Euclid Avenue, Cleveland, OH 44106, USA}
\affiliation{Max Planck Institute for Extraterrestrial Physics, Giessenbachstrasse 1, 85748 Garching, Germany}
\email{mistele@mpe.mpg.de}

\author[0009-0003-1662-5179]{Francis Duey}
\affiliation{Department of Astronomy, Case Western Reserve University, 10900 Euclid Avenue, Cleveland, OH 44106, USA}
\email{fxd113@case.edu}

\author[0009-0007-7808-4653]{Konstantin Haubner} 
\affiliation{Dipartimento di Fisica e Astronomia, Università degli Studi di Firenze, via G. Sansone 1, 50019 Sesto Fiorentino, Firenze, Italy}
\affiliation{INAF --- Arcetri Astrophysical Observatory, Largo Enrico Fermi 5, 50125 Firenze, Italy}
\email{konstantin.haubner@inaf.it}

\author[0000-0002-9024-9883]{Federico Lelli}
\affiliation{INAF --- Arcetri Astrophysical Observatory, Largo Enrico Fermi 5, 50125 Firenze, Italy}
\email{federico.lelli@inaf.it}

\author[0000-0003-2022-1911]{James M. Schombert} 
\affil{Institute for Fundamental Science, University of Oregon, Eugene, OR 97403, USA}
\email{jschombe@uoregon.edu}

\author[0000-0002-6707-2581]{Pengfei Li}
\affil{School of Astronomy and Space Science, Nanjing University, Nanjing, Jiangsu 210023, China}
\email{pli@nju.edu.cn}

\begin{abstract}
We combine data for extragalactic systems to quantify a relation between the observed 
baryonic mass \Mb\ and the enclosed dynamical mass \Mhalo\ inferred from kinematics or gravitational lensing. 
Our sample covers nine orders of magnitude in baryonic mass, including galaxies with kinematic or weak gravitational lensing data and  
groups and clusters of galaxies with new gravitational lensing data. 
For rich clusters with $\Mb > 10^{14}\;\Msun$, the observed baryon fraction is consistent with the cosmic value, $f_b = 0.157$. 
For lower masses, the baryon fraction decreases systematically with mass. The variation is well described by    
$\Mb/\Mhalo = f_b \tanh(\Mb/M_0)^{1/4}$ with $M_0 \approx 5 \times 10^{13}\;\Msun$. 
This relation is qualitatively similar to stellar mass--halo mass relations derived from abundance matching, but exhibits less scatter.
\end{abstract}


\section{Introduction}
\label{sec:intro}

There are a number of distinct missing mass problems in extragalactic astronomy. 
Most commonly, missing mass refers to the need for non-baryonic cold dark matter \citep{CDMPeebles,CDMSteigmanTurner} 
or a modification of the dynamical equations \citep[e.g.,][]{MOND,AeST}.
There are also missing baryon problems, both 
{globally for the whole universe and locally in individual galaxies}.

The global missing baryon problem is a shortfall in the inventory of observed baryons $\sum_i \Omega_i$ \citep[e.g.,][]{fukugita,Shull2012} 
relative to the expectation of Big Bang Nucleosynthesis $\Omega_b$ \citep{BBN,CopiBBN}.
This long-standing problem now appears to be solved, with the bulk of the baryons residing in the intergalactic medium (IGM)
so that $\sum_i \Omega_i \approx \Omega_b$ \citep{Connor2025}. 
However, {the local missing baryon problem persists.}

The local missing baryon problem is the shortfall in the mass of the observed stars and gas $\Mb = \Mst + \Mg$ in any given object relative 
to the mass of baryons available in its dark matter halo, $f_b \Mhalo$ 
\citep[$f_b = \Omega_b/\Omega_m = 0.157$ is the cosmic baryon fraction:][]{PlanckCosmology}. 
Only the most massive clusters of galaxies approach the limit $\Mb \approx f_b \Mhalo$ \citep[e.g.,][]{Gonzalez2013,Lagana2013}.
Lower mass systems have fewer detected baryons than their cosmic fair share, 
often by a large factor \citep{M10,Katz2018,Posti2019,McQuinn2022,Dev2024,ManceraPina2025}. 
In effect there are two missing mass problems in most extragalactic systems: these local missing baryons and the 
presumptively non-baryonic cosmic dark matter. 
This poses obvious questions: where are these unobserved baryons? What form do they take? 

This paper focuses on quantifying the local missing baryon problem. 
We provide a new method to relate the enclosed dynamical mass \Mhalo\ of extragalactic systems to their observed baryonic mass \Mb.
Our approach uses both kinematics and weak gravitational lensing to assess \Mhalo, providing an estimate that is independent from 
abundance matching and the stellar mass--halo mass relations obtained therefrom \citep[e.g.,][]{Moster2013,Behroozi13,KravtsovAM}. 
These different approaches have some qualitative similarities but important quantitative differences.

In section \ref{sec:theory} we lay out the necessary 
conceptual framework while section \ref{sec:data} describes the data we utilize, including novel data for groups of galaxies. 
Section \ref{sec:results} gives our result in the form of a baryonic mass--halo mass relation. 
In section \ref{sec:disc}, we compare this relation to stellar mass--halo mass relations from abundance matching, 
and discuss the implications of the variation of the baryon fraction with mass. Conclusions are given in section \ref{sec:conc}.

The individual galaxies in our samples have direct distance determinations available in the Extragalactic Distance Database \citep{EDD}. 
This is not the case for the groups and clusters, for which we utilize Hubble flow distances \citep{Haubner2025}. 
For these to be on a consistent scale with galaxy data, we adopt $H_0 = 73\;\kms\,\mathrm{Mpc}^{-1}$ \citep{RiessH02022}.

\section{Galactic Mass Components}
\label{sec:theory}


{Galaxies reside at the centers of non-baryonic dark matter halos.
Their baryons exist in multiple components: stars, gas in various phases, and dust.
The galaxies that we see are dominated by stars and cold (atomic and molecular) gas, typically within the central tens of kpc \citep{Hua2025}. 
Their dark matter halos extend much further out, typically hundreds of kpc, as depicted schematically in Fig.\ \ref{fig:schematic}.}

{Beyond the confines of the visible galaxy, there may also be some baryons mixed in with the dark matter: the circumgalactic medium (CGM).
We adopt the working definition that baryonic material within $r_{200}$ not otherwise accounted for is part of the CGM 
while material beyond $r_{200}$ is the IGM.
We discuss each of these mass components below.}

\subsection{Stars} 

{The defining content of galaxies is their stars. 
Stars dominate the baryonic mass of bright galaxies, and stellar mass is the key measurable quantity utilized by abundance matching relations.
For the individual galaxies in our sample (section \ref{sec:data}), stellar masses are estimated from luminosities and colors using stellar
population models \citep{Taylor2011,SML19,SML22}. 
These have become both precise \citep{MS14,MS15} and, for galaxies with direct distance measurements, reasonably accurate \citep{Duey_wiseii}.}

\subsection{Interstellar Medium}

{The space between stars in galaxies contains dust and gas of various phases: molecular, atomic, and ionized. 
The dominant mass component in the ISM of most late type galaxies is atomic gas, which is well traced by the 21 cm line of atomic hydrogen.
The relation of gas mass to 21 cm luminosity follows from the physics of the spin-flip transition \citep{Draine2011}, 
so gas masses can be both precise and accurate. 
We account for metallicity and molecular gas mass using scaling relations \citep{McGaugh2020b}. 
Ionized gas is important in early type galaxies, for which we also employ a scaling relation (see Appendix \ref{sec:appendix:groups}).
The mass of dust is negligibly small relative to the other components of the ISM.}

\subsection{Circumgalactic Medium} 

{Coronal gas has been detected around some bright galaxies in various ways, including  
absorption lines of highly ionized oxygen \citep{Bregman2007,CGMreview}, 
X-ray emission \citep{Li2018ApJ...855L..24L,Zhang2024a,Zhang2024b,Zhang2025a}, 
and the SZ effect \citep{Pratt2021ApJ...920..104P,Bregman2022ApJ...928...14B}.
These detections clearly show that there are baryons beyond the visible extent of galaxies.
The mass in the CGM around $L^*$ galaxies may be comparable to that in stars and other known baryons \citep{Bregman2018}, 
albeit with considerable uncertainty. Most of the CGM mass is inferred to reside at very large radii, 
so much depends on the extrapolation of power-law profiles and various other assumptions \citep{Zhang2025b}. }

\begin{figure}
\includegraphics[width=0.47\textwidth]{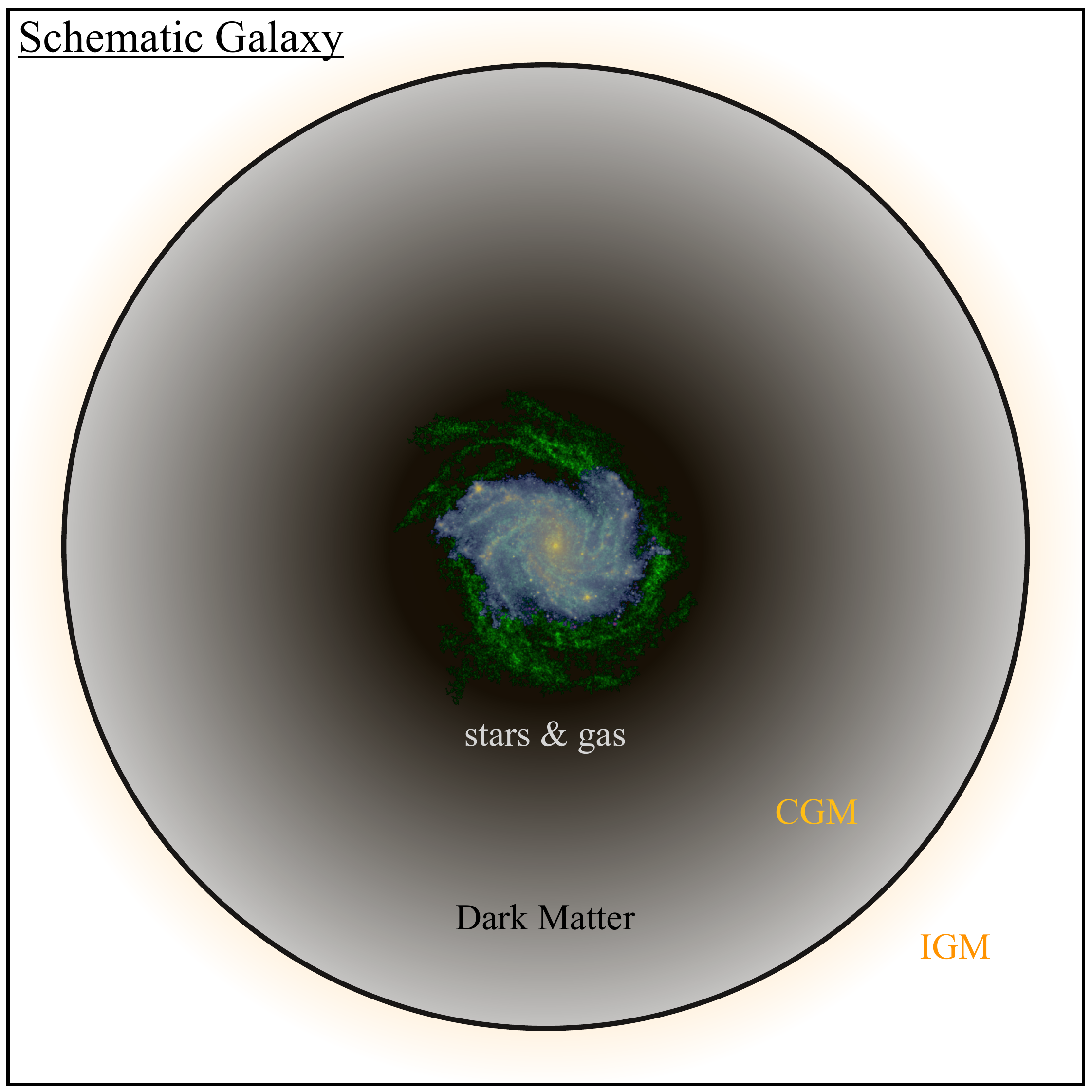}
\caption{Conceptual elements of a galaxy:  
the stars (yellow/blue) and atomic gas (green) of NGC 6946 \citep[Spitzer $3.6\mu$ and 21 cm data:][]{FTHINGS} are shown 
embedded in an extended dark matter halo (black).
The dark matter density decreases continuously with radius so the halo has no hard edge, 
but for convenience we adopt the common convention that the radius $r_{200}$ 
marks the boundary of the dark matter halo and the dividing line between the CGM and IGM (orange).
The stars and gas illustrated here appear within $r < 20\;\mathrm{kpc}$ while $r_{200} \approx 220\;\mathrm{kpc}$ (not shown to scale).
\label{fig:schematic}}
\end{figure}

{Detections of the CGM are limited to bright galaxies \citep[$\Mst \gtrsim 10^{10}\;\Msun$][]{Andreon2017A&A...606A..24A,Zhang2024b}. 
There are many galaxies of lower mass in our sample for which the baryonic mass of the CGM is not constrained.
In analogy with the intracluster medium in rich clusters of galaxies, it is tempting to imagine that the local missing baryonic mass 
is made up by the CGM. However, it is far from clear that this is viable for lower mass galaxies \citep{Bothun1994,Marasco2023}.}

\subsection{Dark Matter Halo}

The total extent and mass of dark matter halos remains a rather notional quantity. 
Kinematic measurements provide strong constraints on the enclosed mass: 
\begin{equation}
M(<r) = r V^2/G
\label{eq:encmass}
\end{equation}
for spherical dark matter halos, but generally provide only lower limits as they do not reach the edge of the halo. 
Observationally, rotation curves become approximately flat at large radii, and persist in remaining flat as far out as measured. 
Weak gravitational lensing observations suggest that this behavior persists to Mpc scales \citep{Brouwer2021,indefinitelyflat}. 
We utilize both kinematic and lensing data to make empirical estimates of the enclosed mass $\Mhalo(r <  r_{200})$. 



Dark matter-only structure formation simulations produce dark matter halos of the NFW form \citep{NFW}. 
This primordial initial condition can be altered by subsequent processes like adiabatic contraction \citep[e.g.,][]{oleg2004,adiabat,Li2022b,Li2022a} and baryonic feedback \citep[e.g.,][]{Dufffeedback,Governato2012,DC2014a,DC2014b,BBKARAA}. 
Here we are concerned only with the halo mass, for which we adopt the common convention of the mass $\Mhalo$ contained within the
radius $r_{200}$ enclosing an overdensity of two hundred times the critical density of the universe \citep[eq.\ 10 of][]{M12}. 
{This choice is somewhat arbitrary and is made to enable comparison with abundance matching relations;
it is straightforward to adjust our results to other choices of halo radius or overdensity \citep[eq.\ 12 of][]{M12}.}

The circular velocity \Vhalo\ of a test particle at $r_{200}$ is related to the enclosed mass by
\begin{equation}
\Mhalo = B\,\Vhalo^3.
\label{eq:NFWMV}
\end{equation}
For $H_0 = 73\;\kms\,\mathrm{Mpc}^{-1}$, $B = 3.3 \times 10^5\;\Msun\,\mathrm{km}^{-3}\,\mathrm{s}^3$ at $z=0$ \citep{SN}.
Note that because of the definition of the critical density, $\Vhalo \propto r_{200}$, so equations \ref{eq:encmass} and \ref{eq:NFWMV} are consistent.
{We will utilize kinematic \citep{LelliTF2019} and lensing measurements \citep{indefinitelyflat} of the flat rotation velocity, \Vf, 
as an indicator of \Vhalo\ and hence \Mhalo\ via equation \ref{eq:NFWMV}.}

\subsection{Baryonic Mass Budget}

{The sum of the baryonic components discussed above sum to the baryonic mass of an extragalactic system, 
which we can compare to the expected mass of available baryons, $f_b \Mhalo$. 
We have assembled a large sample of galaxies and systems of galaxies (section \ref{sec:data}) for which we have accurate measurements
of the mass in stars and cold gas for each individual galaxy, and of the intracluster gas in rich clusters of galaxies. 
Conspicuously missing are accurate measurements of the CGM for individual galaxies.
We therefore split the baryonic mass budget into two terms. The first is for the cold, condensed inner regions, i.e., the galaxies proper, for which}
\begin{equation}
\Mb = \Mst+\Mg.
\label{eq:Mb}
\end{equation}
{This is well measured, but generally falls short of the expected total, $f_b \Mhalo$.
The second term is thus the missing baryons,}
\begin{equation}
\Mmiss = f_b \Mhalo - \Mb.
\label{eq:Mmiss}
\end{equation}
{Some of these `missing' baryons may be in the CGM and not missing at all. 
However, some baryons may have been lost to the IGM, so we do not assume that $\Mmiss = \MCGM$.
Instead, we will obtain an expression for \Mmiss\ given the accurately observed \Mb\ and \Vf, 
and discuss the prospects for these baryons to be part of the CGM or lost to the IGM.}


{It is common to discuss the stellar fraction
\begin{equation}
m_* = \frac{\Mst}{\Mhalo}
\label{eq:mstar}
\end{equation}
in the context of stellar mass--halo mass relations obtained from abundance matching \citep[e.g.,][]{Moster2013,Behroozi13,KravtsovAM}. \
This makes sense when attempting to reconcile the galaxy luminosity function with the halo mass function.
However, all of the cold baryonic mass matters to the kinematics; the stellar mass alone is a poor proxy, especially for low mass galaxies \citep{btforig}.
We therefore consider} the baryonic mass fraction, the ratio of observed baryons to the halo mass:
\begin{equation}
m_b = \frac{\Mb}{\Mhalo}
\label{eq:mb}
\end{equation}
The quantitative variation of $m_b$ with mass is the primary result of this paper.
{We find a variation of $m_b$ with mass that is robust, but its absolute normalization is contingent on the halo mass,
which in turn depends on the relation between the circular velocity of the halo and that which is observed.}

\subsection{The Velocity Factor}

{We utilize measurements of the flat rotation velocity inferred from kinematics or gravitational lensing to estimate the 
dynamical mass \Mhalo\ via eq.\ \ref{eq:NFWMV}. This requires the velocity factor} 
\begin{equation}
f_v = \frac{\Vf}{\Vhalo}.
\label{eq:fv}
\end{equation}
A circular speed curve that persists in remaining flat indefinitely corresponds to $f_v = 1$: the rotation speed at the virial radius is
the same as that observed. The persistence of flat rotation curves to large radii led to the widespread assumption that this was indeed the case,
with asymptotic flatness built into the pseudo-isothermal halo that was frequently used in mass modeling \citep[e.g., equations 1 -- 3 of ][]{dBM97}. 
Empirically, there remains little evidence to contradict this assumption, and some to support its continuation remarkably far out \citep[to $\sim 1$ Mpc:][]{indefinitelyflat}: 

\begin{figure*}
\centering
\includegraphics[width=0.95\textwidth]{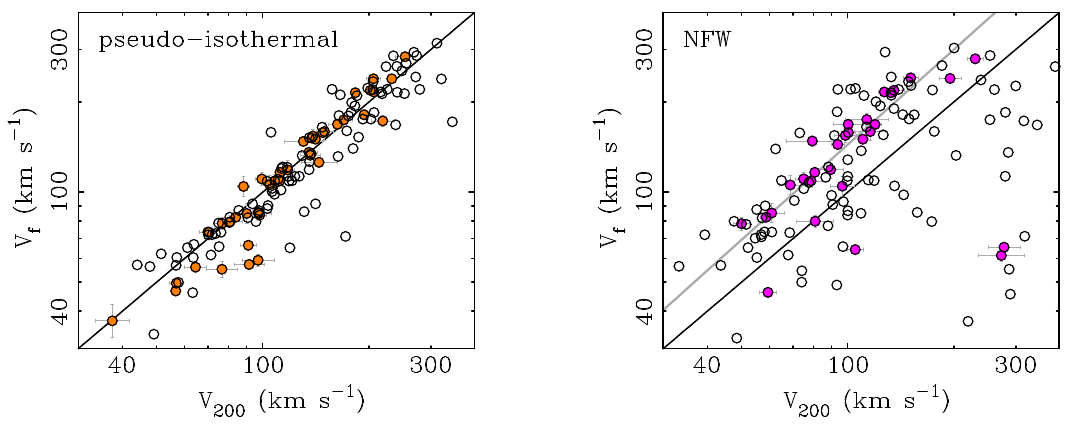}
\caption{The observed flat velocity \Vf\ as it relates to the fitted \Vhalo\ for pseudo-isothermal (left) and NFW (right) halos \citep{Li2020}.
Filled points have formal uncertainties less than 20\% in \Vhalo; open points are less accurate fits. 
The solid line shows $\Vf = \Vhalo$. The gray line in the right panel shows eq.\ 2a of \citet{KatzfV}, which corresponds roughly to $f_v \approx 1.4$.
\label{fig:M200}}
\end{figure*}

The velocity factor $f_v$ cannot differ too much from unity and still explain the flatness of observed rotation curves, but slight differences are possible. 
For halos consistent with CDM simulations, the outer density profile is expected to fall off as $\rho \propto r^{-3}$, so the rotation curve should decline
somewhat as it approaches the virial radius, implying $f_v$ close to but slightly larger than unity. 
Figure \ref{fig:M200} shows the observed \Vf\ vs.\ the fitted \Vhalo\ for two halo models, 
the pseudo-isoithermal \citep{vA1985,dBM97,KormendyFreeman} and NFW \citep{NFW} halos. 
We use the fits\footnote{Note that \Vhalo\ of the fitted model is a parameter of the dark matter halo alone, 
so is subtly different from \Mhalo\ as defined here to include baryons.} of \citet{Li2020} 
that have not been constrained with \LCDM\ priors to illustrate what the data alone can tell us. 
As expected, the projected value of \Vhalo\ for the pseudo-isothermal model closely tracks the observed flat velocity, corresponding to $f_v = 1$. 
For the NFW model, the better data are offset from the one-to-one line by $f_v \approx 1.4$ \citep[see also][]{KatzfV,Posti2019}. 
This goes in the expected direction ($\Vf > \Vhalo$ since rotation curves should decline), but there is an enormous amount of scatter in the value 
of $f_v$ inferred.

The scatter seen in Fig.\ \ref{fig:M200} for the NFW model happens for a number of reasons. For one, 
it is difficult to constrain the rotation speed at the virial radius with rotation curves that do not extend that far out.
Consequently, only the most extended rotation curves provide interesting constraints. 
More generally, NFW halos do not typically provide a good description of the data \citep{M07,Katz2017}. 
One symptom of the cusp-core problem is a propensity for fits to exaggerate the overall size of a halo to compensate for the mismatch in shape
at small radii \citep{deBlok2001}. An example of this is provided by the galaxies in Fig.\ \ref{fig:M200} with $\Vf \approx 60\;\kms$ 
but $\Vhalo \approx 300\;\kms$. These dwarf galaxies, with $\Mb \approx 6 \times 10^8\;\Msun$, presumably do not reside in dark matter halos
with masses $\Mhalo \approx 10^{13}\;\Msun$ that would outweigh the entire Local Group. Clearly we need to exercise caution in attempting to
infer the value of $f_v$, which may also differ for different halo models \citep{Li2020}. 

Despite these difficulties, it is important to note that there is very little scatter in the Tully-Fisher relation \citep{TForig},
especially when \Vf\ is used as the rotation speed measure \citep{LelliTF2019}.
We will accept this as a sign that the observed flat rotation speed is indeed closely related the halo mass, as often assumed.
We thus adopt $f_v = 1$ for illustration, and will explore variations in this parameter in section \ref{sec:disc}. 

\subsection{Constraints from Tully-Fisher}
\label{sec:emp}

For rotationally supported galaxies, \Mb\ and \Vf\ are tightly correlated through the Baryonic Tully-Fisher Relation (BTFR):
\begin{equation}
\Mb = A\,\Vf^4
\label{eq:btfr}
\end{equation}
with $A = 50\;\Aunits$ \citep[][]{M05}.
The best-fit has consistently been in the range $1.65 \le \log A \le 1.75$ for twenty years \citep{M12,Duey_wiseiii}. 
Some of this variation is from fitting uncertainty and some is from differences 
in stellar population models and correction factors for metallicity and molecular gas \citep{metalsandmolecules}. 

The observed slope of the BTFR is indistinguishable from four \citep{MS15,JSH0,Duey_wiseiii}.
This is steeper than the slope three of the theoretical halo mass--circular velocity relation (eq.\ \ref{eq:NFWMV}). 
This difference requires a systematic variation of $m_b$ or $f_v$ or both with mass. The simplest way to reconcile the expected
$\Mhalo \propto \Vhalo^3$ with the observed $\Mb \propto \Vf^4$ is for $m_b$ to increase linearly with \Vf\ \citep{M12}. 
More generally, some combination of variation of both $m_b$ and $f_v$ is possible, but in practice it is difficult to explain much of the
required variation with the velocity factor $f_v$ (Fig.\ \ref{fig:M200}). Consequently, one inevitably infers a strong variation of
the baryon fraction $m_b$ with mass.


Combining the equations above to transform the BTFR into the baryonic mass--halo mass plane, we have
\begin{equation}
m_b = \frac{A}{B} {f_v^3} {\Vf}
\label{eq:mbV}
\end{equation}
which is the linear dependence on \Vf\ with allowance for variation in $f_v$. Since the baryonic mass is generally more accessible observationally 
than the flat rotation speed, we can also write 
\begin{equation}
m_b =  \frac{A^{3/4}}{B} {f_v^3} \Mb^{1/4} 
\label{eq:mbM}
\end{equation}
\citep[see also equations 16 and 17 of][where the detected baryon fraction ${f_d} = m_b/f_b$]{M12}.

These simple relations hold only for rotationally supported galaxies. 
Below we generalize this result to other types of systems over nine decades in baryonic mass. 

\section{Data}
\label{sec:data}

{We measure the baryon fraction in galaxies from dwarfs with $\Mb \approx 10^6\;\Msun$ to giants with $\Mb \approx 4 \times 10^{11}\;\Msun$ 
continuing through groups of galaxies to rich clusters up to $\Mb \approx 5 \times 10^{14}\;\Msun$.
The lensing mass estimates for groups are the novel contribution of this work.
Data for other systems are taken from other papers, as discussed below.
We make no adjustments to published results except to reconcile all data to the same distance scale ($H_0 = 73\;\kms\,\mathrm{Mpc}^{-1}$).} 

{There exist accurate data for the stellar and gas masses for many galaxies. We take the} 
baryonic mass of each object to be the sum of measured stars and gas {(eq.\ \ref{eq:Mb})}, as discussed in the papers describing the data.
{Measurements for the mass of the coronal gas in the CGM are not available for the vast majority of objects with otherwise excellent data,
so we do not include it in the sum. Instead, we obtain a relation between the measured baryonic mass $\Mb = \Mst + \Mg$ and corresponding
baryon fraction $m_b = \Mb/\Mhalo$. This can be used to infer how much mass \Mmiss\ is missing, which in turn provides a constraint on the mass in the
CGM, \MCGM.}

{To estimate the dynamical mass \Mhalo\ including dark matter, 
we utilize equation \ref{eq:NFWMV} and measurements of the flat rotation velocity \Vf.}
The flat rotation speed is estimated from either kinematic or gravitational lensing data. 
We take care to define an equivalent to the flat circular speed for pressure supported systems so that all objects are on a comparable scale. 

\subsection{Kinematic Data}

\subsubsection{Rotationally Supported Galaxies} 

\paragraph{WISE-SPARC Sample}
spiral and irregular galaxies \citep{SPARC} with direct distance determinations \citep{Duey_wiseiii} and stellar mass estimates based on 
WISE near-infrared photometry \citep{Duey_wisei} and detailed stellar population models \citep{Duey_wiseii}. 
The gas mass is is calculated using the observed atomic gas mass and a scaling relation to account for the molecular gas content 
and the trend of metallicity with mass \citep{metalsandmolecules}. 
Flat rotation speeds \citep{LelliTF2019} are measured from resolved 21 cm rotation curves using the algorithm of \citet{LelliTFscatter}.

\paragraph{Gas Rich Galaxy Sample} 
low mass dwarf Irregular galaxies that are not already in the WISE sample of \citet{Duey_wisei}. 
These galaxies are dominated by atomic gas, with $\Mg > \Mst$, making the contribution of stars to the baryonic mass error budget small \citep{M11}. 
The data come from a variety of sources \citep{trach,stark,LeoProt,SHIELD,iorio2017,KK153,Namumba2025}.

There is considerable overlap in mass between these gas rich dwarf irregulars and the gas poor dwarf spheroidals of the Local Group.
The lowest mass examples, Leo P \citep{LeoPdisc,LeoProt} and KK153 \citep{KK153}, have circular velocities $\Vf \approx 15\;\kms$ and stellar masses $\Mst \approx 4 \times 10^5\;\Msun$ that are nearly in the ultrafaint regime.
This places them well below the nominal threshold for quenching from reionization \citep{LeoPunQ}, which had been anticipated to be $\sim 40\;\kms$ \citep{crain,hoeft}. 
Nevertheless, they exist, and are forming stars in disks dominated by cold gas. 

\paragraph{Local Group Rotators} 
Galaxies range from M31 and the Milky Way at the high mass end to DDO 210 and DDO 216 at the low mass end.
The latter two both have $\Vf \approx 15\;\kms$, making them comparable to the lowest mass galaxies in the gas rich sample, 
albeit with large uncertainties. 
The uncertainties are small for the more massive members of the Local Group, which is reflected in their small scatter in Fig.\ \ref{fig:VMb}.
The data are adopted from Table 1 of \citet{McG2021}.
 
\subsubsection{Pressure Supported Local Group Dwarfs}

\paragraph{Classical Dwarfs} form a sequence in luminosity and velocity dispersion that parallels the BTFR \citep{McG2021}. 
We adopt the data from Table 2 of that paper here. For these galaxies, the two sequences are coincident when the factor $\beta = 2$ 
in $\Vf = \beta \sigma$ assuming an average $V$-band mass-to-light ratio of $2\;\MLsun$. This value of $\beta$ is only slightly larger than the nominal $\beta = \sqrt{3}$ expected for isotropic orbits, and may follow from the 
fact that \Vf\ is typically measured at larger radii than $\sigma$ \citep[see discussion in][]{McG2021}.

\paragraph{Ultrafaint Dwarfs} with $L_V < 10^5\;\Lsun$ \citep{simon2019} do not parallel the BTFR as classical dwarfs do \citep{OneLaw}. 
To a crude approximation, a tolerable depiction of the trend in the data is a constant $\Vf \approx 10\;\kms$ for $\Mb < 5 \times 10^5\;\Msun$.
Allowing for a slope, the \Mb-\Vf\ relation obtained by \citet{M10} for $\Vf < 20\;\kms$ remains consistent with the data despite many newly discovered
dwarfs in the intervening time.

These ultrafaint dwarfs are all deep in the potential of the Milky Way where they are subject to considerable tidal perturbation.
It therefore seems unlikely that their measured velocity dispersions are faithful indicators of their dynamical mass \citep{MWolf}, 
so we will restrict our attention to more massive systems. 
Historically, it is worth noting that the point at which galaxies appear to deviate from the BTFR has
decreased from $\Vf \approx 50\;\kms$ \citep{M05} to $\approx 20\;\kms$ \citep{M10} to $\approx 10\;\kms$ (now). This trend stems largely from
improvements in data quality, so it is possible that it will continue as the data continue to improve. 
There is a clear distinction between a plateau of $\sim 10\;\kms$ for the lowest mass galaxies and the continued decrease 
predicted by the extrapolation of the BTFR. This can hopefully be tested with the discovery of isolated low mass galaxies far from tidal influences that should be enabled by the Rubin Observatory and Square Kilometer Array.

\subsection{Lensing Data}

The equivalent of the flat circular velocity can be inferred from weak gravitational lensing observations with the method described by \citet{lensRAR}. \citet{indefinitelyflat} applied this method to the stacked data for large numbers of individual galaxies. 
\citet{Mistele2024b} extended the method to apply to individual clusters of galaxies \citep{Mistele_CLASH}. 
Here, we apply it on the group scale.

\subsubsection{Galaxies}

\paragraph{Late Type Galaxies} are adopted from the analysis of the KiDS-bright sample \citep{Bilicki2021} by \citet{indefinitelyflat}.
Late-type galaxies are defined as having $u-r < 2.5$ and are split into four bins by baryonic mass with thousands of galaxies per bin.
The flat circular velocity is obtained by averaging the circular velocities at radii from $50 < r < 300\;\mathrm{kpc}$.

\paragraph{Early Type Galaxies} are also adopted from \citet{indefinitelyflat}, taken there to be those galaxies with $u-r > 2.5$. 
They are analyzed in the same way as the late-type galaxies.
{The baryonic mass includes a correction for hot gas based on a scaling relation with stellar mass (see Appendix \ref{sec:appendix:groups}).}
There are only three baryonic mass bins due to the small number of early type galaxies {in the lowest mass bin}.

\subsubsection{Galaxy Systems}

\paragraph{Clusters of Galaxies} \citet{Mistele_CLASH} obtained the flat circular velocity for 16 CLASH clusters from gravitational lensing data 
by averaging the implied circular velocities over radii between $750\,\mathrm{kpc} < r < 3\,\mathrm{Mpc}$.
The baryonic masses are evaluated by integrating the beta profile fits of the hot gas to $r = r_{200}$. 
{This is an extrapolation beyond the limits of direct observation. It integrates to a large mass that results in a reasonable baryon fraction,
where other plausible extrapolations do not. To account for the stellar mass in cluster galaxies,
\citet{Mistele_CLASH} assumed} $\Mst = 0.08 \, \Mb$ {following} \citet{Famaey2024}.
{More elaborate scaling relations \cite[e.g., ][]{Lagana2013} do not change the basic result. 
It would obviously be preferable to directly measure the stellar mass of each cluster, but such data are not readily available. 
The uncertainty in the stellar mass is much smaller than that
in the gas mass obtained from the extrapolation of the beta profile. Our results for clusters are in excellent agreement with \LCDM,
so it would become a problem if either our stellar or gas mass estimates were grossly in error. 
This also applies to dynamical masses, for which less favorable \Mhalo\ can reasonably be inferred \citep{PengfeiClusters}.}

\paragraph{Groups of Galaxies}
We measure the stacked mass profiles of galaxy groups from GAMA-II \citep{Driver2011,Liske2015} using KiDS-DR4 weak-lensing data \citep{Kuijken2019,Wright2020,Giblin2021,Hildebrandt2021}.
Following previous works, we select groups with at least five members \citep[e.g.][]{Rana2022,Viola2015,Li2024,Liu2024} and split the data into bins by baryonic mass.
For each $M_b$ bin, we calculate averaged stellar and baryonic masses by summing up the respective quantities of the member galaxies and averaging over the groups in each bin.
We obtain a stacked halo mass profile $M(r)$ using the non-parametric weak-lensing deprojection method from \citet{lensRAR} and \citet{Mistele2024b}.
We correct for miscentering and contributions from the two-halo term. 
These corrections are somewhat uncertain, but have only a moderate effect in the radial range we consider.
We convert $M(r)$ into a circular velocity profile $V_c (r) \equiv \sqrt{G M(r)/r}$ and take $\Vf$ to be the weighted average of the circular velocities at radii between $800\,\mathrm{kpc}$ and $2\,\mathrm{Mpc}$.
The results are provided in Table~\ref{tab:Group} with details explained in Appendix~\ref{sec:appendix:groups}.

\begin{deluxetable*}{ccccrcc}
\tablewidth{0pt}
\tablecaption{{Weak Lensing Group Masses  \label{tab:Group}}
}
\tablehead{
\colhead{Bin} & \colhead{N\tablenotemark{a}} & \colhead{$z$} & \colhead{\Vf} & \colhead{\Mb} & \colhead{\Mst} & \colhead{$\Mst^{\mathrm{BCG}}$}  \\
& & & \colhead{($\kms$)} &   \multicolumn{3}{c}{($10^{12}\;\Msun$)}
}
\startdata
1 & 521  & 0.1168  & $242\pm 47$ &  0.307 & 0.207 & 0.084  \\
2 & 769  & 0.1732  & $335\pm 31$ &  0.978 & 0.601 & 0.188  \\
3 & 295  & 0.2124  & $402\pm 46$ &  1.894 & 1.095 & 0.283  \\
4 & 456  & 0.2452  & $530\pm 30$ &  2.942 & 1.666 & 0.337  \\
5 & 398  & 0.2878  & $614\pm 33$ &  5.461 & 3.001 & 0.406  \\
6 & 206  & 0.3239  & $871\pm 42$ & 11.164 & 5.851 & 0.540  \\
\enddata
\tablenotetext{a}{The number of groups informing the average quantities for each bin.}
\end{deluxetable*}

\section{Results}
\label{sec:results}

\begin{figure*}
\plotone{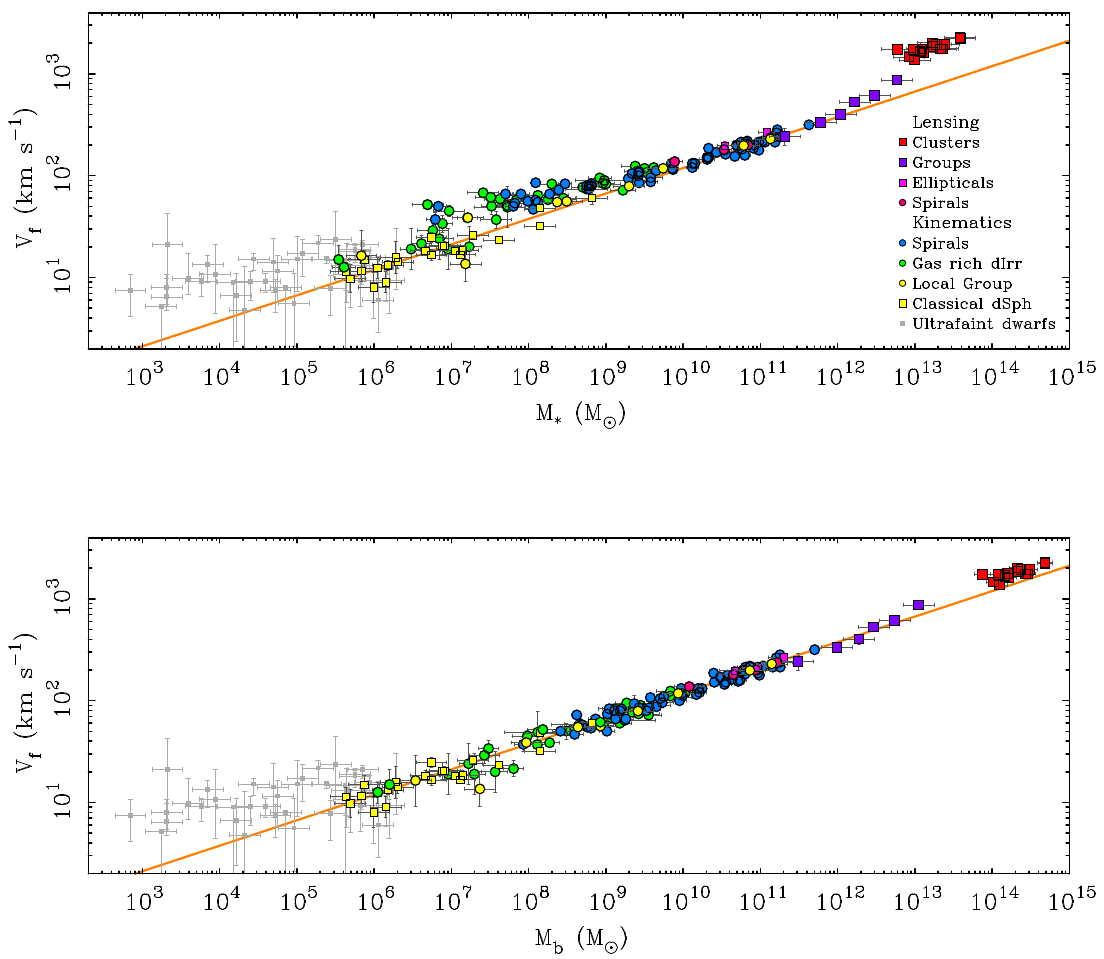}
\caption{The flat-equivalent circular velocity of extragalactic systems as a function of stellar mass (top) and baryonic mass (bottom).
Data for rotationally supported galaxies are depicted by circles; squares represent pressure supported systems. 
The blue circles are galaxies with directly measured distances, \Vf\ from rotation curves, and stellar masses from WISE photometry from \citet[]{Duey_wiseiii}.
Green circles are gas rich galaxies \citep[$\Mg > \Mst$:][]{trach,stark,LeoProt,SHIELD,iorio2017,KK153,Namumba2025} not in \citet{Duey_wiseiii}.
Yellow points are Local Group galaxies, both spirals and dwarfs \citep{McG2021}; gray squares are ultrafaint dwarfs \citep{OneLaw}.
Lensing results for early and late type galaxies \citep{indefinitelyflat} are shown as pink squares and magenta circles,
respectively. Red squares are clusters of galaxies \citep[]{Mistele_CLASH} and purple squares are groups of galaxies (this work). 
The orange line is the BTFR (eq.\ \ref{eq:btfr}). 
\label{fig:VMb}}
\end{figure*}

{The data described above are shown together in Figure \ref{fig:VMb}. 
There is a strong correlation between} the flat rotation speed and both stellar mass and baryonic mass.
{These relations are analogous to the Tully-Fisher relation, but including} extragalactic systems from ultrafaint dwarfs to rich clusters of galaxies.

\begin{figure*}
\plotone{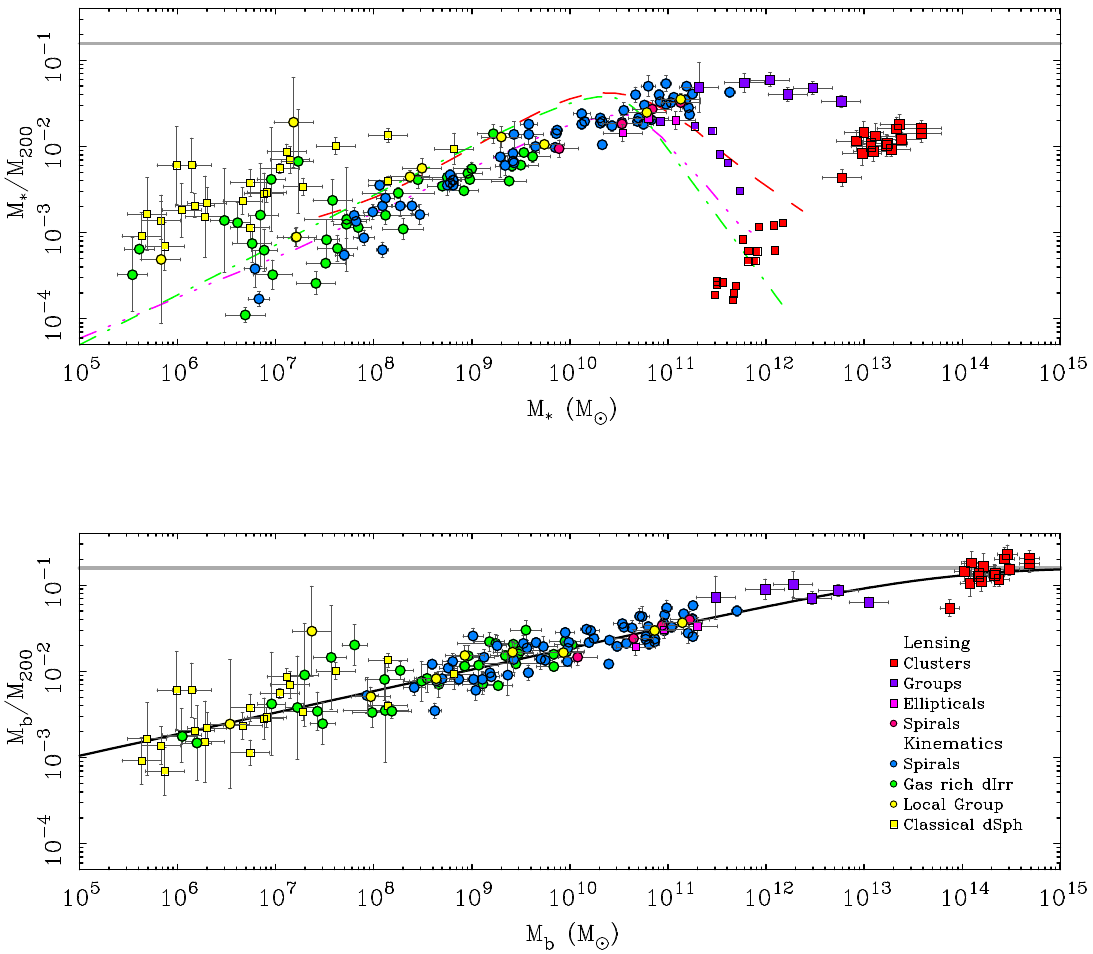}
\caption{The stellar mass fraction as a function of stellar mass (top) 
and the baryonic mass fraction as a function of baryonic mass (bottom). 
Data and symbols as in Fig.\ \ref{fig:VMb} with the additional distinction that large squares in the top panel represent  
the sum of the stellar mass of all galaxies in a group or cluster while small squares are the stellar mass of the brightest galaxy only.
The horizontal line is the cosmic baryon fraction $f_b = 0.157$ \citep{PlanckCosmology}.
The colored lines in the top panels show the stellar mass--halo mass relations from abundance matching given by \citet[dash-dotted green line]{Moster2013},
\citet[dash-triple dotted pink line]{Behroozi13}, and \citet[red dashed line]{KravtsovAM}.
The black line in the lower panel is eq.\ \ref{eq:KMM}.
\label{fig:barfrac}}
\end{figure*}

The stellar fraction $m_* = \Mst/\Mhalo$ and the measured baryon fraction $m_b = \Mb/\Mhalo$ are shown in Figure \ref{fig:barfrac}.
The halo mass is computed using \Vf\ in equation \ref{eq:NFWMV} with $f_v = 1$. 
Some of the groups and clusters are at modest redshift, so we account for the expected evolution of the intercept in the \Mhalo-\Vhalo\ relation \citep{haloevolution}. For the range of observed redshifts, this can be tolerably 
approximated as $\log B(z) = 5.508-0.256 z$. For the highest redshift cluster (MACSJ0744 at $z = 0.686$), this is a $0.18$ dex correction. 
Though modest, this is important to getting the baryon fraction right. 

\subsection{Kinematic Mass Matching Relation}
\label{sec:KMM}

Mathematically, the measured baryon fraction is well described by embedding the BTFR (eq.\ \ref{eq:btfr}) within a hyperbolic tangent that transitions from the monotonic rise of $m_b$ on galaxy scales 
to a constant value at high mass. Specifically, 
\begin{align}
m_b &= f_b \tanh\left(\frac{\Vf}{V_0}\right) \label{eq:KVV} \\
 &= f_b \tanh\left(\frac{\Mb}{M_0}\right)^{1/4} \label{eq:KMM}
\end{align}
where we have obliged the function to asymptote to the cosmic value $f_b = 0.157$ at high mass. 
It need not have been the case that the asymptotic value would agree so well with the cosmic value, but it does,
at least for our choice of \Mhalo\ for the halo mass and the extrapolation of the $\beta$-profile of the X-ray gas.
The turnover region is not well sampled by data, so the precise value of the parameter $M_0$ is not well constrained. 
Equation \ref{eq:KMM} is illustrated in Fig.\ \ref{fig:barfrac} with $\log(M_0/\Msun) = 13.7$, which is equivalent to $V_0 = 1000\;\kms$ in eq.\ \ref{eq:KVV}. 

Equation \ref{eq:KMM} works well for objects with $\Mb > 5 \times 10^5\;\Msun$ irrespective of whether they are individual galaxies or galaxy systems. 
It does not matter whether a galaxy system is a group or a cluster, nor whether an individual galaxy has its baryonic mass dominated by stars or gas. 
The data fall along the same relation with remarkably little scatter. 

{We can find no indication of second-parameter dependences:}
the baryon fraction {depends only on mass. It} does not depend on the star formation history, 
as both red and dead galaxies have the same baryon fraction as active star formers of the same mass,
{contrary to what is seen in some simulations \citep[e.g., Fig.\ 5 of][]{Onorbe2015}.}
This would seem to contradict feedback from star formation {as the driving reason for the variation in baryon fraction.} 
{Neither does the environment seem to play a role, as there is no segregation in circular velocity between 
star-dominated satellite galaxies and isolated gas-rich galaxies of the same baryonic mass (Fig.\ \ref{fig:VMb}).
It is difficult to find a satisfactory explanation for all aspects of the observations: the systematic variation with mass,
the small scatter, and the apparent absence of second-parameter effects.}

\subsection{Variations in Approach}
\label{sec:var}

\begin{figure*}
\centering
\includegraphics[width=0.95\textwidth]{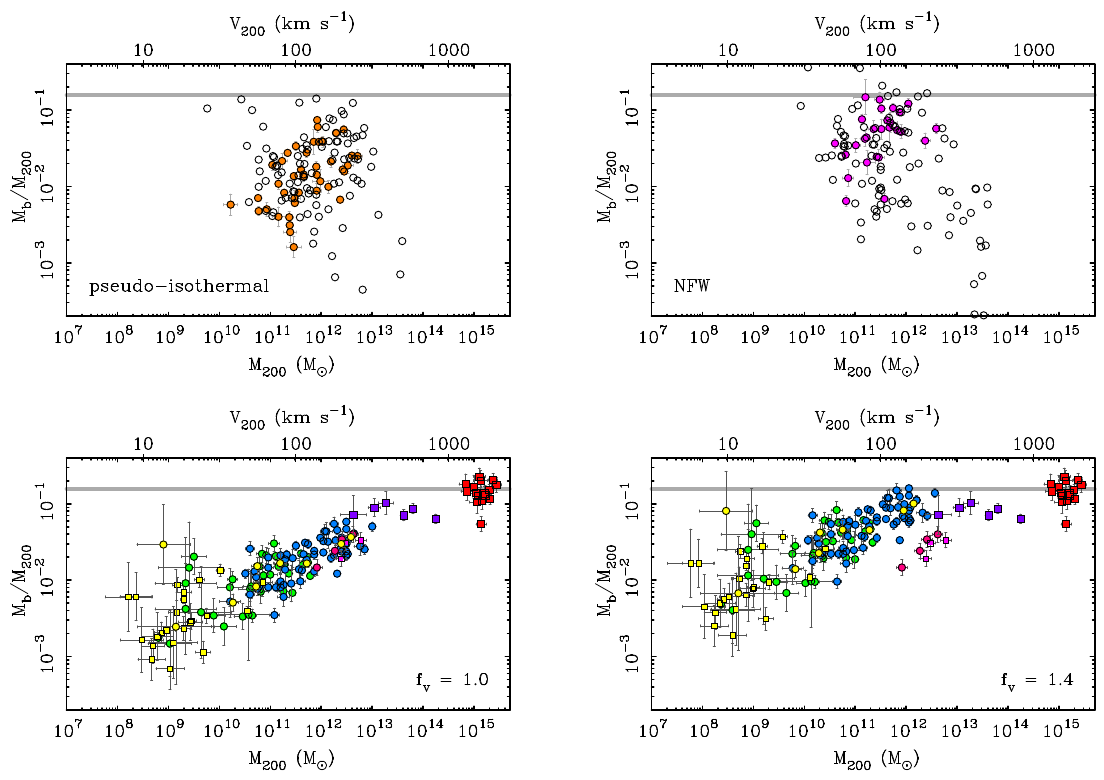}
\caption{The baryonic mass fraction as a function of mass \Mhalo\ with the equivalent \Vhalo\ noted on the top axis. 
Data and symbols are the same as in figures \ref{fig:M200} and \ref{fig:VMb}.
The top panels follow from using rotation curve fits to estimate halo mass as in Fig.\ \ref{fig:M200} for pseudo-isothermal (left) and NFW (right) 
halo models. The bottom panels follow from using the flat rotation speed as the halo 
mass estimator as in Fig.\ \ref{fig:VMb} with $f_v = 1.0$ (left) or $f_v = 1.4$ (right) for the kinematic data; the lensing data are unchanged. 
\label{fig:mbM200}}
\end{figure*}

{We have used the flat circular velocity to obtain the enclosed dynamical mass above, assuming $\Vhalo = \Vf$.
Other approaches are possible. In Fig.\ \ref{fig:mbM200} we illustrate the results if we make use of full rotation curve fits 
or different assumptions about the factor $f_v$ relating \Vf\ to \Vhalo.}

{The top panels of Fig.\ \ref{fig:mbM200} show the baryon fraction using results of fitting halo models to rotation curves \citep{Li2020}. 
As expected, the large scatter in Fig.\ \ref{fig:M200} translates into large scatter in Fig.\ \ref{fig:mbM200}. 
A trend of the baryon fraction with mass is perceptible, but only barely.
The large scatter is an artifact of the procedure: halo masses \Mhalo\ are not well constrained by detailed fits where the halo parameters
are degenerate with the mass-to-light ratio of the stars and with nuisance parameters like distance and inclination.
For the NFW case (top right panel), the baryon fraction sometimes reaches the cosmic value, so we corroborate the result of 
\citet{ManceraPina2025} in this regard, but note that this result depends on the specific halo choice and fitting procedure \citep{Katz2017,Li2020}. 
We do not concur with their interpretation that this explains the so-called diversity of rotation curves,
as the scatter here is an artifact.}

{The bottom panels use \Vf\ as the halo mass indicator. 
The flat rotation speed has long been assumed to be the signature of the dark matter halo; 
the reduction in scatter going from the top panels of Fig.\ \ref{fig:mbM200} to the bottom panels seems to bear that out. 
The lower left panel assumes $f_v = 1$, so is identical to the bottom panel of Fig.\ \ref{fig:barfrac} except for 
the abscissa showing halo properties rather than baryonic mass.
Since we adopt $\Vhalo = \Vf$ here, by construction these data are well described by eq.\ \ref{eq:KVV}.}

{The lower right panel of Fig.\ \ref{fig:mbM200} adopts $f_v = 1.4$ for galaxies with kinematic data as suggested in Fig.\ \ref{fig:M200}.
This provides a scale consistent with NFW halo fits \citep{Katz2017,Li2020} without incurring the penalty of the scatter therein. 
The data shift up relative to the lower left panel because $f_v > 1$ corresponds to a smaller dynamical mass than $f_v = 1$, 
so the baryon fraction $m_b = \Mb/\Mhalo$ goes up.
As in the top right panel, a few galaxies reach the cosmic value, though most do not. 
The turnover scale is lower,  $\log(M_0/\Msun) \approx 11.6$, but the asymptote to the cosmic baryon fraction is less clear.}

{The factor $f_v$ is present to relate the flat rotation speed observed in kinematic data at typical radii of tens of kpc with virial radii 
of hundreds of kpc. This does not apply to the lensing data, which themselves sample hundreds of kpc out \citep{indefinitelyflat}. 
We therefore leave $f_v = 1$ for the lensing data in the lower right panel of Fig.\ \ref{fig:mbM200}. 
This creates an artificial tension with the kinematic data that can be seen as an offset in the points with $\Mhalo \approx 10^{12}\;\Msun$. 
That this offset appears as an artifact of the choice of $f_v$ implies that $f_v \approx 1$ is preferred. }

\section{Discussion}
\label{sec:disc}

\subsection{Comparison with Abundance Matching}
\label{sec:AMcomp}

Equation \ref{eq:KMM} provides a relation between the observed baryonic mass and the halo mass indicated by kinematics and gravitational lensing.
A similar mapping between the visible and invisible is provided by stellar mass
by way of abundance matching \citep[e.g.,][]{Moster2013,Behroozi13,KravtsovAM}.
We compare\footnote{We define \Mhalo\ to include all mass enclosed within $r_{200}$ while many abundance matching relations refer to it as a parameter of the dark matter halo alone. The difference is small, so we make no attempt to reconcile it as this encounters the subtle difficulty that an unknown fraction of the mass in the halo may be the local missing baryons.} the two here.
 
There is reasonable agreement between the kinematic mass indicated by \Vf\ and the stellar mass--halo mass relations from abundance matching
over the range $7.7 \lesssim \Mst \lesssim 10.7\;\Msun$. The modest offset between the kinematic data and any one of the illustrated 
abundance matching relations can be trivially reconciled with a judicious choice of the velocity factor $f_v$ \citep{Posti2019}. 
However, the apparent agreement does not hold outside this mass range.

At high masses ($\Mst > 10^{11}\;\Msun$), the abundance matching relations diverge from the kinematic data. 
This is partly due to a difference in definitions. 
For the stellar mass, we include all the stars in all the galaxies in a group or cluster.
In contrast, abundance matching references the stellar mass of the brightest cluster galaxy (BCG) alone, with the notional identification of the BCG\footnote{\citet{KravtsovAM} adopt a definition that includes intracluster light as part of the central galaxy, which is one reason for its difference from the other abundance matching relations at high masses.} with the ``central'' galaxy of the entire cluster halo, the other cluster galaxies being satellites residing in subhalos. 
However, Fig.\ \ref{fig:barfrac} shows that for larger masses, the stellar mass of an individual galaxy is not a good indicator of the halo mass. 
At $\Mst \approx 4 \times 10^{11}\;\Msun$, a galaxy could be the central of a large cluster, a modest group, or be isolated. 
The corresponding stellar fraction $m_*$ ranges over two orders of magnitude from $m_* \approx 2 \times 10^{-4}$ for the
central of a rich cluster with $\Mhalo \approx 2 \times 10^{15}\;\Msun$ to $m_* \approx 3 \times 10^{-3}$  for a galaxy that is the central of a 
group with $\Mhalo \approx 10^{14}\;\Msun$ to $m_* \approx 0.04$ for an isolated spiral with $\Mhalo \approx 10^{13}\;\Msun$. 
Simply looking at the stellar mass of each of these galaxies does not suffice to indicate the halo mass.
Even around $\Mst = 10^{11}\;\Msun$ there is a large offset between the kinematically indicated mass and that expected from abundance matching \citep{McGvDk}.

At low masses ($\Mst < 10^{7.5}\;\Msun$), there is a bifurcation of $m_*$ between star-dominated and gas-dominated galaxies. 
The stellar fraction of the gas poor, early type dwarfs of the Local Group differs from that of gas rich star-forming dwarfs by a large factor at the same stellar mass. 
The latter are in approximate agreement with abundance matching relations based on field galaxy luminosity functions. 
However, application of such abundance matching relations 
to the classical dwarfs of the Local Group will result in halo mass estimates differing from the kinematic value by an order of magnitude. 

The bifurcation seen in $m_*$ between star-dominated and gas-dominated galaxies does not occur for $m_b$. 
Whatever physics drives this relation, it depends on the mass of baryons in galaxies, not whether those baryons are in the form of stars or gas. 
Neither does it depend on the details of the star formation history of each galaxy, as one might expect if feedback is the physics that determines $m_b$. 
Galaxies that formed their stars long ago and contain no gas now have the same baryon fraction as actively star forming, gas dominated galaxies.

\begin{deluxetable*}{lllccccl}
\tablewidth{0pt}
\tablecaption{{Matched Pairs of Early and Late Type Dwarfs  \label{tab:matched}}
}
\tablehead{
\colhead{Pair} & \colhead{Type} & \colhead{Galaxy} & \colhead{\Mst} & \colhead{$\Mg$} & \colhead{\Mb} & \colhead{\Vf} & Ref.  \\
& & &  \multicolumn{3}{c}{($10^{6}\;\Msun$)}  & \colhead{($\kms$)} &
}
\startdata
1 & ETG & NGC 185 & 140 & \dots & 140 & 48 & 1 \\
  & LTG & AGC 749237 & 53 & 77 & 130 & 49 & 2 \\
  \tableline
2 & ETG  & Fornax  &  41 & \dots & 41 & 23  & 1  \\
  & LTG & AGC 111946 & 17 & 20 & 37 & 20 & 2 \\
  \tableline
3 & ETG & And VII & 19 & \dots & 19 & 26 & 1 \\
  & LTG & AGC 112521 & 7 & 10 & 17 & 24 & 2 \\
  \tableline
4 & ETG & And XXXI & 8.1 & \dots & 8.1 & 21 & 1 \\
  & LTG & AGC 748778 & 3.0 & 6.1 & 9.1 & 19 & 2 \\
  \tableline
5 & ETG & Sculptor & 4.6 & \dots & 4.6 & 18 & 1 \\
  & LTG & DDO 210 & 0.7 & 2.7 & 3.4 & 16 & 3 \\
  \tableline
6 & ETG & Leo II & 1.5 & \dots & 1.5 & 13 & 1 \\
  & LTG & Leo P & 0.3 & 1.2 & 1.5 & 15 & 4 \\
  \tableline
7 & ETG & Tucana & 1.1 & \dots & 1.1 & 12 & 1 \\
  & LTG & KK153 & 0.4 & 0.7 & 1.1 & 13 & 5 \\
\enddata
\tablerefs{1.\ \citet{McG2021}. 2.\ \citet{SHIELD} 3. \citet{FIGGS}. 4.\ \citet{LeoProt}. 5.\ \citet{KK153}.}
\end{deluxetable*}

Indeed, there are examples of pairs of early and late type dwarfs that have indistinguishable baryonic masses and circular velocities, 
within the uncertainties. 
They occupy the same location in the bottom panel of Fig.\ \ref{fig:barfrac} despite their very different locations in the top panel. 
Example of such matched pairs are given in Table \ref{tab:matched}. 
Each of these pairs has essentially the same baryon fraction $m_b$ despite having very different stellar fractions $m_*$.
Regardless of the physical reason for the observed behavior, baryonic mass is a more fundamental quantity than is the stellar mass alone,
and a more reliable indicator of the dynamical mass.

\subsection{Comparison with Stars and Gas in the EAGLE Simulations}
\label{sec:eagle}

{In addition to establishing a kinematic relation analogous to abundance matching, 
we can also test the results of simulations that track baryons through their various phases.
Evaluating all simulations is beyond the scope of this work, so as an example we focus here on the the EAGLE simulations \citep{Schaye2015EAGLE}.
\citet{MitchellSchaye2022} have quantified the variation of the mass in various baryonic components as a function of halo mass.}

{The accounting of \citet{MitchellSchaye2022} includes stars, gas in the ISM, gas in the CGM, 
baryons that are ejected from their halos, and baryons that are prevented by feedback from accreting in the first place.
The CGM dominates in massive halos ($\Mhalo > 10^{13}\;\Msun$), the vast majority of baryons are prevented from accreting onto
low mass halos ($\Mhalo < 10^{11}\;\Msun$), and the mass of ejected baryons dominates in between.
At no point do stars dominate the simulated baryonic mass budget, and the gas of the ISM is always less than the stars.
These latter are the components we measure directly.}

\begin{figure*}
\centering
\includegraphics[width=0.95\textwidth]{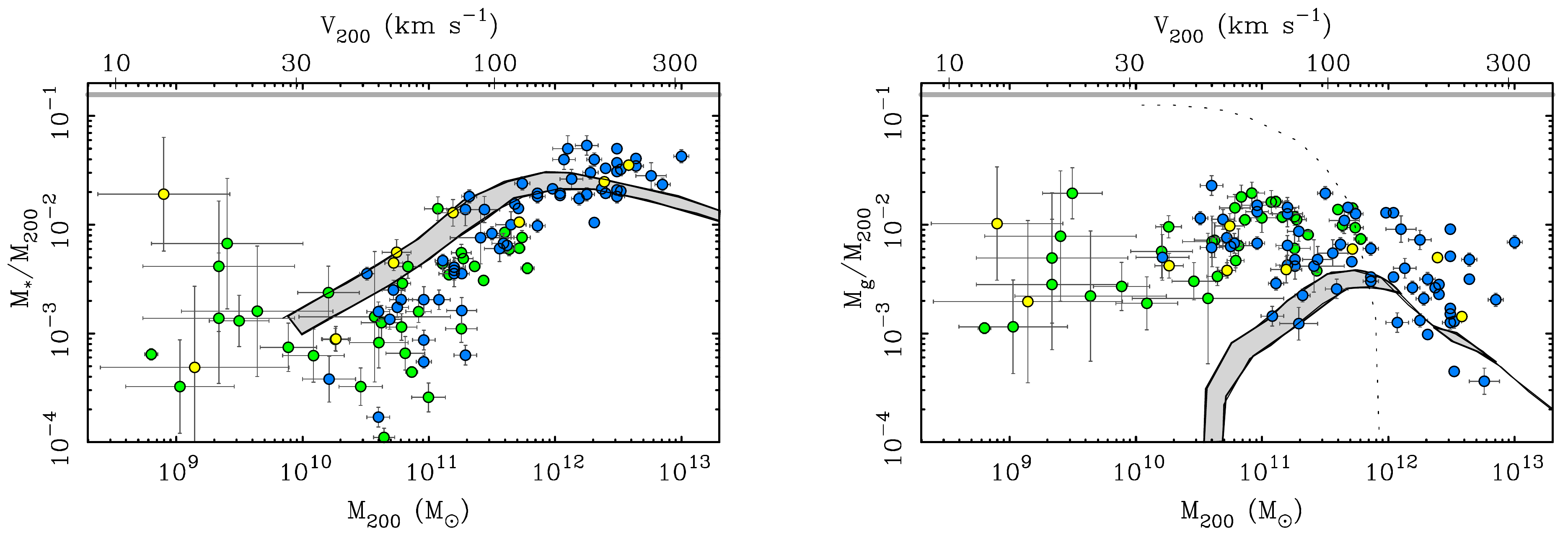}
\caption{The stellar mass fraction (left) and gas mass fraction (right) as a function of mass \Mhalo\ with the equivalent \Vhalo\ on the top axis. 
Data and symbols are the same as in Fig.\ \ref{fig:VMb}; 
the sample is restricted to objects for which both stars and gas are directly measured for each galaxy. 
The lines show the expectation for central subhalos in the EAGLE simulations \citep{MitchellSchaye2022} with the width of the gray
bands representing the range of $f_v$ from 1 (bottom edge) to 1.4 (top edge). 
The dotted line in the right panel denotes the limit where gas is precluded from accreting onto halos in the EAGLE simulations.
\label{fig:eagle}}
\end{figure*}

{The data and simulations are compared in Fig.\ \ref{fig:eagle}.
The stellar mass follows the same qualitative trend in both.
Quantitiatively, real galaxies have relatively more stars than anticipated at high mass. 
At intermediate masses the situation is reversed, with fewer stars per unit dark matter. 
The scale in Fig.\ \ref{fig:eagle} is logarithmic, so this is not a small difference, 
and the difference cannot be rectified with a variable $f_v$ within the plausible range of that parameter. 
The EAGLE simulations do not probe low masses ($\Mhalo < 10^{10}\;\Msun$). }

{As noted by \citet{Korsaga2023}, the gas mass-to-halo mass ratio is approximately constant in the data.
This behavior is not seen in the simulations, where the fraction of cold gas in the ISM peaks around $\Mhalo \sim 10^{12}\;\Msun$,
similar to the stars. Unlike the data, the gas mass fraction in the simulations declines rapidly to lower masses, 
and truncates abruptly at $\Mhalo \sim 10^{10.5}\;\Msun$.
This scale corresponds to where most accretion is prevented by feedback in the EAGLE simulations (the dotted line in Fig.\ \ref{fig:eagle}). 
This suppression of cold gas in simulated low mass objects is the opposite of what is seen in reality, 
as the majority of field galaxies are gas rich \citep[with $\Mg > \Mst$:][]{SFMS}.
Very low mass, gas rich galaxies exist, though it would seem that they should not \citep{LeoPunQ}. 
Feedback in the real universe \citep{Lelli2014feedback,McQuinn2019,Marasco2023} appears to be less prolific at expelling or excluding baryons 
than the implementation of feedback in EAGLE.}

{We have restricted Fig.\ \ref{fig:eagle} to the case of the EAGLE simulations \citep{MitchellSchaye2022} for clarity. 
Improvements in the simulation of gas physics as implemented in the successor simulations to EAGLE, COLIBRE \citep{Schaye2025}, 
might lead to a different answer, but a similar analysis to that of \citet{MitchellSchaye2022} has not yet appeared as of this writing. 
On a related topic, COLIBRE produces disequilibrium rotation curves in dwarf galaxies \citep{Dado2025} 
that are manifestly inconsistent with observational reality \citep{KdN09,KdN2011}, 
so it is not possible to compare these simulated entities with observed galaxies that have well-measured \Vf.}

{It is beyond the scope of this work to attempt an exhaustive exploration of all simulations.
These differ in the amount of mass that is expected in the CGM \citep{Wright2024MNRAS.532.3417W}, 
anticipate more scatter than is observed \citep{Medlock2025ApJ...980...61M}, 
and struggle with other aspects of gas in galaxies \citep{Marasco2025}. 
We thus return to what we can infer empirically from the data.}

\subsection{Missing Baryons}
\label{sec:missing}

The obvious initial expectation was that the observed baryon content would trace the halo mass, so $m_b \sim$ constant \citep[e.g.,][]{MMW98}. 
This is manifestly not the case in the data (Fig.\ \ref{fig:barfrac}). 
Where do the local baryons currently reside?
{Why is the variation with mass so systematic?}

\begin{figure*}
\plotone{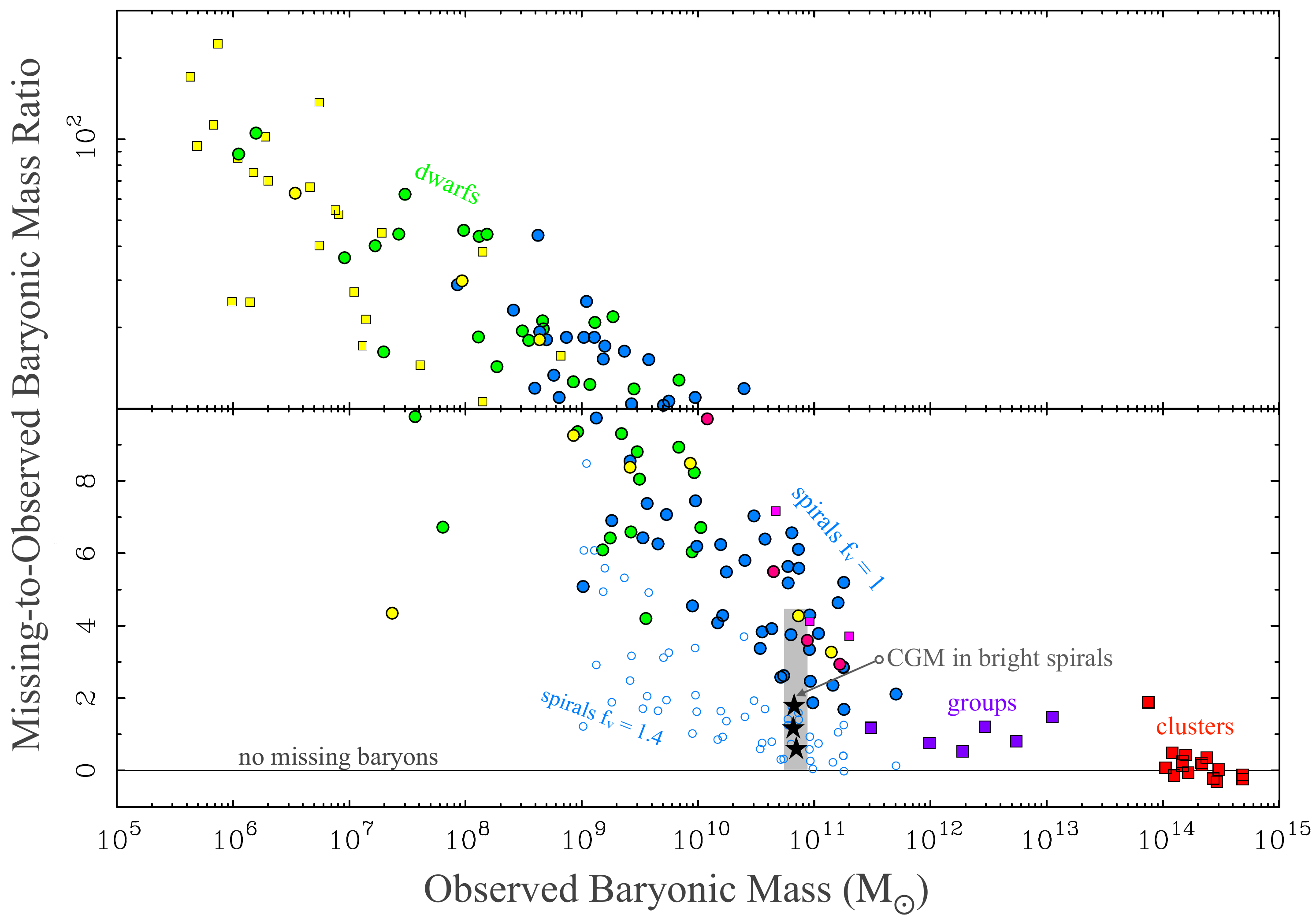}
\caption{The ratio of missing-to-observed baryonic mass as a function of baryonic mass. Data and symbols are the same as in Fig.\ \ref{fig:VMb}.
The ratio is linear in the bottom half of the diagram, then switches to logarithmic in the top half. 
Spiral galaxies of the WISE-SPARC sample are shown twice: once with $f_v = 1.0$ (solid blue circles) and again with $f_v = 1.4$ (small open circles).
The Milky Way is the yellow point at the top of the gray band, 
which shows the range from zero CGM to that required to explain all of the locally missing baryons when $f_v = 1$. 
Stars represent the CGM measurements of Milky Way mass galaxies 
by \citet{coronabaryons}, \citet{Bregman2022ApJ...928...14B}, and \citet{Zhang2025b} from bottom to top.
\label{fig:MCGM}}
\end{figure*}

Rich clusters of galaxies are the one type of system that meets the ideal $\Mb \approx f_b \Mhalo$. 
Averaging over the CLASH clusters analyzed by \citet{Mistele_CLASH}, we obtain $m_b = 0.154 \pm 0.038$, which is perfectly consistent with the cosmic baryon fraction \citep[$f_b = 0.157$:][]{PlanckCosmology}. 
This ideal is not realized in lower mass objects. 
The local baryon deficit is modest for groups and bright galaxies \citep{Bregman2018,Dev2024,ManceraPina2025,Zhang2025b}, 
{and might plausibly be explained by the CGM. This is not the case} for smaller masses. 
For every baryon that is detected in a galaxy with $\Mb \approx 10^9\;\Msun$, $\sim 10$ baryons are missing. 
For dwarfs with $\Mb \approx 10^7\;\Msun$, $\sim 50$ baryons are missing for every one that is observed (Fig.\ \ref{fig:MCGM}).

{To quantify the mass of missing baryons, we combine equations \ref{eq:Mmiss} and \ref{eq:KMM} to obtain}
\begin{equation}
\frac{\Mmiss}{\Mb} = \frac{1}{m_b} -1 = \frac{1}{f_b \tanh\left(\frac{\Mb}{M_0}\right)^{1/4}} -1
\label{eq:MXratio}
\end{equation}
{This ratio of missing-to-observed baryonic mass is plotted in Fig.\ \ref{fig:MCGM}. 
Subtracting quantities and plotting the data as a linear ratio amplifies the appearance of the scatter.
This is an artifact of error propagation; the source data themselves (the bottom panel of Fig.\ \ref{fig:VMb}) have remarkably little intrinsic scatter.
The small intrinsic scatter around the expression for $m_b$ in eq.\ \ref{eq:KMM} conveys to the ratio $\Mmiss/\Mb$ in eq.\ \ref{eq:MXratio}:
the mass in directly observed baryons \Mb\ is eerily predictive of the expected mass in missing baryons \Mmiss.}

There are {four possible solutions} for the locally missing baryon problem:
{(i) they are present but not accounted for, (ii) they are absent, having been ejected or precluded from accreting in the first place,
(iii) some systematic afflicts our mass estimates, or (iv) the \LCDM\ paradigm is incorrect. In the first case,} 
the unaccounted baryons may be lurking in the CGM of their parent dark matter halos \citep{Bregman2007,CGMreview}. 
{In the second case}, they would now reside in the IGM {and not be associated with any particular galaxy}. 
{In the} third, there may be something wrong with {how we relate observed velocity to halo mass} that leads us to infer missing baryons.
{Possibilities (i) -- (iii) might all play a role, occurring in some fine-tuned combination. 
The fourth case admits the possibility that the problem arises because the necessary cosmic assumptions are incorrect.} 
We discuss {each of these} possibilities {in turn}.

\subsubsection{CGM}
\label{sec:CGM}

{Perhaps the most obvious solution to the local missing baryon problem is for them not to be missing at all.
There is no guarantee that all the baryons in a halo cool and condense into the visible galaxy for which \Mb\ is well-measured.
Some may remain mixed in with the dark matter halo out to (or even a bit beyond) the nominal virial radius, 
perhaps in the form of the warm-hot gas of the circumgalactic medium \citep{Bregman2007,CGMreview,Bregman2018}.}

{Considerable progress has been made in detecting hot gas in the CGM of massive 
spiral galaxies \citep[e.g.,][]{Bregman2022ApJ...928...14B,Zhang2024a}.
The uncertainties in the integrated masses are large, so it is unclear whether the CGM is enough to make up the observed shortfall.
We illustrate the case of $\Mmiss = \MCGM$ in Fig.\ \ref{fig:MCGM} for three published measurements of Milky Way mass galaxies:
(1) the Milky Way coronal mass measurement of \citet{coronabaryons}, $\MCGM = 4.3 \times 10^{10}\;\Msun$, 
(2) the average of a dozen $L^*$ galaxies by \citet{Bregman2022ApJ...928...14B}, $\langle \MCGM \rangle = 9.8 \pm 2.8 \times 10^{10}\;\Msun$, and 
(3) the fiducial estimate of the CGM mass of Milky Way analogs from \citet{Zhang2025b}, $\MCGM = 1.26 \times 10^{11}\;\Msun$.
These would be sufficient to explain all of the missing baryons in the Milky Way if $f_v$ were in the range 1.25 -- 1.5.
None suffice if $f_v = 1.0$, for which $\Mmiss \approx 3.3 \times 10^{11}\;\Msun$.
The factor $f_v$ thus plays a critical role in determining whether the local missing baryon problem
appears to be solved by the CGM or not \citep[see also][]{McGvDk}.}

{The need for missing baryons in massive galaxies (Fig.\ \ref{fig:MCGM}) is greatly alleviated if $f_v \approx 1.4$.
This eliminates the need entirely in some cases \citep[][]{ManceraPina2025}, 
but for most spiral galaxies with $\Mb \approx 10^{11}\;\Msun$, we infer $\Mmiss \approx \Mb$. 
That is, roughly as many baryons are missing as observed.}

{If there are as many baryons in the CGM as in the central galaxy itself, as seems plausible, 
then this might also explain the offset of our group measurements: 
the missing mass in groups might simply be the CGM of the galaxies that are members of the group.
This interpretation seems satisfactory, but is not without issues. 
One is the offset between kinematic and lensing data that occurs when $f_v = 1.4$ is adopted (Fig.\ \ref{fig:mbM200}). 
Another is that the equivalence of the CGM in groups and individual galaxies assumes $f_v = 1.0$ for the former but $f_v = 1.4$ for the latter.
While this is possible in principle, there is no discontinuity in the data (Fig.\ \ref{fig:VMb}) before the notional factor $f_v$ is introduced, 
so this would simply inject an artifact.}

{Much of the discussion has focussed on bright galaxies with $\Mb > 10^{10}\;\Msun$.
For lower mass galaxies, $\Mmiss = \MCGM \approx \Mb$ does not suffice to explain} the local missing baryon problem. 
The {amplitude of the problem} is severe: galaxies with $\Mb = 10^{9}\;\Msun$ have $\Mmiss \gtrsim 10\, \Mb$.
{It gets worse to lower masses, sometimes reaching $\Mmiss \approx 100\, \Mb$ (Fig.\ \ref{fig:MCGM}).} 
It strains credulity to suppose that every dwarf galaxy hosts a CGM an order of magnitude more massive than itself, 
regardless of whether it is energetically possible \citep{Katz2018}. 

\subsubsection{IGM}
\label{sec:IGM}

{Another possibility for the locally missing baryons is that they have been} 
expelled from their host galaxies by feedback processes \citep{Silk2003MNRAS.343..249S},
{or precluded from accreting in the first place \citep{MitchellSchaye2022}. 
In either case, the missing baryons we infer would reside in the IGM.}
\citet{Connor2025} estimate that the fraction of baryons in the IGM is $\sim 3/4$ while only $\sim 1/2$ of the dark matter is there. 
This is rather uncertain, but implies that some baryons have been expelled {or excluded} from galaxies. 
{Moreover, some expulsion is required to account for the metallicity of the IGM \citep{Danforth2006,Scannapieco2006}.}

The {observed} trend of {decreasing baryon fraction with decreasing mass} 
is {suggestive of a role for binding energy:} baryons {may} escape from smaller potential wells {with} increasing ease. 
This sounds plausible, but does not, by itself, provide an explanation for the particular trend that we observe. 
Supernova feedback is often invoked in this context, 
but the stochastic explosions of massive stars seem more likely to induce scatter than a smooth variation. 
{Indeed, there is a great deal of scatter in some simulations \citep[e.g.,][]{Medlock2025ApJ...980...61M}.
In the EAGLE simulations \citep{MitchellSchaye2022}, the majority of baryons are ejected from galaxy mass ($11 < \Mhalo < 13\;\Msun$) halos, 
but with little mass dependence: the depth of the potential wells is not the relevant effect.}

Searches for evidence for feedback in dwarf galaxies have routinely obtained results \citep{Bothun1994,Lelli2014feedback,Concas2019} 
that contradict the expectation of vigorous outflows that could liberate the majority of baryons from their host halos \citep{Leroy2015,McQuinn2019}.
To be sure there is some feedback, but {rather than strong} winds, {there is} more of a light breeze \citep{Concas2017,Marasco2023} that 
does not suffice to unbind the entrained baryons and deliver them to the IGM \citep{Katz2018}. 
The same is observed at cosmic noon when star formation rates were much higher \citep{Concas2022}.

The mass dependence of the baryon deficit is very regular: $m_b \propto \Vf \sim \Mb^{1/4}$ with remarkably little scatter.
{This poses} a fine-tuning problem.
{How do halos of the same mass always eject the same fraction of baryons?}

It is, of course, possible to have the missing baryons reside in some combination of both the CGM and the IGM. 
{However, invoking a combination of mechanisms exacerbates the amount of fine-tuning that is required.
It does not suffice to say} that some baryons remain in the CGM while others get ejected to in the IGM;
{we must also explain the variation with mass and the small scatter. 
\citet{RuanBrooks2025} appear to do this over a modest mass range, 
but their results is contingent on an abrupt change in the behavior of $f_v$ around $\log\Mb \sim 8.5$ (their Fig.\ 7). 
This is not apparent in the data, which are smooth and continuous across this mass (Fig.\ \ref{fig:VMb}), so it appears that they
have simply transferred the fine-tuning from one parameter ($m_b$) to another ($f_v$).}

\subsubsection{The Velocity Factor}
\label{sec:fv}

So far, we have inferred variation in the baryon fraction assuming the observed rotation speed is a faithful indicator of the halo mass. 
Here we consider whether variation in the velocity fraction $f_v$ could avoid the inference of the local missing baryon problem. 

\begin{figure*}
\plotone{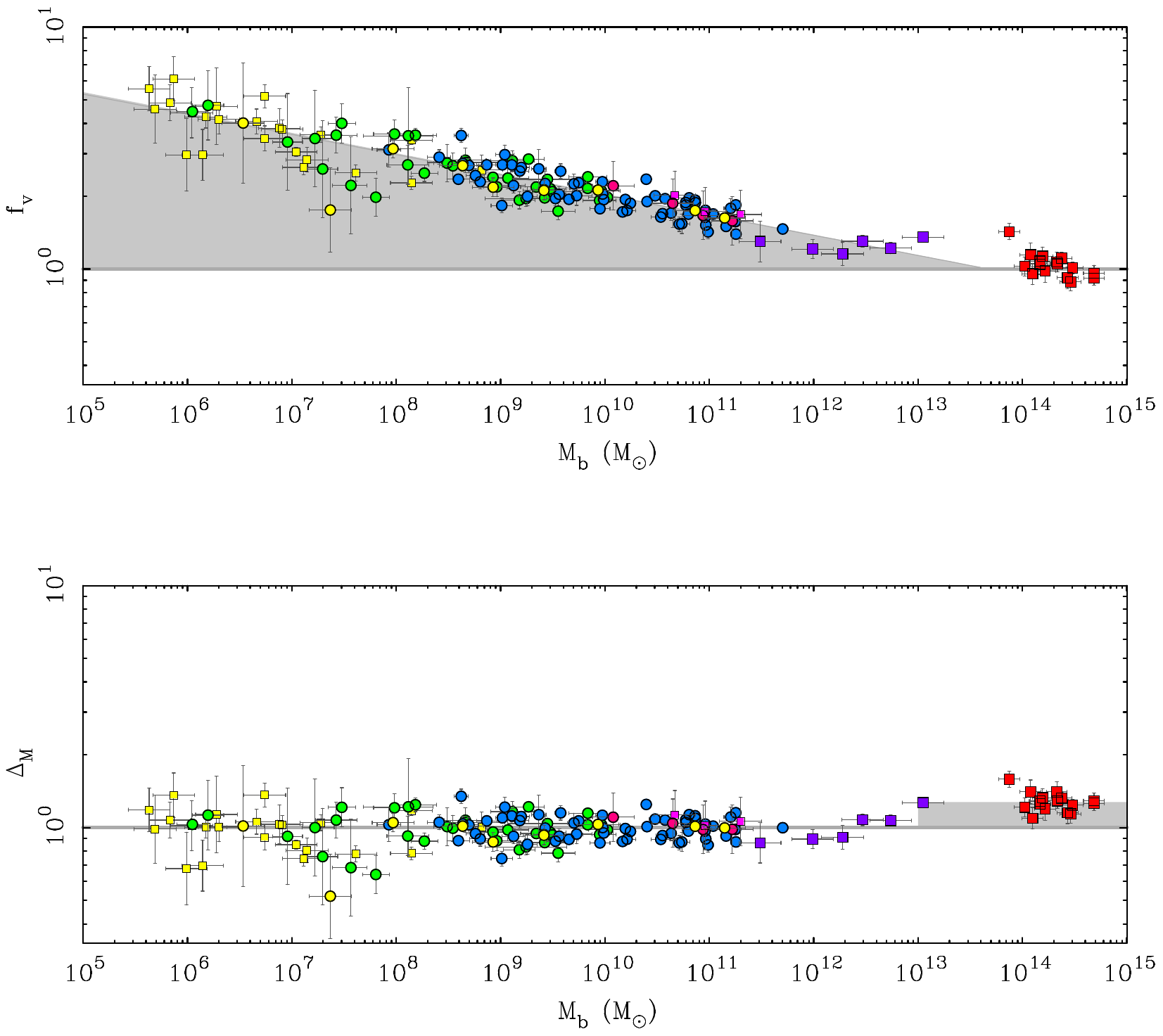}
\caption{The velocity factor in \LCDM\ (top) and the residual velocity in MOND (bottom) as a function of baryonic mass. 
Data and symbols as in Fig.\ \ref{fig:VMb}. 
The gray region illustrates where each theory gets it wrong. In \LCDM, rich clusters of galaxies behave as expected, 
but explaining other systems requires $f_v \propto \Mb^{-1/12}$ (eq.\ \ref{eq:Mfv}). In MOND, we see the opposite effect: 
the observed baryons predict the correct velocity over the range $5 \times 10^5 \le \Mb < 10^{13}\;\Msun$, 
but falls short for larger masses, explaining only $\sim 80\%$ of the velocity in clusters. 
\label{fig:linbarfrac}}
\end{figure*}

For illustration, we fix $m_b = f_b$ so that only $f_v$ varies.
Rewriting equations \ref{eq:mbV} and \ref{eq:mbM}, inserting the adopted values of $\log A = 1.7$ and $\log B = 5.5$ and setting $m_b = f_b = 0.157$, we have
\begin{align}
\log f_v &= 1 - \frac{1}{3} \log \Vf \label{eq:Vfv} \\
 &= 1.14 - \frac{1}{12} \log \Mb. \label{eq:Mfv}
\end{align}
These hold for $5.7 < \log\Mb < 13.68$. 
For $\log \Mb \ge 13.68$, $f_v = 1$. 
Note that we could adopt a broken power law like this for equations \ref{eq:KVV} and \ref{eq:KMM} or instead employ a hyperbolic tangent here as done there;
lacking sufficient data to constrain the shape of the transition, either choice is valid at present. 

In this fashion, it is mathematically possible to make the local baryon problem go away 
if and only if the velocity factor varies in the particular way given by eq.\ \ref{eq:Mfv}. 
This is illustrated in Fig.\ \ref{fig:linbarfrac}, which shows how $f_v$ must increase as mass decreases. 
The gray region illustrates the deficit that we must attribute to missing baryons if $f_v$ is constant. 

Attributing all the variation in the data to $f_v$ is not viable \citep{Posti2019}. 
The magnitude of the required variation is too large, reaching $f_v \approx 4$ at low mass ($\Mb \approx 3 \times 10^6\;\Msun$).
Such large $f_v$ are inconsistent with rotation curve fits: $f_v < 2.4$ for all NFW fits (Fig.\ \ref{fig:M200}), and is usually much less. 
Worse, large $f_v$ means that the observed velocity greatly exceeds that of the dark matter halo, 
contradicting the basic explanation of flat rotation curves. 
Some of the variation might be due to $f_v$, but it seems unlikely to be the dominant effect.
{Having both $m_b$ and $f_v$ vary implies a fine-tuning between them in order to maintain the smooth observed variation.} 

\subsubsection{Paradigm Failure}
\label{sec:mond}

{The fourth possibility is that the local missing baryon problem arises because of a faulty assumption in the underlying paradigm.
\LCDM\ predicts a mass--velocity relation of the form $\Mhalo \propto \Vhalo^3$ \citep{MMW98,SN} while we observe something closer to
$\Mb \propto \Vf^4$ (Fig.\ \ref{fig:VMb}). The inference that the fraction of missing baryons grows systematically greater with declining mass follows from 
the mismatch in slopes. This mismatch induces a curious result in the context of hierarchical galaxy assembly.}

{In the cold dark matter cosmogony, small galaxies merge to form larger ones.} 
Building large galaxies from small ones seems inconsistent with the observation that larger objects have higher baryon fractions. 
If baryons are ejected from small galaxies, they are not available to subsequently build larger galaxies. 
This incurs a teleological dilemma: 
a small halo needs to `know' if it will remain unmerged so that it should eject the `right' fraction of baryons for its mass 
or if it will later merge into a larger object so that it should retain the right amount to contribute to that higher baryon fraction object. 

{Ejection of baryons is not the only possible mechanism, but the concern holds regardless: 
there is a fine-tuning with mass that is not natural to hierarchical structure formation.
Galaxies are the result of many mergers, so this puzzling fine-tuning occurs repeatedly,
along every branch of every merger tree, each of which has many nodes. 
It is not obvious that} a complex hierarchical merger history {can} lead to the uniform cold baryon fraction we observe at low redshift.


{There is a disturbing reason underlying the mismatch in power-law slopes: the bulk of the data look like MOND \citep{MOND}.}
MOND is a modification of the force law suggested as an alternative to dark matter. 
It has had a remarkable number of a priori predictions {corroborated by subsequent observations} \citep{MdB98b,SMmond,LivRev}, 
many of which were surprising \citep{MdB98a,banikzhao2022}.
These often lack a satisfactory explanation in terms of dark matter \citep{M20}. 
{Strangely}, the data do exactly as MOND predicts over seven decades in mass, $5.7 < \log(\Mb/\Msun) < 13$ (Fig.\ \ref{fig:linbarfrac}).

There is no missing baryon problem in MOND for individual galaxies.  
What you see is what you get. 
This works so well that the mass of stars and gas within the 
confines\footnote{See Appendix \ref{sec:appendix:cgminmond} for a discussion of the CGM in MOND.} 
of the visible galaxy can be predicted by observing \Vf, or vice-versa (Fig.\ \ref{fig:VMb}).
There is no dark matter model that performs so well \citep{LivRev}.

{The success of MOND persists for groups of galaxies analyzed via dynamics \citep{MotiGroups2018,MotiGroups2019} and
for the groups of galaxies analyzed here via lensing, with the exception of the most massive bin with $\Mb \approx 10^{13}\;\Msun$
(which is also an outlier in \LCDM). 
It does not hold for groups with X-ray emission \citep{angusbuote} or for rich} clusters \citep{sanders2003,PengfeiClusters,Mistele_CLASH}. 
For the highest mass decade in Fig.\ \ref{fig:linbarfrac}, $14 < \log\Mb < 15$, MOND underpredicts the velocity: 
the observed baryons only explain $79 \pm 7\%$ of what is observed. 
Since $\Mb \propto \Vf^4$ in MOND, this offset in velocity is equivalent to a discrepancy of a factor of 2.3 in mass.
This residual baryon discrepancy in MOND has been persistent \citep{SMmond,KelleherLelli2024}, 
rather like the persistent absence of dark matter detections \citep{LUX2025wimplimit}.

It is tempting to assert that MOND is falsified for its failure in clusters, and that may be correct \citep{McGaugh_2024}. 
However, it is hard to preclude the discovery of additional mass there, 
just as it is hard to preclude the existence of dark baryons in the halos of individual galaxies, as needed in \LCDM.
Both theories suffer a missing baryon problem; Fig.\ \ref{fig:linbarfrac} {illustrates how they compare}. 

More disturbing philosophically is that the local missing baryon problem arises in \LCDM\ \textit{because} the data look like MOND. 
The baryon fraction in galaxies could have been pretty much anything less than the cosmic fraction. 
It could increase or decrease with mass, or not vary with mass at all, as we originally expected. 
The scatter from galaxy to galaxy could be huge, 
and arguably should be if driven by stochastic feedback processes \citep{Medlock2025ApJ...980...61M}. 
In contrast, MOND predicts a unique behavior, and the data do exactly the one thing that MOND predicts over seven decades in mass. 
Why does so much of a \LCDM\ universe look like MOND? 

\section{Conclusions}
\label{sec:conc}

{We have combined kinematic and gravitational lensing data for galaxies and systems of galaxies
to quantify the variation of the baryon fraction with mass.
The data extend over a large dynamic range, from tiny dwarf galaxies to the largest clusters. 
We present novel gravitational lensing measurements for galaxy groups that help to inform the ill-constrained   
mass range between individual galaxies and rich clusters of galaxies.} 

{We find a strong correlation between observed mass and the flat circular velocity \Vf\ of the gravitational potential for objects with 
$\Vf > 10\;\kms$ (Fig. \ref{fig:VMb}). This is a Tully-Fisher-like relation that extends to galaxy groups and clusters of galaxies.
The stellar mass \Mst--\Vf\ relation is strong for high mass galaxies but exhibits scatter at low mass 
and a bifurcation between gas-rich and gas-poor galaxies (Fig. \ref{fig:VMb}). 
The baryonic (stars plus cold gas) mass \Mb--\Vf\ relation exhibits no bifurcation and little intrinsic scatter.
Indeed, there are no second-parameter effects apparent in the \Mb--\Vf\ relation. In this respect, 
the observed baryonic mass of a galaxy is a more fundamental quantity than its stellar mass.}

{The data are well described by $\Mb/\Mhalo = f_b \tanh (\Mb/M_0)^{1/4}$ 
where $f_b$ is the cosmic baryon fraction and $M_0 \approx 5 \times 10^{13}\;\Msun$ (eq.\ \ref{eq:KMM}). 
This kinematic mass matching relation provides a more precise estimator 
of dynamical mass than do stellar mass--halo mass relations obtained from abundance matching.
There is some qualitative agreement between kinematic mass matching and abundance matching, but also significant
quantitative tension between the two at both low and high mass (Fig.\ \ref{fig:barfrac}).} 


{The data imply a local missing baryon problem ($m_b < f_b$) for most systems. 
Rich clusters of galaxies are the exception, with $m_b = 0.154 \pm 0.038$, in good agreement with the cosmic baryon fraction, $f_b = 0.157$.
Less massive systems, from groups through giant galaxies to dwarfs, all have fewer detected baryons than the cosmic fraction.
The amplitude of this discrepancy grows with decreasing mass. 
For galaxy groups and bright ($\Mb \approx 10^{11}\;\Msun$) galaxies, the discrepancy is modest (a factor of two or three),
so might plausibly be explained by the mass of gas in the CGM.
However, the problem exceeds an order of magnitude for low mass ($\Mb \lesssim 10^9\;\Msun$) galaxies where 
it strains credulity to imagine that $> 90\%$ of the local missing baryons reside in the CGM while evading detection.}

{Galaxies with very different star formation histories have indistinguishable baryon fractions at the same mass (Table \ref{tab:matched}). 
This is not naturally explained by feedback that is driven by star formation for which one would expect a dependence on the current star 
formation rate or some time-integral over its variation. Simulations that include feedback have yet to successfully match the data (Fig.\ \ref{fig:eagle}). 
Indeed, the increase of baryon fraction with increasing mass seems contradictory to hierarchical galaxy formation, as merging 
low mass galaxies with low baryon fractions would lead to larger galaxies with the same baryon fractions, 
not the higher baryon fractions that are observed.
Fine-tuning thus seems unavoidable for \LCDM\ models to simultaneously explain the observed variation of the baryon fraction with mass, 
its small intrinsic scatter, and its lack of dependence on star formation rate, gas fraction, or environment.}

{Where \LCDM\ suffers a local missing baryon problem in galaxies, MOND suffers one in rich clusters with $\Mb > 10^{13}\;\Msun$
(Fig.\ \ref{fig:linbarfrac}).
However, the data over the range $5 \times 10^5 < \Mb < 10^{13} \;\Msun$ are more naturally explained by MOND.
No tuning is necessary: a relation of this form was predicted a priori \citep{milgrom83}. 
We are not aware of a viable explanation for why so many of the predictions of MOND \citep{LivRev,banikzhao2022} are realized in a universe made of dark matter.}

\begin{acknowledgements} 
{We thank the referee for comments that improved the presentation of these results. 
We are grateful to Joel Bregman for many detailed insights into the issues concerning the CGM.}
This work is based in part on observations made with ESO Telescopes at the La Silla Paranal Observatory under programme IDs 177.A-3016, 177.A-3017, 177.A-3018 and 179.A-2004, and on data products produced by the KiDS consortium. The KiDS production team acknowledges support from: Deutsche Forschungsgemeinschaft, ERC, NOVA and NWO-M grants; Target; the University of Padova, and the University Federico II (Naples).
GAMA is a joint European-Australasian project based around a spectroscopic campaign using the Anglo-Australian Telescope. The GAMA input catalogue is based on data taken from the Sloan Digital Sky Survey and the UKIRT Infrared Deep Sky Survey. Complementary imaging of the GAMA regions is being obtained by a number of independent survey programmes including GALEX MIS, VST KiDS, VISTA VIKING, WISE, Herschel-ATLAS, GMRT and ASKAP providing UV to radio coverage. GAMA is funded by the STFC (UK), the ARC (Australia), the AAO, and the participating institutions. The GAMA website is https://www.gama-survey.org/~. 
\end{acknowledgements} 

\appendix
\label{sec:App}

\section{Groups}
\label{sec:appendix:groups}

We consider groups from GAMA-II \citep{Driver2011,Liske2015}, specifically from version 10 of the 
`G3CFoFGroup' and `G3CGal' tables.
Following \citet{Rana2022,Li2024,Liu2024,Viola2015}, we select groups with at least 5 members.
We further enforce that the `BCG' of each group has a stellar mass estimate in the `StellarMassesLambdar' table.
This leaves a total of 2751 groups.
We split these into 6 bins by $M_b$, with bin edges $\log_{10} M_b/M_\odot = [11.0, 11.75, 12.2, 12.35, 12.6, 12.9, 13.3]$.

\subsection{Baryonic and stellar masses}

We assign a stellar and baryonic mass to each group using the following procedure.
We start with the stellar masses of the member galaxies from the `StellarMassesLambdar' table version 24 \citep{Taylor2011}, applying the flux corrections listed in version 20 of that table, if these are not smaller than $1$.
Following \citet{lensRAR,indefinitelyflat}, we correct the stellar masses of early-type galaxies by a factor of $1.4$, where we define early-type galaxies as having $u-r > 2.2$ \citep{Strateva2001}, using the columns `fitflux\_u' and `fitflux\_r' of the `StellarMassesLambdar' table.

Following \citet{lensRAR,indefinitelyflat}, we convert the stellar masses of the member galaxies into baryonic masses by adding gas according to two simple scaling relations.
For early-types, we add a hot gas component \citep{ChaeEFE},
\begin{align}
 \label{eq:fhot}
 \frac{M_{g,\mathrm{hot}}}{M_\ast} = 10^{-5.414} \cdot \left(\frac{M_\ast}{M_\odot}\right)^{0.47} \,.
\end{align}
For late-type galaxies, we add a cold gas component \citep{Lelli2016,McGaugh2020b}
\begin{align}
 \label{eq:fcold}
 \frac{M_{g,\mathrm{cold}}}{M_\ast} = \frac{1}{X} \left(11550 \left(\frac{M_\ast}{M_\odot}\right)^{-0.46} + 0.07 \right) \,,
\end{align}
where
\begin{align}
 \label{eq:metal}
 X = 0.75  - 38.2 \left(\frac{M_\ast}{1.5\cdot10^{24}M_\odot}\right)^{0.22} \,.
\end{align}
The first term in eq.~\eqref{eq:fcold} represents atomic gas, the second term takes into account molecular gas.
Equation~\eqref{eq:metal} accounts for the variation of the hydrogen fraction $X$ as metallicity varies with stellar mass \citep{McGaugh2020b}.

A few member galaxies do not have a stellar mass listed in `StellarMassesLambdar'.
For simplicity, we set their stellar and baryonic masses to zero.
This occurs in only a handful of cases when many hundreds of groups inform each mass bin, so the effect on our results is negligible.

We then sum up the members' stellar masses to obtain the stellar mass of the groups.
We apply the flux correction from \citet{Robotham2011} for $N_{\mathrm{FoF}} \geq 5$.
This is only a small correction in the lower $M_b$ bins (due to the relatively small redshift), but becomes a factor of a few in the highest $M_b$ bin (due to the higher redshift).
We follow the same procedure to assign baryonic masses to the groups.
Strictly speaking, the flux correction factor from \citet{Robotham2011} does not apply directly to the baryonic masses, but we expect that it gives at least a first reasonable approximation.

\subsection{$\Vf$ from weak lensing}

We use weak gravitational lensing observations to measure stacked mass profiles $M(r)$ of the galaxy groups and infer a flat circular velocity $\Vf$ from these.

\subsubsection{Excess surface density}

We first use ellipticities from the KiDS-1000 SOM-gold source galaxy catalog \citep{Kuijken2019,Wright2020,Giblin2021,Hildebrandt2021} to measure the so-called excess surface density $\Delta \Sigma$ around each group, i.e. around each lens $l$,
\begin{equation}
\label{eq:ESD_measured}
\Delta \Sigma_l (R) = \frac{1}{1+\mu} \frac{\sum_s W_{ls} \Sigma_{\mathrm{crit},ls} \epsilon_{t,ls}}{\sum_s W_{ls}} \,.
\end{equation}
Here, the $W_{ls}$ are weights, $\epsilon_{t,ls}$ denotes the tangential ellipticity of the source $s$ with respect to the lens $l$, $\Sigma_{\mathrm{crit},ls}$ is the critical surface density, and $1+\mu$ corrects for multiplicative biases.
The sums run over all source galaxies $s$ within a given radial bin around the group.
The critical surface density is calculated following \citet{Brouwer2021,Dvornik2017} and takes into account uncertainties in the source redshifts,
\begin{align}
 \Sigma^{-1}_{\mathrm{crit},ls} = \frac{4 \pi G_{\mathrm{N}}}{c^2} D(z_l) \int_{z_l}^{\infty} d z_s \, n_{ls}(z_s) \cdot \frac{D(z_l, z_s)}{D(z_s)}  \,.
\end{align}
The function $n_{ls}(z)$ is determined as follows.
For a given $z_l$ and a given photometric source redshift $z_{B,s}$, we check in which of the five tomographic bins from \citet{Hildebrandt2021} the $z_{B,s}$ value belongs, get the corresponding redshift distribution function from \citet{Hildebrandt2021}, and normalize this distribution to unity in the interval $[z_l, \infty)$.
The weights $W_{ls}$ are given by $W_{ls} = w_s \Sigma_{\mathrm{crit},ls}^{-2}$ where $w_s$ estimates the precision of the ellipticity measurement of the source $s$ \citep{Brouwer2021,Giblin2021}.
We adopt $1+\mu = 0.98531$ from \citet{Brouwer2021}. 
We use 13 logarithmic radial bins with the smallest bin edge being $0.1\,\mathrm{Mpc}$ and with a logarithmic bin width of $1/7.5$.

\subsubsection{Mass profile}
\label{sec:appendix:groups:mass}

We convert the excess surface density $\Delta \Sigma_l (R)$ into a mass profile $M_l(r)$ using the non-parametric method from \citet{indefinitelyflat,lensRAR,Mistele2024b}, which is based on the following deprojection formula,
\begin{equation}
 \label{eq:M_from_ESD}
 M_l (r) = 4 r^2 \int_0^{\pi/2} d \theta \, \Delta \Sigma_l \left(\frac{r}{\sin \theta}\right) \,.
\end{equation}
To evaluate this integral, we must know $\Delta \Sigma_l (R)$ at all radii, up to $R=\infty$.
In practice, we measure $\Delta \Sigma_l (R)$ only up to a finite maximum radius $R_{\mathrm{max}}$ and only in discrete radial bins, so we need to both interpolate and extrapolate $\Delta \Sigma_l$. 
We interpolate linearly and extrapolate assuming that $\Delta \Sigma_l$ is proportional to $1/R$ beyond $R_{\mathrm{max}}$.

As discussed in \citet{indefinitelyflat,lensRAR}, the precise choices of how to interpolate and extrapolate are relatively unimportant over the bulk of the radial range and become important only close to $R_{\mathrm{max}}$.
This is illustrated by the blue band in Figure~\ref{fig:groupVc}, which shows the range of masses obtained by making the opposite and extreme choices of extrapolating $\Delta \Sigma_l$ assuming its decay is proportional to $1/R^2$ (as fast as for a point mass) and $1/\sqrt{R}$ (slower than for a singular isothermal sphere).
The extrapolation is unimportant in the radial range we use for $\Vf$ (see below).

\subsubsection{Miscentering}
\label{sec:appendix:groups:miscentering}

In practice, the coordinates of the group centers are not known very accurately.
We adopt the luminosity-weighted center of mass (`CenRA' and `CenDEC' in `G3CFoFGroup'), but this does not necessarily reflect the true center of the gravitational potential.
As a result, $\Delta \Sigma_l (R)$, and therefore $M_l (r)$, are often underestimated at small radii.
\citet{Mistele2024b} have derived a formula that allows one to approximately and statistically correct for miscentering, given that one knows the average miscentering offset.
The formula applies at radii larger than the miscentering offset and corresponds to replacing $\Delta \Sigma_l(R)$ by\footnote{
 We assume that the convergence $\kappa$ is negligible for groups, so that we can replace $G_+$ by $\Delta \Sigma$ in the formulas from \citet{Mistele2024b}.
}
\begin{equation}
\label{eq:ESD_miscent_corr}
\Delta \Sigma_l(R) + \frac14 \frac{\langle R_{\mathrm{mc}}^2 \rangle}{R^2} \Delta_{\mathrm{mc}} (R) \,,
\end{equation}
where $R_{\mathrm{mc}}$ is the miscentering offset and
\begin{equation}
 \label{eq:Deltamc}
\Delta_{\mathrm{mc}}(R) \equiv (4 - R \partial_R - R^2 \partial_R^2) \Delta \Sigma_l (R) \,.
\end{equation}
Since we linearly interpolate the observed $\Delta \Sigma_l$ (see Appendix~\ref{sec:appendix:groups:mass}), the second-order derivative in eq.~\eqref{eq:Deltamc} is not well-defined.
However, this second-order derivative is ultimately integrated over in eq.~\eqref{eq:M_from_ESD} so that we can use integration by parts to get rid of it.
Indeed, instead of replacing $\Delta \Sigma_l$ with eq.~\eqref{eq:ESD_miscent_corr} we can keep $\Delta \Sigma_l$ as is and replace eq.~\eqref{eq:M_from_ESD} with the following deprojection formula,
\begin{multline}
 M_l(r) = 4 r^2 \int_0^{\pi/2} d\theta
 \left(
 \Delta \Sigma_l \left(\frac{r}{\sin \theta}\right)
 \right. \\ \left.
  + \frac14 \frac{\langle R_{\mathrm{mc}}^2 \rangle}{r^2} (I_1 + I_2 +I_3)
  \right) \,,
\end{multline} 
where
\begin{align}
I_1 &= 4 s^2_\theta \, \Delta \Sigma_l \left(\frac{r}{s_\theta}\right) \,,
\label{eq:M_from_ESD_miscent_corr} \\
I_2 &= -(t^2_\theta + 2 s^2_\theta) \left(
    \Delta \Sigma_l \left(\frac{r}{s_\theta}\right) - \Delta \Sigma_l \left(r\right)
   \right) \,,
\notag \\
I_3 &= -\left(s_\theta + \frac{t_\theta}{c_\theta}\right) \left(
    r \Delta \Sigma_l' \left(\frac{r}{s_\theta}\right) - r \Delta \Sigma_l' \left(r\right)
   \right) \,.
\notag
\end{align}
Here, a prime denotes a derivative of $\Delta \Sigma_l$ with respect to its argument and we used the shorthand notation $s_\theta = \sin \theta$, $c_\theta = \cos \theta$, and $t_\theta = \tan \theta$.
Importantly, there are no longer any second-order derivatives so that we can apply this formula to our linearly-interpolated $\Delta \Sigma_l$.

As discussed in Appendix~\ref{sec:appendix:groups:mass}, in practice, we extrapolate $\Delta \Sigma_l$ beyond the last data point at $R_{\mathrm{max}}$.
To implement that, we split the integral in eq.~\eqref{eq:M_from_ESD} at $\theta = \arcsin(r/R_{\mathrm{max}})$.
This leads to boundary terms when integrating by parts, resulting in a slightly longer version of eq.~\eqref{eq:M_from_ESD_miscent_corr} that we give in Appendix~\ref{sec:appendix:groups:formulas}.

We adopt the following simple estimate of $\langle R_{\mathrm{mc}}^2 \rangle$.
We assume that the differences between the three different group center coordinates provided by GAMA are representative of the true miscentering offset.
The three different coordinates correspond to the luminosity-weighted center, the BCG, and the so-called iterative center.
For each group, we calculate the three pairwise offsets between these three different centers, square the mean, and then average across all groups within a given $M_b$ bin.
This simple procedure produces $\langle R_{\mathrm{mc}}^2 \rangle$ estimates that are roughly consistent with those implied by the best-fit results from \citet{Rana2022}.
It also reproduces the trend of higher-mass bins having larger miscentering offsets.

\begin{figure*}
\plotone{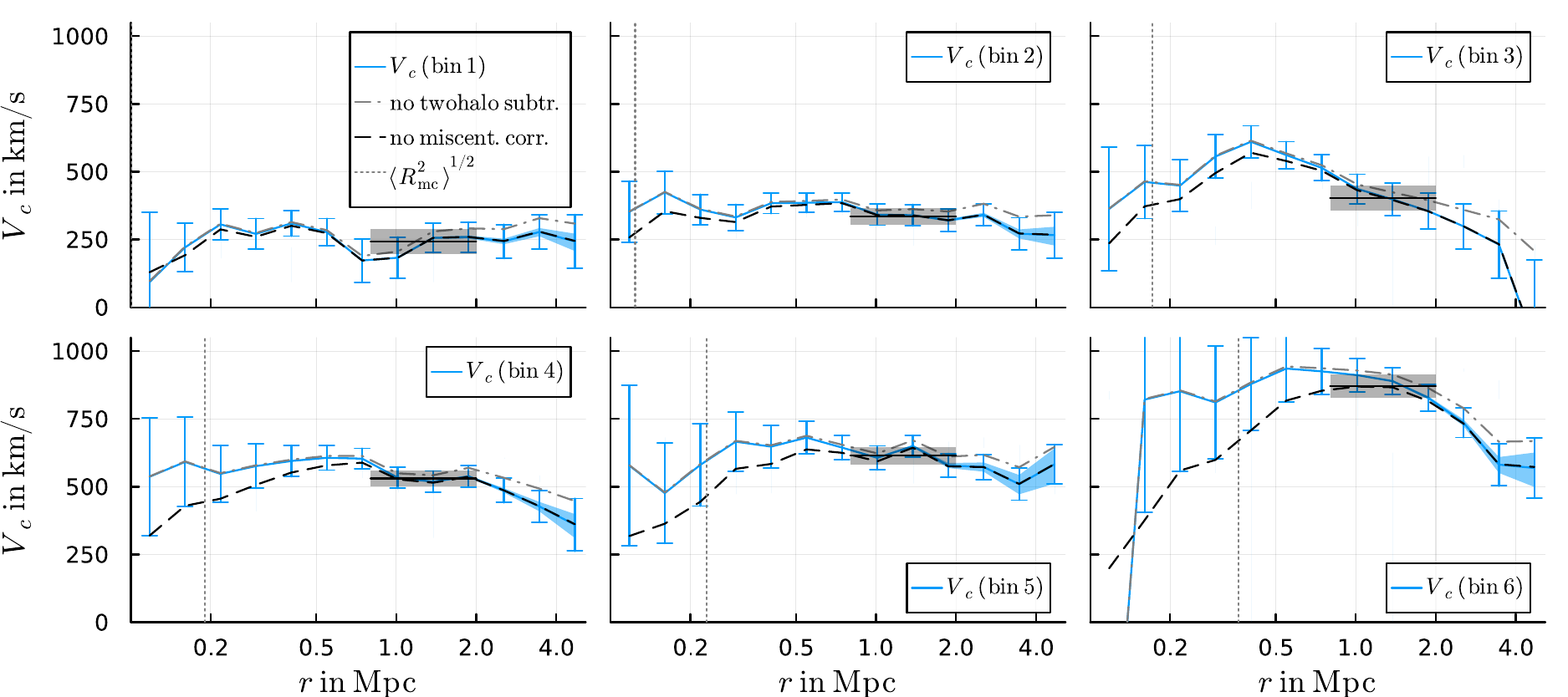}
\caption{The stacked mass profiles $M(r)$ of the GAMA groups in each bin in terms of the implied circular velocity $V_c(r) = \sqrt{G M(r)/r}$ (solid blue lines).
  The blue band indicates the systematic uncertainty from extrapolating $\Delta \Sigma_l$ beyond the last measured data point (Appendix~\ref{sec:appendix:groups:mass}).
  Error bars indicate the statistical uncertainties.
  Dashed vertical lines indicate the estimated miscentering offsets $\sqrt{\langle R_{\mathrm{mc}}^2 \rangle}$ that we correct for.
  Dashed black lines indicate the $V_c$ we obtain without this miscentering correction.
  The miscentering correction is valid only at radii significantly larger than the miscentering offset (Appendix~\ref{sec:appendix:groups:miscentering}).
  Dash-dotted gray lines indicate the $V_c$ we obtain without subtracting the two-halo term (Appendix~\ref{sec:appendix:groups:twohalo}).
  Miscentering is important at small radii, the two-halo term at large radii.
  The shaded black region indicates $\Vf$, its statistical uncertainty, and the radial range from which it is calculated.
\label{fig:groupVc}}
\end{figure*}

Our procedure is illustrated in Fig.~\ref{fig:groupVc} (solid blue vs dashed black lines).
Miscentering effects are generally small at radii much larger than $\sqrt{\langle R_{\mathrm{mc}}^2 \rangle}$ (vertical dashed gray lines in Fig.~\ref{fig:groupVc}), but become significant at smaller radii.
For our main results below, we consider only radii where miscentering is likely only a small effect, i.e. radii larger than a few times $\sqrt{\langle R_{\mathrm{mc}}^2 \rangle}$.

Our miscentering correction formulas originate in a Taylor expansion in $(R_{\mathrm{mc}}/r)^2$ \citep{Mistele2024b} and are therefore only valid at radii significantly larger than $R_{\mathrm{mc}}$.
At smaller radii, these formulas tend to over-correct $M(r)$.
This is another reason to consider only radii larger than a few times $\sqrt{\langle R_{\mathrm{mc}}^2 \rangle}$.

\subsubsection{Stacking}

We stack the mass profiles $M_l(r)$ of all groups in a given $M_b$ bin,
\begin{equation}
 \label{eq:M_stacked}
 M(r) \equiv \frac{\sum_l w_l (r) M_l (r)}{\sum_l w_l (r)} \,,
\end{equation}
with weights $w_l(r)$ given by the inverse square of the uncertainty on $M_l (r)$ 
\begin{equation}
 w_l (r) \equiv \frac{1}{\left. \sigma^2_{M_l (r)}\right|_{\mathrm{from}\;\epsilon_t}} \,.
\end{equation}
Specifically, for $w_l$ we consider only the uncertainty induced by the ellipticities $\epsilon_{t,ls}$, not additional uncertainties in, for example, $\langle R_{\mathrm{mc}}^2 \rangle$ (see Appendix~\ref{sec:appendix:groups:uncertainties}).
This is to not hinder the intrinsic ellipticities of the source galaxies from cancelling out.

We use the same weights to calculate averaged stellar and baryonic masses in each bin.
Following \citet{indefinitelyflat}, we adopt $\langle \sqrt{M_b} \rangle^2$ as the averaged baryonic mass and similarly for the averaged stellar mass.

\subsubsection{Additive biases}

To take into account additive biases, we use the procedure described above to measure the mass profiles around 1 million random points in each of the three GAMA regions G09, G12, and G15.
We use the pre-defined random coordinates from the GAMA table `Randoms' version 2.
This results in three stacked profiles $M_{\mathrm{random},i} (r)$ where $i$ labels the three regions.
For each group, we subtract the $M_{\mathrm{random},i}(r)$ of the corresponding region from that group's $M_l(r)$ in eq.~\eqref{eq:M_stacked}.


\subsubsection{Two-halo term}
\label{sec:appendix:groups:twohalo}

The procedure described above assumes that the observed lensing signal is entirely due to the galaxy groups themselves.
At large radii, however, contributions from the groups' local environments become important.
We follow the procedure from \citet{Mistele_CLASH} to subtract this so-called two-halo term.
This procedure is based on the following estimate of the contribution of the two-halo term to the excess surface density $\Delta \Sigma$ \citep[e.g.][]{Guzik2001,Oguri2011,Covone2014}
\begin{multline}
 \Delta \Sigma_{2\mathrm{h}} (R) = b \frac{\bar{\rho}_{m,0}}{2 \pi D(z_l)^2} \int_0^\infty d \ell \, \ell \\
 \times J_2\left(\frac{\ell R}{D(z_l)}\right) P_m(k_\ell; z_l) \,.
\end{multline}
Here, $J_2$ denotes the second Bessel function of the first kind, $b$ is the bias according to \citet{Tinker2010}, $\bar{\rho}_{m,0}$ is the mean matter density at redshift $z=0$, and $P_m(k_\ell; z_l)$ is the linear matter power spectrum at $k_\ell = \ell/[(1+z) D(z_l)]$.
We compute the contribution of $\Delta \Sigma_{2\mathrm{h}}$ to our inferred mass $M(r)$ and subtract it (see \citet{Mistele_CLASH} for details).
The bias factor $b$ depends on the halo mass $M_{200}$ of the lens under consideration.
For the purpose of calculating $\Delta \Sigma_{2\mathrm{h}}$, we treat each $M_b$ bin as a single lens located at the mean redshift of the bin and with the stacked $M(r)$ as its mass profile.

This simple procedure does not allow for a quantitatively precise subtraction of the two-halo term.
It does, however, serve as a rough first estimate of its size and of the affected radii.
For our main results below, we consider only radii small enough for the two-halo term to have only a minor effect.
This is illustrated in Fig.~\ref{fig:groupVc} (solid blue lines vs dash-dotted gray lines).

\subsubsection{Flat circular velocities}

We convert the stacked $M(r)$ into a stacked circular velocity via $V_c(r)^2 \equiv G M(r) / r$.
We take \Vf\ to be the weighted average of the circular velocities in the radial bins between $800\,\mathrm{kpc}$ and $2\,\mathrm{Mpc}$, with weights given by the inverse squared uncertainty of $V_c$.
This radial range is chosen to be larger than typical miscentering offsets (Appendix~\ref{sec:appendix:groups:miscentering}) but small enough for the two-halo term to have only a moderate effect (Appendix~\ref{sec:appendix:groups:twohalo}).

The flat rotation speed obtained in this way is illustrated by the shaded region in Fig.~\ref{fig:groupVc}.
Most bins have a range of nearly constant rotation speed, but bin 3 declines steadily, and bin 6 truncates with unphysical abruptness. 
The group data look more ragged than that for stacked individual galaxies \citep{indefinitelyflat}, which we attribute largely to the difference in the
number of available systems: there are hundreds of groups per bin but thousands of galaxies. 
The group mass range is also difficult to probe free of external effects, so we see no reason 
to infer a different behavior from the indefinite flatness seen in both lower and higher mass systems. 


\subsubsection{Statistical uncertainties}
\label{sec:appendix:groups:uncertainties}

We take into account the statistical uncertainties $\sigma_\epsilon$ in the source galaxy ellipticities \citep{Giblin2021} and the uncertainty $\sigma_{\langle R_{\mathrm{mc}}^2 \rangle}$ in $\langle R_{\mathrm{mc}}^2 \rangle$.
We adopt  $\sigma_{\langle R_{\mathrm{mc}}^2 \rangle} = \langle R_{\mathrm{mc}}^2 \rangle$, corresponding to a Rayleigh distribution \citep{Johnston2007}.

We propagate these uncertainties into the final result using linear error propagation.
To this end, we first use analytical formulas to calculate the covariance matrix $\mathrm{Cov}(M(r), M(r'))$ for the stacked mass profile before applying the two-halo subtraction.
These analytical formulas are conceptually similar to those given in Appendix~A of \citet{indefinitelyflat}, but additionally correct for miscentering and take into account that a given source $s$ can contribute to multiple groups \citep[see also][]{Viola2015}.
These analytical formulas are given in Appendix~\ref{sec:appendix:groups:formulas}.
In a second step, we propagate this covariance matrix of the un-subtracted mass profile into a covariance matrix for the two-halo subtracted mass profile using automatic differentiation \citep{Mistele2024b,Mistele_CLASH}.

\begin{widetext}
\subsubsection{Analytical formulas for mass profiles and covariance matrices}
\label{sec:appendix:groups:formulas}

For each individual group, before stacking, we use the following deprojection formula, which explicitly shows how we correct for miscentering and how we extrapolate $\Delta \Sigma_l (R)$ beyond $R_{\mathrm{max}}$,
\begin{align}
\begin{split}
 \label{eq:M_from_ESD_with_miscentering_and_extrapolation}
 M_l(r) \cdot (4 r^2)^{-1}
 =
 &\int_{\theta_0}^{\pi/2} d\theta
 \left(
  \Delta \Sigma_l \left(\frac{r}{\sin \theta}\right)
  + \frac14 \frac{R_{\mathrm{mc}}^2}{r^2} (I_1 + I_2 +I_3)
 \right)
 \\
 &- \frac14 \left(\frac{R_{\mathrm{mc}}}{r}\right)^2
    \sin^2 \theta_0 \tan \theta_0 \left(
     \Delta \Sigma_l \left(\frac{r}{\sin \theta_0}\right) - \Delta \Sigma_l(r)
    \right)
 \\
 &- \frac14 \left(\frac{R_{\mathrm{mc}}}{r}\right)^2
    \sin \theta_0 \tan \theta_0 \left(
     r \Delta \Sigma_l'\left(\frac{r}{\sin \theta_0}\right)
     - r \Delta \Sigma_l'(r)
    \right)
 \\
 &+\int_0^{\theta_0} d\theta \Delta \Sigma^{\mathrm{powerlaw}}_l \left(\frac{r}{\sin \theta}\right)
  \,,
\end{split}
\end{align}
where $\theta_0 \equiv \arcsin(r/R_{\mathrm{max}})$ and $\Delta \Sigma_l^{\mathrm{powerlaw}}(R)$ is the extrapolation of $\Delta \Sigma_l (R)$ beyond $R_{\mathrm{max}}$.
We assume a simple power law $1/R^{n_{\mathrm{extrap}}}$ which, at $R = R_{\mathrm{max}}$, is matched to the miscentering-corrected observed $\Delta \Sigma_l$.
The integrands $I_1$, $I_2$, $I_3$ are defined in Appendix~\ref{sec:appendix:groups:miscentering}.

We now list the analytical formulas that we use to numerically evaluate eq.~\eqref{eq:M_from_ESD_with_miscentering_and_extrapolation} and the covariance matrix of the resulting stacked mass profiles.
Let $R_1, R_2, ..., R_N \equiv R_{\mathrm{max}}$ be the discrete radii where we measure $\Delta \Sigma_l$, in increasing order.
Since eq.~\eqref{eq:M_from_ESD_with_miscentering_and_extrapolation} is linear in $\Delta \Sigma_l$ and since we linearly interpolate between the discrete measured $\Delta \Sigma_l$ data points, eq.~\eqref{eq:M_from_ESD_with_miscentering_and_extrapolation} is of the form
\begin{align}
 M_l (R_\alpha) = 4 R_\alpha^2 \sum_{i=\alpha}^N C_{\alpha i} \Delta \Sigma_l (R_i) \,,
\end{align}
for some coefficients $C$ that are independent of the $\Delta \Sigma$ measurements ($\alpha = 1, \dots, N$).
As we will see below, they are of the form $C_{\alpha i} = C^{(0)}_{\alpha i} + R_{\mathrm{mc}}^2 C^{(1)}_{\alpha i}$ where $C^{(0)}$ and $C^{(1)}$ are independent of $R_{\mathrm{mc}}^2$.
That is, the coefficients $C$ are linear in $R_{\mathrm{mc}}^2$.

The covariance matrix for the stacked masses $M(r)$ from eq.~\eqref{eq:M_stacked} can then be written as
\begin{align}
 \mathrm{Cov}(M(R_\alpha), M(R_\beta)) =
  \sum_s \sigma_{\epsilon_s}^2 (
   X_{\alpha s}[\sin] X_{\beta s}[\sin] + 
   X_{\alpha s}[\cos] X_{\beta s}[\cos]
  ) + \sigma_{R_{\mathrm{mc}}^2}^2 Y_\alpha Y_\beta
  \,,
\end{align}
where $\sigma_{\epsilon_s}$ is the ellipticity dispersion of the source $s$ from Table I of \citet{Giblin2021} and
\begin{align}
 X_{\alpha s}[f] &\equiv
   \sum_l \left. \sum_{i=\alpha}^N  \right|_{R_{ls} = R_i}
     4 R_\alpha^2 w_l(R_\alpha)
     \langle C_{\alpha i} \rangle Z_{l i}^{-1} W_{ls}
     \Sigma_{\mathrm{crit},ls} \, f(2 \phi_{\mathrm{ls}})  \,,
     \\
 Y_\alpha &\equiv
  \sum_l \sum_{i=\alpha}^N 4 R_\alpha^2 w_l(R_\alpha) \langle \Delta \Sigma_{\alpha} (R_i) \rangle \frac{\partial C_{\alpha i}}{\partial R_{\mathrm{mc}}^2}
  \,,
\end{align}
where the ``$R_{ls} = R_i$'' indicates that the sum is restricted to lenses $l$ relative to which the source $s$ is located in the $i$-th radial bin.
Further, $\phi_{l s}$ is the position angle of the source with respect to the lens, $Z_{l i} \equiv  \sum_s|_{R_{l s} = R_i} W_{ ls}$, and $\partial C_{\alpha i}/\partial R_{\mathrm{mc}}^2 \equiv C^{(1)}_{\alpha i}$.
For the expectation value $\langle C_{\alpha i} \rangle$ we use $C_{\alpha i}$ with $R_{\mathrm{mc}}^2$ replaced by $\langle R_{\mathrm{mc}}^2 \rangle$ and for the expectation value $\langle \Delta \Sigma_l (R_i) \rangle$ we use the point estimate eq.~\eqref{eq:ESD_measured}.
This formula is conceptually similar to the analytical covariance matrix from \citet{Viola2015} and follows from linear error propagation assuming that $R_{\mathrm{mc}}^2$ is independent of the source ellipticities $\epsilon$.

It remains to give the definition of the coefficients $C$.
We adopt the following shorthand notations:
\begin{align}
 x_{\alpha i} \equiv \frac{R_\alpha}{R_i} \,, \quad
 \theta_{\alpha i} \equiv \arcsin(x_{\alpha i}) \,, \quad
 \Delta \theta_{\alpha i} \equiv \theta_{\alpha i} - \theta_{\alpha,i+1} \,, \quad
 c_{\alpha i} \equiv \sqrt{1 - x_{\alpha i}^2} \,.
\end{align}
We can decompose $C_{\alpha i}$ into three parts corresponding to their respective origin in eq.~\eqref{eq:M_from_ESD_with_miscentering_and_extrapolation},
\begin{align}
 C_{\alpha i} = C^{\mathrm{bulk}}_{\alpha i} + C^{\mathrm{boundary}}_{\alpha i} + C^{\mathrm{tail}}_{\alpha i} \,.
\end{align}
We start with $C_{\alpha i}^{\mathrm{bulk}}$, which we further decompose as
\begin{align}
 C^{\mathrm{bulk}}_{\alpha i} = a_{\alpha i} + b_{\alpha, i-1} + \delta_{i \alpha} A_\alpha + \delta_{i, \alpha+1} B_\alpha \,,
\end{align}
where $\delta$ is the Kronecker delta and the coefficients $a_{\alpha i}$, $b_{\alpha,i-1}$, $A_\alpha$, $B_\alpha$, are identically zero for $\alpha = N$.
Further, $a_{\alpha N} = 0$ and $b_{\alpha,\alpha-1} = 0$.
For $\alpha < N$, we have
\begin{align}
\begin{split}
 B_\alpha &\equiv \frac{R_{\mathrm{mc}}^2}{4 R_\alpha^2} \left[
 \frac{
  x_{\alpha,\alpha+1} ( x_{\alpha, \alpha+1} - 1) - 1
 }{
  \sqrt{
   \frac1{x^2_{\alpha,\alpha+1}} - 1
  }
 }
 + \frac{x_{\alpha,\alpha+1}}{1 - x_{\alpha,\alpha+1}} \left(
  \frac{x^2_{\alpha,\alpha+1}}{c_{\alpha,\alpha+1}} -
  \frac{x^2_{\alpha,N}}{c_{\alpha N}}
  \right)
 \right] \,,
 \\
  A_\alpha &\equiv - B_\alpha + \frac{R_{\mathrm{mc}}^2}{4 R_\alpha^2} \left(
  \frac{x^3_{\alpha,\alpha+1}}{c_{\alpha,\alpha+1}} -
  \frac{x^3_{\alpha,N}}{c_{\alpha N}}
  \right) \,.
\end{split}
\end{align}
Further, we have for $\alpha < N$ and $i \geq \alpha$,
\begin{align}
a_{\alpha i} = \frac{a^{(0)}_{\alpha i} + a^{(1)}_{ \alpha i} + a^{(2)}_{ \alpha i} + a^{(3)}_{ \alpha i}}{x_{\alpha,i+1} - x_{\alpha i}}
\,, \quad
b_{\alpha i} = \frac{b^{(0)}_{ \alpha i} + b^{(1)}_{ \alpha i } + b^{(2)}_{ \alpha i} + b^{(3)}_{ \alpha i}}{x_{\alpha,i+1} - x_{\alpha i}} \,,
\end{align}
with
\begin{align}
\begin{split}
 a^{(0)}_{\alpha i} &\equiv
  - x_{\alpha i} \Delta \theta_{\alpha i} + x_{\alpha i} x_{\alpha, i+1}\left(
   \arctanh\left(c_{\alpha,i+1}\right) -
   \arctanh\left(c_{\alpha i}\right)
  \right)
  \,,
 \\
 - b^{(0)}_{\alpha i} &\equiv
  - x_{\alpha,i+1} \Delta \theta_{\alpha i} + x_{\alpha i} x_{\alpha, i+1}\left(
   \arctanh\left(c_{\alpha,i+1}\right) -
   \arctanh\left(c_{\alpha i}\right)
  \right)
 \,,
 \\
 a^{(1)}_{\alpha i} &\equiv \frac{R_{\mathrm{mc}}^2}{4 R_\alpha^2} \left[
  x_{\alpha i} x_{\alpha,i+1} \left(
   2 c_{\alpha,i+1}
   - 4 c_{\alpha i}
  \right)
  -  2 x_{\alpha i} \Delta \theta_{\alpha i}
   + x_{\alpha i} \sin(2 \theta_{\alpha i})
 \right]
 \,,
 \\
 -b^{(1)}_{\alpha i} &\equiv \frac{R_{\mathrm{mc}}^2}{4 R_\alpha^2} \left[
  x_{\alpha i} x_{\alpha,i+1} \left(
   4 c_{\alpha,i+1}
   - 4 c_{\alpha i}
  \right)
  -  2 x_{\alpha,i+1} \Delta \theta_{\alpha i}
   + x_{\alpha,i+1} \left(
    \sin(2 \theta_{\alpha i}) -
    \sin(2 \theta_{\alpha,i+1})
    \right)
 \right] \,,
 \\
 a^{(2)}_{\alpha i} &\equiv
  \frac{R_{\mathrm{mc}}^2}{4 R_\alpha^2} \cdot
  \begin{cases}
   \frac{x_{\alpha i} x_{\alpha,i+1}}{c_{\alpha i}} \left(
    1
    - c_{\alpha i} c_{\alpha,i+1}
    + x_{\alpha i}^2 \left( \frac{x_{\alpha i}}{x_{\alpha,i+1}} - 2\right)
   \right)
  \,,
  &\alpha \neq i\,, \\
  0\,,
  &\alpha = i \,,
 \end{cases}
 \\
 -b^{(2)}_{\alpha i} &\equiv - \left.a^{(2)}_{\alpha i}\right|_{i \leftrightarrow i+1}
 \,,
 \\
 a^{(3)}_{\alpha i} &\equiv
 \frac{R_{\mathrm{mc}}^2}{4 R_\alpha^2} \cdot
 \begin{cases}
  x_{\alpha i} x_{\alpha,i+1} \left(
   \frac{x_{\alpha,i+1}^2}{c_{\alpha,i+1}} -
   \frac{x_{\alpha i}^2}{c_{\alpha i}}
  \right)
  \,, & \alpha \neq i \,, \\
 0 \,, & \alpha = i \,,
 \end{cases}
 \\
 -b^{(3)}_{\alpha i} &\equiv a^{(3)}_{ \alpha i} \,.
\end{split}
\end{align}
The ``$i \leftrightarrow i+1$'' next to $a^{(2)}_{\alpha i}$ indicates to first take the expression for $a^{(2)}_{\alpha i}$ and then swap all $i$ and $i+1$ indices in that expression.
For $C_{\alpha i}^{\mathrm{boundary}}$, we have $C_{N i}^{\mathrm{boundary}} = 0$ and, for $\alpha < N$,
\begin{align}
\begin{split}
 C_{\alpha i}^{\mathrm{boundary}}
 =
  &-\frac14 \left(\frac{R_{\mathrm{mc}}}{R_\alpha}\right)^2
    \sin^2 \theta_{\alpha N} \tan \theta_{\alpha N}
    (\delta_{i N} - \delta_{i \alpha})
  \\
  &-\frac14 \left(\frac{R_{\mathrm{mc}}}{R_\alpha}\right)^2
    \sin \theta_{\alpha N} \tan \theta_{\alpha N}
    R_\alpha
    \left(
     \frac{\delta_{i N} - \delta_{i,N-1}}{R_N - R_{N-1}} -
     \frac{\delta_{i,\alpha+1} - \delta_{i \alpha}}{R_{\alpha+1} - R_\alpha}
    \right)
 \,.
\end{split}
\end{align}
Finally, $C_{\alpha i}^{\mathrm{tail}}$ is given by 
\begin{align}
 C_{\alpha i}^{\mathrm{tail}} = \delta_{i N} \, \frac{1 - c_{\alpha N}}{x_{\alpha N}} \left(
   1 + \frac14 \left(\frac{R_{\mathrm{mc}}}{R_N}\right)^2 (4 - n_{\mathrm{extrap}}^2)
  \right)\,.
\end{align}

\end{widetext}

\section{The Role of the CGM in MOND}
\label{sec:appendix:cgminmond}

{\LCDM\ and MOND are such different theories that it can be difficult to compare their predictions, even with the same data.
The visible galaxy composed of stars and cold gas (HI and H$_2$) is the same in both cases. 
In \LCDM, it is natural to consider the CGM that is within the virial radius of the dark matter halo to be part of each galaxy.
There is no dark matter in MOND, so this concept is meaningless: there is no dark matter halo (Fig.\ \ref{fig:schematic}) 
with an artificial boundary at $r_{200}$ or any other overdensity.
Any gas in the vicinity of a galaxy but beyond its apparent limits is merely circum-galactic: 
in the same neighborhood, but not necessarily distinct from the IGM.
There is no expectation for its mass or radial extent and no baryon budget for it to aspirationally fill.}

{In MOND, the observed baryons $\Mb = \Mst + \Mg$ in the central galaxy are the appropriate quantity to compare to the measured flat
rotation speed \Vf\ that encloses them. 
There is no need for an intermediary factor $f_v$ to relate \Vf\ to some notional halo quantity at larger radius. 
However, there is a subtle geometric effect as we transition from the edge of the disk where the kinematic \Vf\ is measured to 
much larger radii where gravitational lensing provides an estimate of \Vf\ \citep{indefinitelyflat}. }

{MOND predicts an absolute relation between mass and flat rotation speed \citep{MOND}:
\begin{equation}
a_0 G \Mb = \Vf^4.
\label{eq:MASR}
\end{equation}
This is the origin of the empirical BTFR in MOND, but strictly applies at very large radii where the gravitational potential is effectively spherical. 
Since a flattened mass rotates faster than the equivalent spherical distribution \citep{BT}, 
the intercept $A$ of the BTFR (eq.\ \ref{eq:btfr}) is related to the acceleration scale $a_0$ of MOND by the factor $\zeta$ such that
\begin{equation}
A = \frac{\zeta}{a_0 G}.
\label{eq:zeta}
\end{equation}
The geometric factor $\zeta = 1$ for spheres but is slighlty different for other mass distributions.
A razor thin exponential disk has $\zeta \approx 0.76$ at four scale lengths \citep{MdB98b}, typical of where \Vf\ is measured. 
We can infer the effective value of $\zeta$ for finite thickness disks by fitting the BTFR: $A = 50\;\Aunits$ when
$(a_0 G)^{-1} = 63\;\Aunits$ \citep{M05}, so $\zeta \approx 0.8$. The exact value of $\zeta$ can vary with the individual thickness of each galaxy, 
but this effect is too subtle to detect at present.}

{For the kinematic data, \Vf\ is measured where $\zeta \approx 0.8$ is appropriate.
As we probe to much larger radii with gravitational lensing, all galaxies begin to resemble point masses, so we expect $\zeta \rightarrow 1$.
If there were no mass in the CGM, this should result in slowly declining rotation curves and an offset between the kinematic and lensing BTFR.
No such offset is observed \citep{indefinitelyflat}, though the uncertainties may preclude its notice \citep{lensRAR}. }

{In principle, this provides yet another test of MOND. In practice, it depends on the mass in the CGM. 
For a rotation curve to remain exactly flat from tens to hundreds of kpc implies additional mass to counterbalance
the transition in $\zeta$ such that $\Mb \rightarrow \zeta^{-1} \Mb$. This extra velocity would be accounted for if $\MCGM \approx 0.25\, \Mb$. }

{There already exists a similar effect in the transition from stars to atomic gas. 
By itself, an exponential stellar disk in MOND predicts a slight dip in rotation speed in the outskirts of galaxies \citep{milgrom83} 
that is not observed \citep{Kent1987}. This is an artifact of neglecting the atomic gas, 
which has a more extended mass distribution \citep{Milgrom1988}.} 

{Rotation curves need not be exactly flat; their shapes depend on the mass distribution \citep{milgrom83}.
This is not well-measured for the CGM, but to provide an example, \citet{M2018} built models of the Milky Way both with and without 
a CGM component. For the CGM component, the mass distribution suggested by
\citet [$\MCGM = 4.3 \times 10^{10}\;\Msun$ out to 250 kpc]{coronabaryons} was adopted ($\MCGM \approx 0.6\,\Mb$).
This increases the MOND-predicted velocity from $V(R = 100\;\mathrm{kpc}) = 190\;\kms$ without any CGM to $196\;\kms$ with it, 
so the effect is subtle. In the case of the Milky Way, it may be impossible to accurately measure the circular velocity of the potential
at such large radii due to the perturbation of the LMC \citep{bradawarp,Besla2007ApJ...668..949B,Oehm2024Univ...10..143O}.}

{The effect of the CGM on the extended rotation curves of galaxies in MOND depends on each individual mass distribution, 
but in general additional mass would increase \Vf\ (eq.\ \ref{eq:MASR}). 
However, the presence of other galaxies also matters.
Rotation curves cannot remain flat to infinity; at some point the influence of other objects comes to dominate. 
This external field effect causes the rotation velocity at very large radii to decline \citep{ChaeEFE}, 
acting in the opposite sense of additional mass in any CGM. 
These offsetting effects can be difficult to disentangle \citep{ChaeEFEII}.}

\bibliography{AMredux}

@ARTICLE{KormendyFreeman,
       author = {{Kormendy}, John and {Freeman}, K.~C.},
        title = "{Scaling Laws for Dark Matter Halos in Late-type and Dwarf Spheroidal Galaxies}",
      journal = {\apj},
         year = 2016,
        month = feb,
       volume = {817},
       number = {2},
          eid = {84},
        pages = {84},
          doi = {10.3847/0004-637X/817/2/84},
archivePrefix = {arXiv},
       eprint = {1411.2170},
 primaryClass = {astro-ph.GA},
       adsurl = {https://ui.adsabs.harvard.edu/abs/2016ApJ...817...84K}
}

@ARTICLE{Onorbe2015,
       author = {{O{\~n}orbe}, Jose and {Boylan-Kolchin}, Michael and {Bullock}, James S. and {Hopkins}, Philip F. and {Kere{\v{s}}}, Du{\v{s}}an and {Faucher-Gigu{\`e}re}, Claude-Andr{\'e} and {Quataert}, Eliot and {Murray}, Norman},
        title = "{Forged in FIRE: cusps, cores and baryons in low-mass dwarf galaxies}",
      journal = {\mnras},
         year = 2015,
        month = dec,
       volume = {454},
       number = {2},
        pages = {2092-2106},
          doi = {10.1093/mnras/stv2072},
archivePrefix = {arXiv},
       eprint = {1502.02036},
 primaryClass = {astro-ph.GA},
       adsurl = {https://ui.adsabs.harvard.edu/abs/2015MNRAS.454.2092O}
}

@ARTICLE{banikzhao2022,
       author = {{Banik}, Indranil and {Zhao}, Hongsheng},
        title = "{From Galactic Bars to the Hubble Tension: Weighing Up the Astrophysical Evidence for Milgromian Gravity}",
      journal = {Symmetry},
         year = 2022,
        month = jun,
       volume = {14},
       number = {7},
          eid = {1331},
        pages = {1331},
          doi = {10.3390/sym14071331},
archivePrefix = {arXiv},
       eprint = {2110.06936},
 primaryClass = {astro-ph.CO},
       adsurl = {https://ui.adsabs.harvard.edu/abs/2022Symm...14.1331B}
}

@ARTICLE{LUX2025wimplimit,
       author = {{Aalbers}, J. and {Akerib}, D.~S. and {Musalhi}, A.~K. Al and {Alder}, F. and {Amarasinghe}, C.~S. and {Ames}, A. and {Anderson}, T.~J. and {Angelides}, N. and {Ara{\'u}jo}, H.~M. and {Armstrong}, J.~E. and {Arthurs}, M. and {Baker}, A. and {Balashov}, S. and {Bang}, J. and {Bargemann}, J.~W. and {Barillier}, E.~E. and {Bauer}, D. and {Beattie}, K. and {Benson}, T. and {Bhatti}, A. and {Biekert}, A. and {Biesiadzinski}, T.~P. and {Birch}, H.~J. and {Bishop}, E. and {Blockinger}, G.~M. and {Boxer}, B. and {Brew}, C.~A.~J. and {Br{\'a}s}, P. and {Burdin}, S. and {Buuck}, M. and {Carmona-Benitez}, M.~C. and {Carter}, M. and {Chawla}, A. and {Chen}, H. and {Cherwinka}, J.~J. and {Chin}, Y.~T. and {Chott}, N.~I. and {Converse}, M.~V. and {Coronel}, R. and {Cottle}, A. and {Cox}, G. and {Curran}, D. and {Dahl}, C.~E. and {Darlington}, I. and {Dave}, S. and {David}, A. and {Delgaudio}, J. and {Dey}, S. and {de Viveiros}, L. and {Di Felice}, L. and {Ding}, C. and {Dobson}, J.~E.~Y. and {Druszkiewicz}, E. and {Dubey}, S. and {Eriksen}, S.~R. and {Fan}, A. and {Fayer}, S. and {Fearon}, N.~M. and {Fieldhouse}, N. and {Fiorucci}, S. and {Flaecher}, H. and {Fraser}, E.~D. and {Fruth}, T.~M.~A. and {Gaitskell}, R.~J. and {Geffre}, A. and {Genovesi}, J. and {Ghag}, C. and {Ghosh}, A. and {Gibbons}, R. and {Gokhale}, S. and {Green}, J. and {van der Grinten}, M.~G.~D. and {Haiston}, J.~J. and {Hall}, C.~R. and {Hall}, T.~J. and {Han}, S. and {Hartigan-O'Connor}, E. and {Haselschwardt}, S.~J. and {Hernandez}, M.~A. and {Hertel}, S.~A. and {Heuermann}, G. and {Homenides}, G.~J. and {Horn}, M. and {Huang}, D.~Q. and {Hunt}, D. and {Jacquet}, E. and {James}, R.~S. and {Johnson}, J. and {Kaboth}, A.~C. and {Kamaha}, A.~C. and {Meghna K.}, K. and {Khaitan}, D. and {Khazov}, A. and {Khurana}, I. and {Kim}, J. and {Kim}, Y.~D. and {Kingston}, J. and {Kirk}, R. and {Kodroff}, D. and {Korley}, L. and {Korolkova}, E.~V. and {Kraus}, H. and {Kravitz}, S. and {Kreczko}, L. and {Kudryavtsev}, V.~A. and {Lawes}, C. and {Leonard}, D.~S. and {Lesko}, K.~T. and {Levy}, C. and {Lin}, J. and {Lindote}, A. and {Lippincott}, W.~H. and {Lopes}, M.~I. and {Lorenzon}, W. and {Lu}, C. and {Luitz}, S. and {Majewski}, P.~A. and {Manalaysay}, A. and {Mannino}, R.~L. and {Maupin}, C. and {McCarthy}, M.~E. and {McDowell}, G. and {McKinsey}, D.~N. and {McLaughlin}, J. and {McLaughlin}, J.~B. and {McMonigle}, R. and {Mizrachi}, E. and {Monte}, A. and {Monzani}, M.~E. and {Mendoza}, J.~D. Morales and {Morrison}, E. and {Mount}, B.~J. and {Murdy}, M. and {Murphy}, A. St. J. and {Naylor}, A. and {Nelson}, H.~N. and {Neves}, F. and {Nguyen}, A. and {O'Brien}, C.~L. and {Olcina}, I. and {Oliver-Mallory}, K.~C. and {Orpwood}, J. and {Oyulmaz}, K.~Y. and {Palladino}, K.~J. and {Palmer}, J. and {Pannifer}, N.~J. and {Parveen}, N. and {Patton}, S.~J. and {Penning}, B. and {Pereira}, G. and {Perry}, E. and {Pershing}, T. and {Piepke}, A. and {Qie}, Y. and {Reichenbacher}, J. and {Rhyne}, C.~A. and {Richards}, A. and {Riffard}, Q. and {Rischbieter}, G.~R.~C. and {Ritchey}, E. and {Riyat}, H.~S. and {Rosero}, R. and {Rushton}, T. and {Rynders}, D. and {Santone}, D. and {Sazzad}, A.~B.~M.~R. and {Schnee}, R.~W. and {Sehr}, G. and {Shafer}, B. and {Shaw}, S. and {Shutt}, T. and {Silk}, J.~J. and {Silva}, C. and {Sinev}, G. and {Siniscalco}, J. and {Smith}, R. and {Solovov}, V.~N. and {Sorensen}, P. and {Soria}, J. and {Stancu}, I. and {Stevens}, A. and {Stifter}, K. and {Suerfu}, B. and {Sumner}, T.~J. and {Szydagis}, M. and {Tiedt}, D.~R. and {Timalsina}, M. and {Tong}, Z. and {Tovey}, D.~R. and {Tranter}, J. and {Trask}, M. and {Tripathi}, M. and {Us{\'o}n}, A. and {Vacheret}, A. and {Vaitkus}, A.~C. and {Valentino}, O. and {Velan}, V. and {Wang}, A. and {Wang}, J.~J. and {Wang}, Y.},
        title = "{Dark Matter Search Results from <inline-formula><mml:math><mml:mrow><mml:mn>4.2</mml:mn><mml:mtext> </mml:mtext><mml:mtext> </mml:mtext><mml:mtext>Tonne</mml:mtext><mml:mtext>-</mml:mtext><mml:mtext>Years</mml:mtext></mml:mrow></mml:math></inline-formula> of Exposure of the LUX-ZEPLIN (LZ) Experiment}",
      journal = {\prl},
         year = 2025,
        month = jul,
       volume = {135},
       number = {1},
          eid = {011802},
        pages = {011802},
          doi = {10.1103/4dyc-z8zf},
archivePrefix = {arXiv},
       eprint = {2410.17036},
 primaryClass = {hep-ex},
       adsurl = {https://ui.adsabs.harvard.edu/abs/2025PhRvL.135a1802A}
}

@article{McGvDk,
	author = {{McGaugh}, Stacy S. and {van Dokkum}, Pieter},
	title = "{Dark Matter Halo Masses from Abundance Matching and Kinematics:
	         Tensions for the Milky Way and M31}",
	journal = {Research Notes of the American Astronomical Society},
	year = 2021,
	month = feb,
	volume = {5},
	number = {2},
	eid = {23},
	pages = {23},
	doi = {10.3847/2515-5172/abe1ba},
	adsurl = {https://ui.adsabs.harvard.edu/abs/2021RNAAS...5...23M},
}

@article{RiessH02022,
	author = {{Riess}, Adam G. and {Yuan}, Wenlong and {Macri}, Lucas M. and {
	          Scolnic}, Dan and {Brout}, Dillon and {Casertano}, Stefano and {
	          Jones}, David O. and {Murakami}, Yukei and {Anand}, Gagandeep S.
	          and {Breuval}, Louise and {Brink}, Thomas G. and {Filippenko},
	          Alexei V. and {Hoffmann}, Samantha and {Jha}, Saurabh W. and {
	          D'arcy Kenworthy}, W. and {Mackenty}, John and {Stahl}, Benjamin E.
	          and {Zheng}, WeiKang},
	title = "{A Comprehensive Measurement of the Local Value of the Hubble
	         Constant with 1 km s$^{-1}$ Mpc$^{-1}$ Uncertainty from the Hubble
	         Space Telescope and the SH0ES Team}",
	journal = {\apjl},
	year = 2022,
	month = jul,
	volume = {934},
	number = {1},
	eid = {L7},
	pages = {L7},
	doi = {10.3847/2041-8213/ac5c5b},
	archivePrefix = {arXiv},
	eprint = {2112.04510},
	primaryClass = {astro-ph.CO},
	adsurl = {https://ui.adsabs.harvard.edu/abs/2022ApJ...934L...7R},
}

@article{Connor2025,
	author = {{Connor}, Liam and {Ravi}, Vikram and {Sharma}, Kritti and {Ocker}
	          , Stella Koch and {Faber}, Jakob and {Hallinan}, Gregg and {Harnach
	          }, Charlie and {Hellbourg}, Greg and {Hobbs}, Rick and {Hodge},
	          David and {Hodges}, Mark and {Kosogorov}, Nikita and {Lamb}, James
	          and {Law}, Casey and {Rasmussen}, Paul and {Sherman}, Myles and {
	          Somalwar}, Jean and {Weinreb}, Sander and {Woody}, David and {
	          Konietzka}, Ralf M.},
	title = "{A gas-rich cosmic web revealed by the partitioning of the missing
	         baryons}",
	journal = {Nature Astronomy},
	year = 2025,
	month = jun,
	doi = {10.1038/s41550-025-02566-y},
	adsurl = {https://ui.adsabs.harvard.edu/abs/2025NatAs.tmp..131C},
}

@article{Shull2012,
	author = {{Shull}, J. Michael and {Smith}, Britton D. and {Danforth},
	          Charles W.},
	title = "{The Baryon Census in a Multiphase Intergalactic Medium: 30\% of
	         the Baryons May Still be Missing}",
	journal = {\apj},
	year = 2012,
	month = nov,
	volume = {759},
	number = {1},
	eid = {23},
	pages = {23},
	doi = {10.1088/0004-637X/759/1/23},
	archivePrefix = {arXiv},
	eprint = {1112.2706},
	primaryClass = {astro-ph.CO},
	adsurl = {https://ui.adsabs.harvard.edu/abs/2012ApJ...759...23S},
}

@article{Duey_wiseiii,
	author = {{Duey}, Francis and {Schombert}, James and {McGaugh}, Stacy and {
	          Lelli}, Federico},
	title = "{The Baryonic Tully{\textendash}Fisher Relation. III. Calibration
	         with Redshift-Independent Distances}",
	journal = {\aj},
	year = 2025,
	month = aug,
	volume = {xxx},
	number = {x},
	eid = {xxx},
	pages = {xxx},
	doi = {10.3847/1538-3881/xxx},
}

@article{Duey_wiseii,
	author = {{Duey}, Francis and {Schombert}, James and {McGaugh}, Stacy and {
	          Lelli}, Federico},
	title = "{The Baryonic Tully{\textendash}Fisher Relation. II. Stellar Mass
	         Models}",
	journal = {\aj},
	year = 2025,
	month = mar,
	volume = {169},
	number = {3},
	eid = {186},
	pages = {186},
	doi = {10.3847/1538-3881/adaf21},
	archivePrefix = {arXiv},
	eprint = {2501.10919},
	primaryClass = {astro-ph.GA},
	adsurl = {https://ui.adsabs.harvard.edu/abs/2025AJ....169..186D},
}

@article{Duey_wisei,
	author = {{Duey}, Francis and {Schombert}, James and {McGaugh}, Stacy and {
	          Lelli}, Federico},
	title = "{The Baryonic Tully{\textendash}Fisher Relation. I. WISE/Spitzer
	         Photometry}",
	journal = {\aj},
	year = 2024,
	month = jul,
	volume = {168},
	number = {1},
	eid = {19},
	pages = {19},
	doi = {10.3847/1538-3881/ad454c},
	archivePrefix = {arXiv},
	eprint = {2404.02339},
	primaryClass = {astro-ph.GA},
	adsurl = {https://ui.adsabs.harvard.edu/abs/2024AJ....168...19D},
}

@article{Mistele_CLASH,
	author = {{Mistele}, Tobias and {Lelli}, Federico and {McGaugh}, Stacy and {
	          Schombert}, James and {Famaey}, Benoit},
	title = "{Mass models of galaxy clusters from a non-parametric weak-lensing
	         reconstruction}",
	journal = {arXiv e-prints},
	year = 2025,
	month = jun,
	eid = {arXiv:2506.13716},
	pages = {arXiv:2506.13716},
	archivePrefix = {arXiv},
	eprint = {2506.13716},
	primaryClass = {astro-ph.CO},
	adsurl = {https://ui.adsabs.harvard.edu/abs/2025arXiv250613716M},
}

@ARTICLE{MotiGroups2018,
       author = {{Milgrom}, Mordehai},
        title = "{MOND in galaxy groups}",
      journal = {\prd},
         year = 2018,
        month = nov,
       volume = {98},
       number = {10},
          eid = {104036},
        pages = {104036},
          doi = {10.1103/PhysRevD.98.104036},
archivePrefix = {arXiv},
       eprint = {1810.03089},
 primaryClass = {astro-ph.GA},
       adsurl = {https://ui.adsabs.harvard.edu/abs/2018PhRvD..98j4036M}
}

@article{MotiGroups2019,
	author = {{Milgrom}, Mordehai},
	title = "{MOND in galaxy groups: A superior sample}",
	journal = {\prd},
	year = 2019,
	month = feb,
	volume = {99},
	number = {4},
	eid = {044041},
	pages = {044041},
	doi = {10.1103/PhysRevD.99.044041},
	archivePrefix = {arXiv},
	eprint = {1811.12233},
	primaryClass = {astro-ph.GA},
	adsurl = {https://ui.adsabs.harvard.edu/abs/2019PhRvD..99d4041M},
}

@article{lensRAR,
	author = {{Mistele}, T. and {McGaugh}, S. and {Lelli}, F. and {Schombert},
	          J. and {Li}, P.},
	title = "{Radial acceleration relation of galaxies with joint kinematic and
	         weak-lensing data}",
	journal = {\jcap},
	year = 2024,
	month = apr,
	volume = {2024},
	number = {4},
	eid = {020},
	pages = {020},
	doi = {10.1088/1475-7516/2024/04/020},
	archivePrefix = {arXiv},
	eprint = {2310.15248},
	primaryClass = {astro-ph.GA},
	adsurl = {https://ui.adsabs.harvard.edu/abs/2024JCAP...04..020M},
}

@article{indefinitelyflat,
	author = {{Mistele}, Tobias and {McGaugh}, Stacy and {Lelli}, Federico and {
	          Schombert}, James and {Li}, Pengfei},
	title = "{Indefinitely Flat Circular Velocities and the Baryonic Tully{
	         \textendash}Fisher Relation from Weak Lensing}",
	journal = {\apjl},
	year = 2024,
	month = jul,
	volume = {969},
	number = {1},
	eid = {L3},
	pages = {L3},
	doi = {10.3847/2041-8213/ad54b0},
	archivePrefix = {arXiv},
	eprint = {2406.09685},
	primaryClass = {astro-ph.GA},
	adsurl = {https://ui.adsabs.harvard.edu/abs/2024ApJ...969L...3M},
}

@article{Li2022b,
	author = {{Li}, Pengfei and {McGaugh}, Stacy S. and {Lelli}, Federico and {
	          Schombert}, James M. and {Pawlowski}, Marcel S.},
	title = "{Incorporating baryon-driven contraction of dark matter halos in
	         rotation curve fits}",
	journal = {\aap},
	year = 2022,
	month = sep,
	volume = {665},
	eid = {A143},
	pages = {A143},
	doi = {10.1051/0004-6361/202243916},
	archivePrefix = {arXiv},
	eprint = {2208.04326},
	primaryClass = {astro-ph.GA},
	adsurl = {https://ui.adsabs.harvard.edu/abs/2022A&A...665A.143L},
}

@article{Li2022a,
	author = {{Li}, Pengfei and {McGaugh}, Stacy S. and {Lelli}, Federico and {
	          Tian}, Yong and {Schombert}, James M. and {Ko}, Chung-Ming},
	title = "{The Effect of Adiabatic Compression on Dark Matter Halos and the
	         Radial Acceleration Relation}",
	journal = {\apj},
	year = 2022,
	month = mar,
	volume = {927},
	number = {2},
	eid = {198},
	pages = {198},
	doi = {10.3847/1538-4357/ac52aa},
	archivePrefix = {arXiv},
	eprint = {2202.03421},
	primaryClass = {astro-ph.GA},
	adsurl = {https://ui.adsabs.harvard.edu/abs/2022ApJ...927..198L},
}

@article{PengfeiClusters,
	author = {{Li}, Pengfei and {Tian}, Yong and {J{\'u}lio}, Mariana P. and {
	          Pawlowski}, Marcel S. and {Lelli}, Federico and {McGaugh}, Stacy S.
	          and {Schombert}, James M. and {Read}, Justin I. and {Yu}, Po-Chieh
	          and {Ko}, Chung-Ming},
	title = "{Measuring galaxy cluster mass profiles into the low-acceleration
	         regime with galaxy kinematics}",
	journal = {\aap},
	year = 2023,
	month = sep,
	volume = {677},
	eid = {A24},
	pages = {A24},
	doi = {10.1051/0004-6361/202346431},
	archivePrefix = {arXiv},
	eprint = {2303.10175},
	primaryClass = {astro-ph.CO},
	adsurl = {https://ui.adsabs.harvard.edu/abs/2023A&A...677A..24L},
}

@article{McG2021,
	author = {{McGaugh}, Stacy S. and {Lelli}, Federico and {Schombert}, James
	          M. and {Li}, Pengfei and {Visgaitis}, Tiffany and {Parker}, Kaelee
	          S. and {Pawlowski}, Marcel S.},
	title = "{The Baryonic Tully-Fisher Relation in the Local Group and the
	         Equivalent Circular Velocity of Pressure-supported Dwarfs}",
	journal = {\aj},
	year = 2021,
	month = nov,
	volume = {162},
	number = {5},
	eid = {202},
	pages = {202},
	doi = {10.3847/1538-3881/ac2502},
	archivePrefix = {arXiv},
	eprint = {2109.03251},
	primaryClass = {astro-ph.GA},
	adsurl = {https://ui.adsabs.harvard.edu/abs/2021AJ....162..202M},
}

@article{M20,
	author = {{McGaugh}, Stacy},
	title = "{Predictions and Outcomes for the Dynamics of Rotating Galaxies}",
	journal = {Galaxies},
	year = 2020,
	month = apr,
	volume = {8},
	number = {2},
	pages = {35},
	doi = {10.3390/galaxies8020035},
	archivePrefix = {arXiv},
	eprint = {2004.14402},
	primaryClass = {astro-ph.GA},
	adsurl = {https://ui.adsabs.harvard.edu/abs/2020Galax...8...35M},
}

@article{metalsandmolecules,
	author = {{McGaugh}, Stacy S. and {Lelli}, Federico and {Schombert}, James
	          M.},
	title = "{Scaling Relations for Molecular Gas and Metallicity: Impact on the
	         Baryonic Tully-Fisher Relation}",
	journal = {Research Notes of the American Astronomical Society},
	year = 2020,
	month = mar,
	volume = {4},
	number = {4},
	eid = {45},
	pages = {45},
	doi = {10.3847/2515-5172/ab8471},
	adsurl = {https://ui.adsabs.harvard.edu/abs/2020RNAAS...4...45M},
}

@article{simon2019,
	author = {{Simon}, Joshua D.},
	title = "{The Faintest Dwarf Galaxies}",
	journal = {\araa},
	year = 2019,
	month = aug,
	volume = {57},
	pages = {375-415},
	doi = {10.1146/annurev-astro-091918-104453},
	archivePrefix = {arXiv},
	eprint = {1901.05465},
	primaryClass = {astro-ph.GA},
	adsurl = {https://ui.adsabs.harvard.edu/abs/2019ARA&A..57..375S},
}

@article{iorio2017,
	author = {{Iorio}, G. and {Fraternali}, F. and {Nipoti}, C. and {Di Teodoro}
	          , E. and {Read}, J.~I. and {Battaglia}, G.},
	title = "{LITTLE THINGS in 3D: robust determination of the circular velocity
	         of dwarf irregular galaxies}",
	journal = {\mnras},
	year = 2017,
	month = apr,
	volume = {466},
	number = {4},
	pages = {4159-4192},
	doi = {10.1093/mnras/stw3285},
	archivePrefix = {arXiv},
	eprint = {1611.03865},
	primaryClass = {astro-ph.GA},
	adsurl = {https://ui.adsabs.harvard.edu/abs/2017MNRAS.466.4159I},
}

@article{LelliTF2019,
	author = {{Lelli}, Federico and {McGaugh}, Stacy S. and {Schombert}, James
	          M. and {Desmond}, Harry and {Katz}, Harley},
	title = "{The baryonic Tully-Fisher relation for different velocity
	         definitions and implications for galaxy angular momentum}",
	journal = {\mnras},
	year = 2019,
	month = apr,
	volume = {484},
	number = {3},
	pages = {3267-3278},
	doi = {10.1093/mnras/stz205},
	archivePrefix = {arXiv},
	eprint = {1901.05966},
	primaryClass = {astro-ph.GA},
	adsurl = {https://ui.adsabs.harvard.edu/abs/2019MNRAS.484.3267L},
}

@article{Li2020,
	author = {{Li}, Pengfei and {Lelli}, Federico and {McGaugh}, Stacy and {
	          Schombert}, James},
	title = "{A Comprehensive Catalog of Dark Matter Halo Models for SPARC
	         Galaxies}",
	journal = {\apjs},
	year = 2020,
	month = mar,
	volume = {247},
	number = {1},
	eid = {31},
	pages = {31},
	doi = {10.3847/1538-4365/ab700e},
	archivePrefix = {arXiv},
	eprint = {2001.10538},
	primaryClass = {astro-ph.GA},
	adsurl = {https://ui.adsabs.harvard.edu/abs/2020ApJS..247...31L},
}

@article{Katz2018,
	author = {{Katz}, Harley and {Desmond}, Harry and {Lelli}, Federico and {
	          McGaugh}, Stacy and {Di Cintio}, Arianna and {Brook}, Chris and {
	          Schombert}, James},
	title = "{Stellar feedback and the energy budget of late-type Galaxies:
	         missing baryons and core creation}",
	journal = {\mnras},
	year = 2018,
	month = nov,
	volume = {480},
	number = {4},
	pages = {4287-4301},
	doi = {10.1093/mnras/sty2129},
	archivePrefix = {arXiv},
	eprint = {1808.00971},
	primaryClass = {astro-ph.GA},
	adsurl = {https://ui.adsabs.harvard.edu/abs/2018MNRAS.480.4287K},
}

@article{Katz2017,
	author = {{Katz}, Harley and {Lelli}, Federico and {McGaugh}, Stacy S. and {
	          Di Cintio}, Arianna and {Brook}, Chris B. and {Schombert}, James M.
	          },
	title = "{Testing feedback-modified dark matter haloes with galaxy rotation
	         curves: estimation of halo parameters and consistency with {
	         \ensuremath{\Lambda}}CDM scaling relations}",
	journal = {\mnras},
	year = 2017,
	month = apr,
	volume = {466},
	number = {2},
	pages = {1648-1668},
	doi = {10.1093/mnras/stw3101},
	archivePrefix = {arXiv},
	eprint = {1605.05971},
	primaryClass = {astro-ph.GA},
	adsurl = {https://ui.adsabs.harvard.edu/abs/2017MNRAS.466.1648K},
}

@article{KatzfV,
	author = {{Katz}, Harley and {Desmond}, Harry and {McGaugh}, Stacy and {
	          Lelli}, Federico},
	title = "{The tight empirical relation between dark matter halo mass and
	         flat rotation velocity for late-type galaxies}",
	journal = {\mnras},
	year = 2019,
	month = feb,
	volume = {483},
	number = {1},
	pages = {L98-L103},
	doi = {10.1093/mnrasl/sly203},
	archivePrefix = {arXiv},
	eprint = {1810.12347},
	primaryClass = {astro-ph.GA},
	adsurl = {https://ui.adsabs.harvard.edu/abs/2019MNRAS.483L..98K},
}

@article{Posti2019,
	author = {{Posti}, Lorenzo and {Marasco}, Antonino and {Fraternali}, Filippo
	          and {Famaey}, Benoit},
	title = "{Galaxy disc scaling relations: A tight linear galaxy-halo
	         connection challenges abundance matching}",
	journal = {\aap},
	year = "2019",
	month = "Sep",
	volume = {629},
	eid = {A59},
	pages = {A59},
	doi = {10.1051/0004-6361/201935982},
	archivePrefix = {arXiv},
	eprint = {1909.01344},
	primaryClass = {astro-ph.GA},
	adsurl = {https://ui.adsabs.harvard.edu/abs/2019A&A...629A..59P},
}

@article{BBKARAA,
	author = {{Bullock}, James S. and {Boylan-Kolchin}, Michael},
	title = "{Small-Scale Challenges to the {\ensuremath{\Lambda}}CDM Paradigm}",
	journal = {\araa},
	year = "2017",
	month = "Aug",
	volume = {55},
	number = {1},
	pages = {343-387},
	doi = {10.1146/annurev-astro-091916-055313},
	archivePrefix = {arXiv},
	eprint = {1707.04256},
	primaryClass = {astro-ph.CO},
	adsurl = {https://ui.adsabs.harvard.edu/abs/2017ARA&A..55..343B},
}

@article{Moster2013,
	author = {{Moster}, Benjamin P. and {Naab}, Thorsten and {White}, Simon D.~
	          M.},
	title = "{Galactic star formation and accretion histories from matching
	         galaxies to dark matter haloes}",
	journal = {\mnras},
	year = "2013",
	month = "Feb",
	volume = {428},
	number = {4},
	pages = {3121-3138},
	doi = {10.1093/mnras/sts261},
	archivePrefix = {arXiv},
	eprint = {1205.5807},
	primaryClass = {astro-ph.CO},
	adsurl = {https://ui.adsabs.harvard.edu/abs/2013MNRAS.428.3121M},
}

@article{Behroozi13,
	author = {{Behroozi}, P.~S. and {Wechsler}, R.~H. and {Conroy}, C.},
	title = "{On the Lack of Evolution in Galaxy Star Formation Efficiency}",
	journal = {\apjl},
	archivePrefix = "arXiv",
	eprint = {1209.3013},
	primaryClass = "astro-ph.CO",
	year = 2013,
	month = jan,
	volume = 762,
	eid = {L31},
	pages = {L31},
	doi = {10.1088/2041-8205/762/2/L31},
	adsurl = {https://ui.adsabs.harvard.edu/abs/2013ApJ...762L..31B},
}

@article{KravtsovAM,
	author = {{Kravtsov}, A.~V. and {Vikhlinin}, A.~A. and {Meshcheryakov}, A.~
	          V.},
	title = "{Stellar Mass{\textemdash}Halo Mass Relation and Star Formation
	         Efficiency in High-Mass Halos}",
	journal = {Astronomy Letters},
	year = "2018",
	month = "Jan",
	volume = {44},
	number = {1},
	pages = {8-34},
	doi = {10.1134/S1063773717120015},
	archivePrefix = {arXiv},
	eprint = {1401.7329},
	primaryClass = {astro-ph.CO},
	adsurl = {https://ui.adsabs.harvard.edu/abs/2018AstL...44....8K},
}

@article{DC2014b,
	author = {{Di Cintio}, Arianna and {Brook}, Chris B. and {Dutton}, Aaron A.
	          and {Macci{\`o}}, Andrea V. and {Stinson}, Greg S. and {Knebe},
	          Alexander},
	title = "{A mass-dependent density profile for dark matter haloes including
	         the influence of galaxy formation}",
	journal = {\mnras},
	year = 2014,
	month = jul,
	volume = {441},
	number = {4},
	pages = {2986-2995},
	doi = {10.1093/mnras/stu729},
	archivePrefix = {arXiv},
	eprint = {1404.5959},
	primaryClass = {astro-ph.CO},
	adsurl = {https://ui.adsabs.harvard.edu/abs/2014MNRAS.441.2986D},
}

@article{DC2014a,
	author = {{Di Cintio}, Arianna and {Brook}, Chris B. and {Macci{\`o}},
	          Andrea V. and {Stinson}, Greg S. and {Knebe}, Alexander and {Dutton
	          }, Aaron A. and {Wadsley}, James},
	title = "{The dependence of dark matter profiles on the stellar-to-halo mass
	         ratio: a prediction for cusps versus cores}",
	journal = {\mnras},
	year = 2014,
	month = jan,
	volume = {437},
	number = {1},
	pages = {415-423},
	doi = {10.1093/mnras/stt1891},
	archivePrefix = {arXiv},
	eprint = {1306.0898},
	primaryClass = {astro-ph.CO},
	adsurl = {https://ui.adsabs.harvard.edu/abs/2014MNRAS.437..415D},
}

@article{M2018,
	author = {{McGaugh}, S.~S.},
	title = "{A Precise Milky Way Rotation Curve Model for an Accurate
	         Galactocentric Distance}",
	journal = {Research Notes of the American Astronomical Society},
	archivePrefix = "arXiv",
	eprint = {1808.09435},
	year = 2018,
	month = aug,
	volume = 2,
	number = 3,
	eid = {156},
	pages = {156},
	doi = {10.3847/2515-5172/aadd4b},
	adsurl = {http://adsabs.harvard.edu/abs/2018RNAAS...2c.156M},
}

@article{coronabaryons,
	author = {{Miller}, M.~J. and {Bregman}, J.~N.},
	title = "{Constraining the Milky Way's Hot Gas Halo with O VII and O VIII
	         Emission Lines}",
	journal = {\apj},
	archivePrefix = "arXiv",
	eprint = {1412.3116},
	year = 2015,
	month = feb,
	volume = 800,
	eid = {14},
	pages = {14},
	doi = {10.1088/0004-637X/800/1/14},
	adsurl = {http://adsabs.harvard.edu/abs/2015ApJ...800...14M},
}

@article{SPARC,
	author = {{Lelli}, Federico and {McGaugh}, Stacy S. and {Schombert}, James
	          M.},
	title = "{SPARC: Mass Models for 175 Disk Galaxies with Spitzer Photometry
	         and Accurate Rotation Curves}",
	journal = {\aj},
	year = 2016,
	month = dec,
	volume = {152},
	number = {6},
	eid = {157},
	pages = {157},
	doi = {10.3847/0004-6256/152/6/157},
	archivePrefix = {arXiv},
	eprint = {1606.09251},
	primaryClass = {astro-ph.GA},
	adsurl = {https://ui.adsabs.harvard.edu/abs/2016AJ....152..157L},
}

@article{OneLaw,
	author = {{Lelli}, F. and {McGaugh}, S.~S. and {Schombert}, J.~M. and {
	          Pawlowski}, M.~S.},
	title = "{One Law to Rule Them All: The Radial Acceleration Relation of
	         Galaxies}",
	journal = {\apj},
	archivePrefix = "arXiv",
	eprint = {1610.08981},
	year = 2017,
	month = feb,
	volume = 836,
	eid = {152},
	pages = {152},
	doi = {10.3847/1538-4357/836/2/152},
	adsurl = {http://adsabs.harvard.edu/abs/2017ApJ...836..152L},
}

@article{LelliTFscatter,
	author = {{Lelli}, Federico and {McGaugh}, Stacy S. and {Schombert}, James
	          M.},
	title = "{The Small Scatter of the Baryonic Tully-Fisher Relation}",
	journal = {\apjl},
	year = 2016,
	month = jan,
	volume = {816},
	number = {1},
	eid = {L14},
	pages = {L14},
	doi = {10.3847/2041-8205/816/1/L14},
	archivePrefix = {arXiv},
	eprint = {1512.04543},
	primaryClass = {astro-ph.GA},
	adsurl = {https://ui.adsabs.harvard.edu/abs/2016ApJ...816L..14L},
}

@article{M07,
	author = {{McGaugh}, S.~S. and {de Blok}, W.~J.~G. and {Schombert}, J.~M.
	          and {Kuzio de Naray}, R. and {Kim}, J.~H.},
	title = "{The Rotation Velocity Attributable to Dark Matter at Intermediate
	         Radii in Disk Galaxies}",
	journal = {\apj},
	eprint = {astro-ph/0612410},
	year = 2007,
	month = apr,
	volume = 659,
	pages = {149-161},
	doi = {10.1086/511807},
	adsurl = {http://adsabs.harvard.edu/abs/2007ApJ...659..149M},
}

@ARTICLE{vA1985,
       author = {{van Albada}, T.~S. and {Bahcall}, J.~N. and {Begeman}, K. and {Sancisi}, R.},
        title = "{Distribution of dark matter in the spiral galaxy NGC 3198.}",
      journal = {\apj},
         year = 1985,
        month = aug,
       volume = {295},
        pages = {305-313},
          doi = {10.1086/163375},
       adsurl = {https://ui.adsabs.harvard.edu/abs/1985ApJ...295..305V}
}

@article{crain,
	author = {{Crain}, R.~A. and {Eke}, V.~R. and {Frenk}, C.~S. and {Jenkins},
	          A. and {McCarthy}, I.~G. and {Navarro}, J.~F. and {Pearce}, F.~R.},
	title = "{The baryon fraction of {$\Lambda$}CDM haloes}",
	journal = {\mnras},
	eprint = {arXiv:astro-ph/0610602},
	year = 2007,
	month = may,
	volume = 377,
	pages = {41-49},
	doi = {10.1111/j.1365-2966.2007.11598.x},
	adsurl = {http://adsabs.harvard.edu/abs/2007MNRAS.377...41C},
}

@article{hoeft,
	author = {{Hoeft}, M. and {Gottloeber}, S.},
	title = "{Dwarf Galaxies in Voids: Dark Matter Halos and Gas Cooling}",
	journal = {arXiv:1001.4721},
	archivePrefix = "arXiv",
	eprint = {1001.4721},
	year = 2010,
	month = jan,
	adsurl = {http://adsabs.harvard.edu/abs/2010arXiv1001.4721H},
}

@article{bradawarp,
	author = {{Brada}, R. and {Milgrom}, M.},
	title = "{The Modified Dynamics is Conducive to Galactic Warp Formation}",
	journal = {\apjl},
	eprint = {arXiv:astro-ph/0001146},
	year = 2000,
	month = mar,
	volume = 531,
	pages = {L21-L24},
	doi = {10.1086/312510},
	adsurl = {http://adsabs.harvard.edu/abs/2000ApJ...531L..21B},
}

@article{LivRev,
	author = {{Famaey}, Beno{\^\i}t and {McGaugh}, Stacy S.},
	title = "{Modified Newtonian Dynamics (MOND): Observational Phenomenology
	         and Relativistic Extensions}",
	journal = {Living Reviews in Relativity},
	year = "2012",
	month = "Sep",
	volume = {15},
	number = {1},
	eid = {10},
	pages = {10},
	doi = {10.12942/lrr-2012-10},
	archivePrefix = {arXiv},
	eprint = {1112.3960},
	primaryClass = {astro-ph.CO},
	adsurl = {https://ui.adsabs.harvard.edu/abs/2012LRR....15...10F},
}

@article{MdB98a,
	author = {{McGaugh}, S.~S. and {de Blok}, W.~J.~G.},
	title = "{Testing the Dark Matter Hypothesis with Low Surface Brightness
	         Galaxies and Other Evidence}",
	journal = {\apj},
	eprint = {arXiv:astro-ph/9801123},
	year = 1998,
	month = may,
	volume = 499,
	pages = {41-+},
	adsurl = {http://adsabs.harvard.edu/abs/1998ApJ...499...41M},
}

@article{MdB98b,
	author = {{McGaugh}, S.~S. and {de Blok}, W.~J.~G.},
	title = "{Testing the Hypothesis of Modified Dynamics with Low Surface
	         Brightness Galaxies and Other Evidence}",
	journal = {\apj},
	eprint = {arXiv:astro-ph/9801102},
	year = 1998,
	month = may,
	volume = 499,
	pages = {66-+},
	adsurl = {http://adsabs.harvard.edu/abs/1998ApJ...499...66M},
}

@article{dBM97,
	author = {{de Blok}, W.~J.~G. and {McGaugh}, S.~S.},
	title = "{The dark and visible matter content of low surface brightness disc
	         galaxies}",
	journal = {\mnras},
	year = 1997,
	month = sep,
	volume = {290},
	number = {3},
	pages = {533-552},
	doi = {10.1093/mnras/290.3.533},
	archivePrefix = {arXiv},
	eprint = {astro-ph/9704274},
	primaryClass = {astro-ph},
	adsurl = {https://ui.adsabs.harvard.edu/abs/1997MNRAS.290..533D},
}

@article{CDMSteigmanTurner,
	author = {{Steigman}, Gary and {Turner}, Michael S.},
	title = "{Cosmological constraints on the properties of weakly interacting
	         massive particles}",
	journal = {Nuclear Physics B},
	year = 1985,
	month = jan,
	volume = {253},
	pages = {375-386},
	doi = {10.1016/0550-3213(85)90537-1},
	adsurl = {https://ui.adsabs.harvard.edu/abs/1985NuPhB.253..375S},
}

@article{CDMPeebles,
	author = {{Peebles}, P.~J.~E.},
	title = "{Dark matter and the origin of galaxies and globular star clusters}
	         ",
	journal = {\apj},
	year = 1984,
	month = feb,
	volume = {277},
	pages = {470-477},
	doi = {10.1086/161714},
	adsurl = {https://ui.adsabs.harvard.edu/abs/1984ApJ...277..470P},
}

@article{MWolf,
	author = {{McGaugh}, S.~S. and {Wolf}, J.},
	title = "{Local Group Dwarf Spheroidals: Correlated Deviations from the
	         Baryonic Tully-Fisher Relation}",
	journal = {\apj},
	archivePrefix = "arXiv",
	eprint = {1003.3448},
	primaryClass = "astro-ph.CO",
	year = 2010,
	month = oct,
	volume = 722,
	pages = {248-261},
	doi = {10.1088/0004-637X/722/1/248},
	adsurl = {http://adsabs.harvard.edu/abs/2010ApJ...722..248M},
}

@article{JSH0,
	author = {{Schombert}, James and {McGaugh}, Stacy and {Lelli}, Federico},
	title = "{Using the Baryonic Tully-Fisher Relation to Measure H$_{o}$}",
	journal = {\aj},
	year = 2020,
	month = aug,
	volume = {160},
	number = {2},
	eid = {71},
	pages = {71},
	doi = {10.3847/1538-3881/ab9d88},
	archivePrefix = {arXiv},
	eprint = {2006.08615},
	primaryClass = {astro-ph.CO},
	adsurl = {https://ui.adsabs.harvard.edu/abs/2020AJ....160...71S},
}

@article{FTHINGS,
	author = {{Walter}, F. and {Brinks}, E. and {de Blok}, W.~J.~G. and {Bigiel}
	          , F. and {Kennicutt}, R.~C. and {Thornley}, M.~D. and {Leroy}, A.},
	title = "{THINGS: The H I Nearby Galaxy Survey}",
	journal = {\aj},
	archivePrefix = "arXiv",
	eprint = {0810.2125},
	year = 2008,
	month = dec,
	volume = 136,
	pages = {2563-2647},
	doi = {10.1088/0004-6256/136/6/2563},
	adsurl = {http://adsabs.harvard.edu/abs/2008AJ....136.2563W},
}

@ARTICLE{deBlok2001,
       author = {{de Blok}, W.~J.~G. and {McGaugh}, Stacy S. and {Bosma}, Albert and {Rubin}, Vera C.},
        title = "{Mass Density Profiles of Low Surface Brightness Galaxies}",
      journal = {\apjl},
         year = 2001,
        month = may,
       volume = {552},
       number = {1},
        pages = {L23-L26},
          doi = {10.1086/320262},
archivePrefix = {arXiv},
       eprint = {astro-ph/0103102},
 primaryClass = {astro-ph},
       adsurl = {https://ui.adsabs.harvard.edu/abs/2001ApJ...552L..23D}
}

@ARTICLE{Bregman2018,
       author = {{Bregman}, Joel N. and {Anderson}, Michael E. and {Miller}, Matthew J. and {Hodges-Kluck}, Edmund and {Dai}, Xinyu and {Li}, Jiang-Tao and {Li}, Yunyang and {Qu}, Zhijie},
        title = "{The Extended Distribution of Baryons around Galaxies}",
      journal = {\apj},
         year = 2018,
        month = jul,
       volume = {862},
       number = {1},
          eid = {3},
        pages = {3},
          doi = {10.3847/1538-4357/aacafe},
archivePrefix = {arXiv},
       eprint = {1803.08963},
 primaryClass = {astro-ph.GA},
       adsurl = {https://ui.adsabs.harvard.edu/abs/2018ApJ...862....3B}
}

@ARTICLE{Bregman2007,
       author = {{Bregman}, Joel N.},
        title = "{The Search for the Missing Baryons at Low Redshift}",
      journal = {\araa},
         year = 2007,
        month = sep,
       volume = {45},
       number = {1},
        pages = {221-259},
          doi = {10.1146/annurev.astro.45.051806.110619},
archivePrefix = {arXiv},
       eprint = {0706.1787},
 primaryClass = {astro-ph},
       adsurl = {https://ui.adsabs.harvard.edu/abs/2007ARA&A..45..221B}
}

@ARTICLE{CGMreview,
       author = {{Tumlinson}, Jason and {Peeples}, Molly S. and {Werk}, Jessica K.},
        title = "{The Circumgalactic Medium}",
      journal = {\araa},
         year = 2017,
        month = aug,
       volume = {55},
       number = {1},
        pages = {389-432},
          doi = {10.1146/annurev-astro-091916-055240},
archivePrefix = {arXiv},
       eprint = {1709.09180},
 primaryClass = {astro-ph.GA},
       adsurl = {https://ui.adsabs.harvard.edu/abs/2017ARA&A..55..389T}
}

@article{CopiBBN,
	author = {{Copi}, Craig J. and {Schramm}, David N. and {Turner}, Michael S.},
	title = "{Big-Bang Nucleosynthesis and the Baryon Density of the Universe}",
	journal = {Science},
	year = 1995,
	month = jan,
	volume = {267},
	number = {5195},
	pages = {192-199},
	doi = {10.1126/science.7809624},
	archivePrefix = {arXiv},
	eprint = {astro-ph/9407006},
	primaryClass = {astro-ph},
	adsurl = {https://ui.adsabs.harvard.edu/abs/1995Sci...267..192C},
}

@article{BBN,
	author = {{Walker}, T.~P. and {Steigman}, G. and {Kang}, H.-S. and {Schramm}
	          , D.~M. and {Olive}, K.~A.},
	title = "{Primordial nucleosynthesis redux}",
	journal = {\apj},
	year = 1991,
	month = jul,
	volume = 376,
	pages = {51-69},
	doi = {10.1086/170255},
	adsurl = {http://adsabs.harvard.edu/abs/1991ApJ...376...51W},
}

@ARTICLE{EDD,
       author = {{Tully}, R. Brent and {Rizzi}, Luca and {Shaya}, Edward J. and
         {Courtois}, H{\'e}l{\`e}ne M. and {Makarov}, Dmitry I. and
         {Jacobs}, Bradley A.},
        title = "{The Extragalactic Distance Database}",
      journal = {\aj},
         year = 2009,
        month = "Aug",
       volume = {138},
       number = {2},
        pages = {323-331},
          doi = {10.1088/0004-6256/138/2/323},
       adsurl = {https://ui.adsabs.harvard.edu/abs/2009AJ....138..323T}
}

@article{KdN09,
	author = {{Kuzio de Naray}, R. and {McGaugh}, S.~S. and {Mihos}, J.~C. },
	title = "{Constraining the NFW Potential with Observations and Modeling of
	         Low Surface Brightness Galaxy Velocity Fields}",
	journal = {\apj},
	archivePrefix = "arXiv",
	eprint = {0810.5118},
	year = 2009,
	month = feb,
	volume = 692,
	pages = {1321-1332},
	doi = {10.1088/0004-637X/692/2/1321},
	adsurl = {http://adsabs.harvard.edu/abs/2009ApJ...692.1321K},
}

@ARTICLE{haloevolution,
       author = {{Mu{\~n}oz-Cuartas}, J.~C. and {Macci{\`o}}, A.~V. and {Gottl{\"o}ber}, S. and {Dutton}, A.~A.},
        title = "{The redshift evolution of {\ensuremath{\Lambda}} cold dark matter halo parameters: concentration, spin and shape}",
      journal = {\mnras},
         year = 2011,
        month = feb,
       volume = {411},
       number = {1},
        pages = {584-594},
          doi = {10.1111/j.1365-2966.2010.17704.x},
archivePrefix = {arXiv},
       eprint = {1007.0438},
 primaryClass = {astro-ph.CO},
       adsurl = {https://ui.adsabs.harvard.edu/abs/2011MNRAS.411..584M}
}

@ARTICLE{RuanBrooks2025,
       author = {{Ruan}, Dilys and {Brooks}, Alyson M. and {Cruz}, Akaxia and {Peter}, Annika H.~G. and {Keller}, Benjamin W. and {Quinn}, Thomas and {Wadsley}, James and {Adams}, Elizabeth A.~K.},
        title = "{Predictions for detecting a turndown in the baryonic Tully{\textendash}Fisher relation}",
      journal = {\mnras},
         year = 2025,
        month = aug,
       volume = {541},
       number = {3},
        pages = {2180-2196},
          doi = {10.1093/mnras/staf1099},
archivePrefix = {arXiv},
       eprint = {2503.16607},
 primaryClass = {astro-ph.GA},
       adsurl = {https://ui.adsabs.harvard.edu/abs/2025MNRAS.541.2180R}
}

@article{Namumba2025,
	author = {{Namumba}, Brenda and {Ianjamasimanana}, Roger and {Koribalski}, B
	          {\"a}rbel and {Bosma}, Albert and {Athanassoula}, Evangelia and {
	          Carignan}, Claude and {J{\'o}zsa}, Gyula I.~G. and {Kamphuis},
	          Peter and {Deane}, Roger P. and {Sikhosana}, Sinenhlanhla P. and {
	          Verdes-Montenegro}, Lourdes and {Sorgho}, Amidou and {Ndaliso},
	          Xola and {Amram}, Philippe and {Brinks}, Elias and {Chemin},
	          Laurent and {Combes}, Francoise and {de Blok}, Erwin and {Deg},
	          Nathan and {English}, Jayanne and {Healy}, Julia and {Kurapati},
	          Sushma and {Marasco}, Antonino and {McGaugh}, Stacy and {Oman},
	          Kyle and {Spekkens}, Kristine and {Veronese}, Simone and {Wong}, O.
	          Ivy},
	title = "{Investigating the HI distribution and kinematics of ESO444-G084
	         and [KKS2000]23: New insights from the MHONGOOSE survey}",
	journal = {arXiv e-prints},
	year = 2025,
	month = jun,
	eid = {arXiv:2506.04101},
	pages = {arXiv:2506.04101},
	doi = {10.48550/arXiv.2506.04101},
	archivePrefix = {arXiv},
	eprint = {2506.04101},
	primaryClass = {astro-ph.GA},
	adsurl = {https://ui.adsabs.harvard.edu/abs/2025arXiv250604101N},
}

@article{KK153,
       author = {{Xu}, Jin-Long and {Zhu}, Ming and {Yu}, Nai-Ping and {Zhang}, Chuan-Peng and {Liu}, Xiao-Lan and {Ai}, Mei and {Jiang}, Peng},
        title = "{FAST Discovery of a Gas-rich and Ultrafaint Dwarf Galaxy: KK153}",
      journal = {\apjl},
         year = 2025,
        month = apr,
       volume = {982},
       number = {2},
          eid = {L36},
        pages = {L36},
          doi = {10.3847/2041-8213/adbe7e},
archivePrefix = {arXiv},
       eprint = {2503.08999},
 primaryClass = {astro-ph.GA},
       adsurl = {https://ui.adsabs.harvard.edu/abs/2025ApJ...982L..36X}
}

@article{LeoProt,
	author = {{Bernstein-Cooper}, Elijah Z. and {Cannon}, John M. and {Elson},
	          Edward C. and {Warren}, Steven R. and {Chengular}, Jayaram and {
	          Skillman}, Evan D. and {Adams}, Elizabeth A.~K. and {Bolatto},
	          Alberto D. and {Giovanelli}, Riccardo and {Haynes}, Martha P. and {
	          McQuinn}, Kristen B.~W. and {Pardy}, Stephen A. and {Rhode},
	          Katherine L. and {Salzer}, John J.},
	title = "{ALFALFA Discovery of the Nearby Gas-rich Dwarf Galaxy Leo P. V.
	         Neutral Gas Dynamics and Kinematics}",
	journal = {\aj},
	year = 2014,
	month = aug,
	volume = {148},
	number = {2},
	eid = {35},
	pages = {35},
	doi = {10.1088/0004-6256/148/2/35},
	archivePrefix = {arXiv},
	eprint = {1404.5298},
	primaryClass = {astro-ph.GA},
	adsurl = {https://ui.adsabs.harvard.edu/abs/2014AJ....148...35B},
}

@article{LeoPdisc,
	author = {{Giovanelli}, Riccardo and {Haynes}, Martha P. and {Adams},
	          Elizabeth A.~K. and {Cannon}, John M. and {Rhode}, Katherine L. and
	          {Salzer}, John J. and {Skillman}, Evan D. and {Bernstein-Cooper},
	          Elijah Z. and {McQuinn}, Kristen B.~W.},
	title = "{ALFALFA Discovery of the Nearby Gas-rich Dwarf Galaxy Leo P. I. H
	         I Observations}",
	journal = {\aj},
	year = 2013,
	month = jul,
	volume = {146},
	number = {1},
	eid = {15},
	pages = {15},
	doi = {10.1088/0004-6256/146/1/15},
	archivePrefix = {arXiv},
	eprint = {1305.0272},
	primaryClass = {astro-ph.CO},
	adsurl = {https://ui.adsabs.harvard.edu/abs/2013AJ....146...15G},
}

@article{LeoPunQ,
	author = {{McQuinn}, Kristen B.~W. and {Skillman}, Evan D. and {Dolphin},
	          Andrew and {Cannon}, John M. and {Salzer}, John J. and {Rhode},
	          Katherine L. and {Adams}, Elizabeth A.~K. and {Berg}, Danielle and
	          {Giovanelli}, Riccardo and {Girardi}, L{\'e}o and {Haynes}, Martha
	          P.},
	title = "{Leo P: An Unquenched Very Low-mass Galaxy}",
	journal = {\apj},
	year = 2015,
	month = oct,
	volume = {812},
	number = {2},
	eid = {158},
	pages = {158},
	doi = {10.1088/0004-637X/812/2/158},
	archivePrefix = {arXiv},
	eprint = {1506.05495},
	primaryClass = {astro-ph.GA},
	adsurl = {https://ui.adsabs.harvard.edu/abs/2015ApJ...812..158M},
}

@article{FIGGS,
	author = {{Begum}, A. and {Chengalur}, J.~N. and {Karachentsev}, I.~D. and {
	          Sharina}, M.~E. and {Kaisin}, S.~S.},
	title = "{FIGGS: Faint Irregular Galaxies GMRT Survey - overview,
	         observations and first results}",
	journal = {\mnras},
	archivePrefix = "arXiv",
	eprint = {0802.3982},
	year = 2008,
	month = may,
	volume = 386,
	pages = {1667-1682},
	doi = {10.1111/j.1365-2966.2008.13150.x},
	adsurl = {http://adsabs.harvard.edu/abs/2008MNRAS.386.1667B},
}

@article{trach,
	author = {{Trachternach}, C. and {de Blok}, W.~J.~G. and {McGaugh}, S.~S.
	          and {van der Hulst}, J.~M. and {Dettmar}, {R.-J.}},
	title = "{The baryonic Tully-Fisher relation and its implication for dark
	         matter halos}",
	journal = {\aap},
	archivePrefix = "arXiv",
	eprint = {0907.5533},
	year = 2009,
	month = oct,
	volume = 505,
	pages = {577-587},
	doi = {10.1051/0004-6361/200811136},
	adsurl = {http://adsabs.harvard.edu/abs/2009A26A...505..577T},
}

@article{stark,
	author = {{Stark}, D.~V. and {McGaugh}, S.~S. and {Swaters}, R.~A.},
	title = "{A First Attempt to Calibrate the Baryonic Tully-Fisher Relation
	         with Gas-Dominated Galaxies}",
	journal = {\aj},
	archivePrefix = "arXiv",
	eprint = {},
	year = 2009,
	month = aug,
	volume = 138,
	pages = {392-401},
	doi = {10.1088/0004-6256/138/2/392},
	adsurl = {http://adsabs.harvard.edu/abs/2009AJ....138..392S},
}

@article{M05,
	author = {{McGaugh}, S.~S.},
	title = "{The Baryonic Tully-Fisher Relation of Galaxies with Extended
	         Rotation Curves and the Stellar Mass of Rotating Galaxies}",
	journal = {\apj},
	eprint = {},
	year = 2005,
	month = oct,
	volume = 632,
	pages = {859-871},
	doi = {10.1086/432968},
	adsurl = {http://adsabs.harvard.edu/abs/2005ApJ...632..859M},
}

@article{M11,
	author = {{McGaugh}, Stacy S.},
	title = "{Novel Test of Modified Newtonian Dynamics with Gas Rich Galaxies}",
	journal = {\prl},
	year = 2011,
	month = mar,
	volume = {106},
	number = {12},
	eid = {121303},
	pages = {121303},
	doi = {10.1103/PhysRevLett.106.121303},
	archivePrefix = {arXiv},
	eprint = {1102.3913},
	primaryClass = {astro-ph.CO},
	adsurl = {https://ui.adsabs.harvard.edu/abs/2011PhRvL.106l1303M},
}

@article{M12,
	author = {{McGaugh}, Stacy S.},
	title = "{The Baryonic Tully-Fisher Relation of Gas-rich Galaxies as a Test
	         of {\ensuremath{\Lambda}}CDM and MOND}",
	journal = {\aj},
	year = 2012,
	month = feb,
	volume = {143},
	number = {2},
	eid = {40},
	pages = {40},
	doi = {10.1088/0004-6256/143/2/40},
	archivePrefix = {arXiv},
	eprint = {1107.2934},
	primaryClass = {astro-ph.CO},
	adsurl = {https://ui.adsabs.harvard.edu/abs/2012AJ....143...40M},
}

@article{btforig,
	author = {{McGaugh}, S.~S. and {Schombert}, J.~M. and {Bothun}, G.~D. and {
	          de Blok}, W.~J.~G.},
	title = "{The Baryonic Tully-Fisher Relation}",
	journal = {\apjl},
	eprint = {arXiv:astro-ph/0003001},
	year = 2000,
	month = apr,
	volume = 533,
	pages = {L99-L102},
	doi = {10.1086/312628},
	adsurl = {http://adsabs.harvard.edu/abs/2000ApJ...533L..99M},
}

@ARTICLE{Governato2012,
       author = {{Governato}, F. and {Zolotov}, A. and {Pontzen}, A. and {Christensen}, C. and {Oh}, S.~H. and {Brooks}, A.~M. and {Quinn}, T. and {Shen}, S. and {Wadsley}, J.},
        title = "{Cuspy no more: how outflows affect the central dark matter and baryon distribution in {\ensuremath{\Lambda}} cold dark matter galaxies}",
      journal = {\mnras},
         year = 2012,
        month = may,
       volume = {422},
       number = {2},
        pages = {1231-1240},
          doi = {10.1111/j.1365-2966.2012.20696.x},
archivePrefix = {arXiv},
       eprint = {1202.0554},
 primaryClass = {astro-ph.CO},
       adsurl = {https://ui.adsabs.harvard.edu/abs/2012MNRAS.422.1231G}
}

@ARTICLE{Bothun1994,
       author = {{Bothun}, Gregory D. and {Eriksen}, James and {Schombert}, James M.},
        title = "{ROSAT Observations of Quiescent Low Mass Disk Galaxies: No Evidence of Baryonic Blow Out}",
      journal = {\aj},
         year = 1994,
        month = sep,
       volume = {108},
        pages = {913},
          doi = {10.1086/117121},
       adsurl = {https://ui.adsabs.harvard.edu/abs/1994AJ....108..913B}
}

@ARTICLE{Dufffeedback,
       author = {{Duffy}, Alan R. and {Schaye}, Joop and {Kay}, Scott T. and {Dalla Vecchia}, Claudio and {Battye}, Richard A. and {Booth}, C.~M.},
        title = "{Impact of baryon physics on dark matter structures: a detailed simulation study of halo density profiles}",
      journal = {\mnras},
         year = 2010,
        month = jul,
       volume = {405},
       number = {4},
        pages = {2161-2178},
          doi = {10.1111/j.1365-2966.2010.16613.x},
archivePrefix = {arXiv},
       eprint = {1001.3447},
 primaryClass = {astro-ph.CO},
       adsurl = {https://ui.adsabs.harvard.edu/abs/2010MNRAS.405.2161D}
}

@ARTICLE{oleg2004,
       author = {{Gnedin}, Oleg Y. and {Kravtsov}, Andrey V. and {Klypin}, Anatoly A. and {Nagai}, Daisuke},
        title = "{Response of Dark Matter Halos to Condensation of Baryons: Cosmological Simulations and Improved Adiabatic Contraction Model}",
      journal = {\apj},
         year = 2004,
        month = nov,
       volume = {616},
       number = {1},
        pages = {16-26},
          doi = {10.1086/424914},
archivePrefix = {arXiv},
       eprint = {astro-ph/0406247},
 primaryClass = {astro-ph},
       adsurl = {https://ui.adsabs.harvard.edu/abs/2004ApJ...616...16G}
}

@article{adiabat,
	author = {{Sellwood}, J.~A. and {McGaugh}, S.~S.},
	title = "{The Compression of Dark Matter Halos by Baryonic Infall}",
	journal = {\apj},
	eprint = {arXiv:astro-ph/0507589},
	year = 2005,
	month = nov,
	volume = 634,
	pages = {70-76},
	doi = {10.1086/491731},
	adsurl = {http://adsabs.harvard.edu/abs/2005ApJ...634...70S},
}

@article{M10,
	author = {{McGaugh}, S.~S. and {Schombert}, J.~M. and {de Blok}, W.~J.~G.
	          and {Zagursky}, M.~J.},
	title = "{The Baryon Content of Cosmic Structures}",
	journal = {\apjl},
	archivePrefix = "arXiv",
	eprint = {0911.2700},
	year = 2010,
	month = jan,
	volume = 708,
	pages = {L14-L17},
	doi = {10.1088/2041-8205/708/1/L14},
	adsurl = {http://adsabs.harvard.edu/abs/2010ApJ...708L..14M},
}

@article{MS14,
	author = {{McGaugh}, Stacy S. and {Schombert}, James M.},
	title = "{Color-Mass-to-light-ratio Relations for Disk Galaxies}",
	journal = {\aj},
	year = 2014,
	month = nov,
	volume = {148},
	number = {5},
	eid = {77},
	pages = {77},
	doi = {10.1088/0004-6256/148/5/77},
	archivePrefix = {arXiv},
	eprint = {1407.1839},
	primaryClass = {astro-ph.GA},
	adsurl = {https://ui.adsabs.harvard.edu/abs/2014AJ....148...77M},
}

@article{MS15,
	author = {{McGaugh}, Stacy S. and {Schombert}, James M.},
	title = "{Weighing Galaxy Disks With the Baryonic Tully-Fisher Relation}",
	journal = {\apj},
	year = 2015,
	month = mar,
	volume = {802},
	number = {1},
	eid = {18},
	pages = {18},
	doi = {10.1088/0004-637X/802/1/18},
	archivePrefix = {arXiv},
	eprint = {1501.06826},
	primaryClass = {astro-ph.GA},
	adsurl = {https://ui.adsabs.harvard.edu/abs/2015ApJ...802...18M},
}

@article{SML19,
	author = {{Schombert}, James and {McGaugh}, Stacy and {Lelli}, Federico},
	title = "{The mass-to-light ratios and the star formation histories of disc
	         galaxies}",
	journal = {\mnras},
	year = 2019,
	month = feb,
	volume = {483},
	number = {2},
	pages = {1496-1512},
	doi = {10.1093/mnras/sty3223},
	archivePrefix = {arXiv},
	eprint = {1811.10579},
	primaryClass = {astro-ph.GA},
	adsurl = {https://ui.adsabs.harvard.edu/abs/2019MNRAS.483.1496S},
}

@ARTICLE{SML22,
       author = {{Schombert}, James and {McGaugh}, Stacy and {Lelli}, Federico},
        title = "{Stellar Mass-to-light Ratios: Composite Bulge+Disk Models and the Baryonic Tully-Fisher Relation}",
      journal = {\aj},
         year = 2022,
        month = apr,
       volume = {163},
       number = {4},
          eid = {154},
        pages = {154},
          doi = {10.3847/1538-3881/ac5249},
archivePrefix = {arXiv},
       eprint = {2202.02290},
 primaryClass = {astro-ph.GA},
       adsurl = {https://ui.adsabs.harvard.edu/abs/2022AJ....163..154S}
}

@article{SFMS,
	author = {{McGaugh}, Stacy S. and {Schombert}, James M. and {Lelli},
	          Federico},
	title = "{The Star-forming Main Sequence of Dwarf Low Surface Brightness
	         Galaxies}",
	journal = {\apj},
	year = 2017,
	month = dec,
	volume = {851},
	number = {1},
	eid = {22},
	pages = {22},
	doi = {10.3847/1538-4357/aa9790},
	archivePrefix = {arXiv},
	eprint = {1710.11236},
	primaryClass = {astro-ph.GA},
	adsurl = {https://ui.adsabs.harvard.edu/abs/2017ApJ...851...22M},
}

@article{PlanckCosmology,
	author = {{Planck Collaboration} and {Aghanim}, N. and {Akrami}, Y. and {
	          Ashdown}, M. and {Aumont}, J. and {Baccigalupi}, C. and {Ballardini
	          }, M. and {Banday}, A.~J. and {Barreiro}, R.~B. and {Bartolo}, N.
	          and {Basak}, S. and {Battye}, R. and {Benabed}, K. and {Bernard},
	          J. -P. and {Bersanelli}, M. and {Bielewicz}, P. and {Bock}, J.~J.
	          and {Bond}, J.~R. and {Borrill}, J. and {Bouchet}, F.~R. and {
	          Boulanger}, F. and {Bucher}, M. and {Burigana}, C. and {Butler}, R.
	          ~C. and {Calabrese}, E. and {Cardoso}, J. -F. and {Carron}, J. and
	          {Challinor}, A. and {Chiang}, H.~C. and {Chluba}, J. and {Colombo},
	          L.~P.~L. and {Combet}, C. and {Contreras}, D. and {Crill}, B.~P.
	          and {Cuttaia}, F. and {de Bernardis}, P. and {de Zotti}, G. and {
	          Delabrouille}, J. and {Delouis}, J. -M. and {Di Valentino}, E. and
	          {Diego}, J.~M. and {Dor{\'e}}, O. and {Douspis}, M. and {Ducout},
	          A. and {Dupac}, X. and {Dusini}, S. and {Efstathiou}, G. and {
	          Elsner}, F. and {En{\ss}lin}, T.~A. and {Eriksen}, H.~K. and {
	          Fantaye}, Y. and {Farhang}, M. and {Fergusson}, J. and {
	          Fernandez-Cobos}, R. and {Finelli}, F. and {Forastieri}, F. and {
	          Frailis}, M. and {Fraisse}, A.~A. and {Franceschi}, E. and {Frolov}
	          , A. and {Galeotta}, S. and {Galli}, S. and {Ganga}, K. and {G{\'e}
	          nova-Santos}, R.~T. and {Gerbino}, M. and {Ghosh}, T. and {Gonz{\'a
	          }lez-Nuevo}, J. and {G{\'o}rski}, K.~M. and {Gratton}, S. and {
	          Gruppuso}, A. and {Gudmundsson}, J.~E. and {Hamann}, J. and {
	          Handley}, W. and {Hansen}, F.~K. and {Herranz}, D. and {Hildebrandt
	          }, S.~R. and {Hivon}, E. and {Huang}, Z. and {Jaffe}, A.~H. and {
	          Jones}, W.~C. and {Karakci}, A. and {Keih{\"a}nen}, E. and {
	          Keskitalo}, R. and {Kiiveri}, K. and {Kim}, J. and {Kisner}, T.~S.
	          and {Knox}, L. and {Krachmalnicoff}, N. and {Kunz}, M. and {
	          Kurki-Suonio}, H. and {Lagache}, G. and {Lamarre}, J. -M. and {
	          Lasenby}, A. and {Lattanzi}, M. and {Lawrence}, C.~R. and {Le Jeune
	          }, M. and {Lemos}, P. and {Lesgourgues}, J. and {Levrier}, F. and {
	          Lewis}, A. and {Liguori}, M. and {Lilje}, P.~B. and {Lilley}, M.
	          and {Lindholm}, V. and {L{\'o}pez-Caniego}, M. and {Lubin}, P.~M.
	          and {Ma}, Y. -Z. and {Mac{\'\i}as-P{\'e}rez}, J.~F. and {Maggio},
	          G. and {Maino}, D. and {Mandolesi}, N. and {Mangilli}, A. and {
	          Marcos-Caballero}, A. and {Maris}, M. and {Martin}, P.~G. and {
	          Martinelli}, M. and {Mart{\'\i}nez-Gonz{\'a}lez}, E. and {Matarrese
	          }, S. and {Mauri}, N. and {McEwen}, J.~D. and {Meinhold}, P.~R. and
	          {Melchiorri}, A. and {Mennella}, A. and {Migliaccio}, M. and {
	          Millea}, M. and {Mitra}, S. and {Miville-Desch{\^e}nes}, M. -A. and
	          {Molinari}, D. and {Montier}, L. and {Morgante}, G. and {Moss}, A.
	          and {Natoli}, P. and {N{\o}rgaard-Nielsen}, H.~U. and {Pagano}, L.
	          and {Paoletti}, D. and {Partridge}, B. and {Patanchon}, G. and {
	          Peiris}, H.~V. and {Perrotta}, F. and {Pettorino}, V. and {
	          Piacentini}, F. and {Polastri}, L. and {Polenta}, G. and {Puget},
	          J. -L. and {Rachen}, J.~P. and {Reinecke}, M. and {Remazeilles}, M.
	          and {Renzi}, A. and {Rocha}, G. and {Rosset}, C. and {Roudier}, G.
	          and {Rubi{\~n}o-Mart{\'\i}n}, J.~A. and {Ruiz-Granados}, B. and {
	          Salvati}, L. and {Sandri}, M. and {Savelainen}, M. and {Scott}, D.
	          and {Shellard}, E.~P.~S. and {Sirignano}, C. and {Sirri}, G. and {
	          Spencer}, L.~D. and {Sunyaev}, R. and {Suur-Uski}, A. -S. and {
	          Tauber}, J.~A. and {Tavagnacco}, D. and {Tenti}, M. and {Toffolatti
	          }, L. and {Tomasi}, M. and {Trombetti}, T. and {Valenziano}, L. and
	          {Valiviita}, J. and {Van Tent}, B. and {Vibert}, L. and {Vielva},
	          P. and {Villa}, F. and {Vittorio}, N. and {Wandelt}, B.~D. and {
	          Wehus}, I.~K. and {White}, M. and {White}, S.~D.~M. and {Zacchei},
	          A. and {Zonca}, A.},
	title = "{Planck 2018 results. VI. Cosmological parameters}",
	journal = {\aap},
	year = 2020,
	month = sep,
	volume = {641},
	eid = {A6},
	pages = {A6},
	doi = {10.1051/0004-6361/201833910},
	archivePrefix = {arXiv},
	eprint = {1807.06209},
	primaryClass = {astro-ph.CO},
	adsurl = {https://ui.adsabs.harvard.edu/abs/2020A&A...641A...6P},
}

@article{Gonzalez2013,
	author = {{Gonzalez}, Anthony H. and {Sivanandam}, Suresh and {Zabludoff},
	          Ann I. and {Zaritsky}, Dennis},
	title = "{Galaxy Cluster Baryon Fractions Revisited}",
	journal = {\apj},
	year = 2013,
	month = nov,
	volume = {778},
	number = {1},
	eid = {14},
	pages = {14},
	doi = {10.1088/0004-637X/778/1/14},
	archivePrefix = {arXiv},
	eprint = {1309.3565},
	primaryClass = {astro-ph.CO},
	adsurl = {https://ui.adsabs.harvard.edu/abs/2013ApJ...778...14G},
}

@article{Lagana2013,
	author = {{Lagan{\'a}}, T.~F. and {Martinet}, N. and {Durret}, F. and {Lima
	          Neto}, G.~B. and {Maughan}, B. and {Zhang}, Y. -Y.},
	title = "{A comprehensive picture of baryons in groups and clusters of
	         galaxies}",
	journal = {\aap},
	year = 2013,
	month = jul,
	volume = {555},
	eid = {A66},
	pages = {A66},
	doi = {10.1051/0004-6361/201220423},
	archivePrefix = {arXiv},
	eprint = {1304.6061},
	primaryClass = {astro-ph.CO},
	adsurl = {https://ui.adsabs.harvard.edu/abs/2013A&A...555A..66L},
}

@article{angusbuote,
	author = {{Angus}, G.~W. and {Famaey}, B. and {Buote}, D.~A.},
	title = "{X-ray group and cluster mass profiles in MOND: unexplained mass on
	         the group scale}",
	journal = {\mnras},
	archivePrefix = "arXiv",
	eprint = {0709.0108},
	year = 2008,
	month = jul,
	volume = 387,
	pages = {1470-1480},
	doi = {10.1111/j.1365-2966.2008.13353.x},
	adsurl = {http://adsabs.harvard.edu/abs/2008MNRAS.387.1470A},
}

@article{SMmond,
	author = {{Sanders}, R.~H. and {McGaugh}, S.~S.},
	title = "{Modified Newtonian Dynamics as an Alternative to Dark Matter}",
	journal = {\araa},
	eprint = {arXiv:astro-ph/0204521},
	year = 2002,
	volume = 40,
	pages = {263-317},
	doi = {10.1146/annurev.astro.40.060401.093923},
	adsurl = {http://adsabs.harvard.edu/abs/2002ARA26A..40..263S},
}

@article{fukugita,
	author = {{Fukugita}, M. and {Hogan}, C.~J. and {Peebles}, P.~J.~E.},
	title = "{The Cosmic Baryon Budget}",
	journal = {\apj},
	year = 1998,
	month = aug,
	volume = 503,
	pages = {518-+},
	adsurl = {
	          http://adsabs.harvard.edu/cgi-bin/nph-bib_query?bibcode=1998ApJ...503..518F&db_key=AST
	          },
}

@article{SN,
	author = {{Steinmetz}, M. and {Navarro}, J.~F.},
	title = "{The Cosmological Origin of the Tully-Fisher Relation}",
	journal = {\apj},
	year = 1999,
	month = mar,
	volume = 513,
	pages = {555-560},
	adsurl = {
	          http://adsabs.harvard.edu/cgi-bin/nph-bib_query?bibcode=1999ApJ...513..555S&db_key=AST
	          },
}

@article{MMW98,
	author = {{Mo}, H.~J. and {Mao}, S. and {White}, S.~D.~M.},
	title = "{The formation of galactic discs}",
	journal = {\mnras},
	eprint = {arXiv:astro-ph/9707093},
	year = 1998,
	month = apr,
	volume = 295,
	pages = {319-336},
	doi = {10.1046/j.1365-8711.1998.01227.x},
	adsurl = {http://adsabs.harvard.edu/abs/1998MNRAS.295..319M},
}

@article{NFW,
	author = {{Navarro}, J.~F. and {Frenk}, C.~S. and {White}, S.~D.~M.},
	title = "{A Universal Density Profile from Hierarchical Clustering}",
	journal = {\apj},
	eprint = {},
	year = 1997,
	month = dec,
	volume = 490,
	pages = {493-508},
	doi = {10.1086/304888},
	adsurl = {http://adsabs.harvard.edu/abs/1997ApJ...490..493N},
}

@article{AeST,
	author = {{Skordis}, Constantinos and {Z{\l}o{\'s}nik}, Tom},
	title = "{New Relativistic Theory for Modified Newtonian Dynamics}",
	journal = {\prl},
	year = 2021,
	month = oct,
	volume = {127},
	number = {16},
	eid = {161302},
	pages = {161302},
	doi = {10.1103/PhysRevLett.127.161302},
	archivePrefix = {arXiv},
	eprint = {2007.00082},
	primaryClass = {astro-ph.CO},
	adsurl = {https://ui.adsabs.harvard.edu/abs/2021PhRvL.127p1302S},
}

@article{MOND,
	author = {{Milgrom}, M.},
	title = "{A modification of the Newtonian dynamics as a possible alternative
	         to the hidden mass hypothesis.}",
	journal = {\apj},
	year = 1983,
	month = jul,
	volume = {270},
	pages = {365-370},
	doi = {10.1086/161130},
	adsurl = {https://ui.adsabs.harvard.edu/abs/1983ApJ...270..365M},
}

@article{milgrom83,
	author = {{Milgrom}, M.},
	title = "{A modification of the Newtonian dynamics - Implications for
	         galaxies}",
	journal = "{\apj}",
	year = 1983,
	month = jul,
	volume = 270,
	pages = {371-389},
	adsurl = {
	          http://adsabs.harvard.edu/cgi-bin/nph-bib_query?bibcode=1983ApJ...270..371M&db_key=AST
	          },
}

@article{sanders2003,
	author = {{Sanders}, R.~H.},
	title = "{Clusters of galaxies with modified Newtonian dynamics}",
	journal = {\mnras},
	eprint = {arXiv:astro-ph/0212293},
	year = 2003,
	month = jul,
	volume = 342,
	pages = {901-908},
	doi = {10.1046/j.1365-8711.2003.06596.x},
	adsurl = {http://adsabs.harvard.edu/abs/2003MNRAS.342..901S},
}

@article{TForig,
	author = {{Tully}, R.~B. and {Fisher}, J.~R.},
	title = "{A new method of determining distances to galaxies}",
	journal = {\aap},
	year = 1977,
	month = feb,
	volume = 54,
	pages = {661-673},
	adsurl = {
	          http://adsabs.harvard.edu/cgi-bin/nph-bib_query?bibcode=1977A26A....54..661T&db_key=AST
	          },
}

@article{ChaeEFE,
	author = {{Chae}, Kyu-Hyun and {Lelli}, Federico and {Desmond}, Harry and {
	          McGaugh}, Stacy S. and {Li}, Pengfei and {Schombert}, James M.},
	title = "{Testing the Strong Equivalence Principle: Detection of the
	         External Field Effect in Rotationally Supported Galaxies}",
	journal = {\apj},
	year = 2020,
	month = nov,
	volume = {904},
	number = {1},
	eid = {51},
	pages = {51},
	doi = {10.3847/1538-4357/abbb96},
	archivePrefix = {arXiv},
	eprint = {2009.11525},
	primaryClass = {astro-ph.GA},
	adsurl = {https://ui.adsabs.harvard.edu/abs/2020ApJ...904...51C},
}

@ARTICLE{ChaeEFEII,
       author = {{Chae}, Kyu-Hyun and {Desmond}, Harry and {Lelli}, Federico and {McGaugh}, Stacy S. and {Schombert}, James M.},
        title = "{Testing the Strong Equivalence Principle. II. Relating the External Field Effect in Galaxy Rotation Curves to the Large-scale Structure of the Universe}",
      journal = {\apj},
         year = 2021,
        month = nov,
       volume = {921},
       number = {2},
          eid = {104},
        pages = {104},
          doi = {10.3847/1538-4357/ac1bba},
archivePrefix = {arXiv},
       eprint = {2109.04745},
 primaryClass = {astro-ph.GA},
       adsurl = {https://ui.adsabs.harvard.edu/abs/2021ApJ...921..104C}
}

@article{SHIELD,
	author = {{McNichols}, Andrew T. and {Teich}, Yaron G. and {Nims}, Elise and
	          {Cannon}, John M. and {Adams}, Elizabeth A.~K. and {
	          Bernstein-Cooper}, Elijah Z. and {Giovanelli}, Riccardo and {Haynes
	          }, Martha P. and {J{\'o}zsa}, Gyula I.~G. and {McQuinn}, Kristen B.
	          ~W. and {Salzer}, John J. and {Skillman}, Evan D. and {Warren},
	          Steven R. and {Dolphin}, Andrew and {Elson}, E.~C. and {Haurberg},
	          Nathalie and {Ott}, J{\"u}rgen and {Saintonge}, Amelie and {Cave},
	          Ian and {Hagen}, Cedric and {Huang}, Shan and {Janowiecki}, Steven
	          and {Marshall}, Melissa V. and {Thomann}, Clara M. and {Van Sistine
	          }, Angela},
	title = "{SHIELD: Neutral Gas Kinematics and Dynamics}",
	journal = {\apj},
	year = 2016,
	month = nov,
	volume = {832},
	number = {1},
	eid = {89},
	pages = {89},
	doi = {10.3847/0004-637X/832/1/89},
	archivePrefix = {arXiv},
	eprint = {1609.05376},
	primaryClass = {astro-ph.GA},
	adsurl = {https://ui.adsabs.harvard.edu/abs/2016ApJ...832...89M},
}

@book{BT,
	author = {{Binney}, J. and {Tremaine}, S.},
	title = "{Galactic Dynamics}",
	booktitle = {Galactic Dynamics},
	publisher = {Princeton, NJ, Princeton University Press},
	year = 1987,
	adsurl = {http://adsabs.harvard.edu/abs/1987gady.book.....B},
}

@article{Robotham2011,
	adsurl = {https://ui.adsabs.harvard.edu/abs/2011MNRAS.416.2640R},
	archiveprefix = {arXiv},
	author = {{Robotham}, A.~S.~G. and {Norberg}, P. and {Driver}, S.~P. and {
	          Baldry}, I.~K. and {Bamford}, S.~P. and {Hopkins}, A.~M. and {Liske
	          }, J. and {Loveday}, J. and {Merson}, A. and {Peacock}, J.~A. and {
	          Brough}, S. and {Cameron}, E. and {Conselice}, C.~J. and {Croom},
	          S.~M. and {Frenk}, C.~S. and {Gunawardhana}, M. and {Hill}, D.~T.
	          and {Jones}, D.~H. and {Kelvin}, L.~S. and {Kuijken}, K. and {
	          Nichol}, R.~C. and {Parkinson}, H.~R. and {Pimbblet}, K.~A. and {
	          Phillipps}, S. and {Popescu}, C.~C. and {Prescott}, M. and {Sharp},
	          R.~G. and {Sutherland}, W.~J. and {Taylor}, E.~N. and {Thomas}, D.
	          and {Tuffs}, R.~J. and {van Kampen}, E. and {Wijesinghe}, D.},
	doi = {10.1111/j.1365-2966.2011.19217.x},
	eprint = {1106.1994},
	journal = {\mnras},
	month = oct,
	number = {4},
	pages = {2640--2668},
	primaryclass = {astro-ph.CO},
	title = {{Galaxy and Mass Assembly (GAMA): the GAMA galaxy group catalogue
	         (G$^{3}$Cv1)}},
	volume = {416},
	year = 2011,
}

@article{Rana2022,
	adsurl = {https://ui.adsabs.harvard.edu/abs/2022MNRAS.510.5408R},
	archiveprefix = {arXiv},
	author = {{Rana}, Divya and {More}, Surhud and {Miyatake}, Hironao and {
	          Nishimichi}, Takahiro and {Takada}, Masahiro and {Robotham}, Aaron
	          S.~G. and {Hopkins}, Andrew M. and {Holwerda}, Benne W.},
	doi = {10.1093/mnras/stac007},
	eprint = {2107.05641},
	journal = {\mnras},
	month = mar,
	number = {4},
	pages = {5408--5425},
	primaryclass = {astro-ph.CO},
	title = {{The Subaru HSC weak lensing mass-observable scaling relations of
	         spectroscopic galaxy groups from the GAMA survey}},
	volume = {510},
	year = 2022,
}

@article{Mistele2024b,
	adsurl = {https://ui.adsabs.harvard.edu/abs/2024OJAp....7E.120M},
	archiveprefix = {arXiv},
	author = {{Mistele}, Tobias and {Durakovic}, Amel},
	doi = {10.33232/001c.127612},
	eid = {120},
	eprint = {2408.07026},
	journal = {The Open Journal of Astrophysics},
	month = dec,
	pages = {120},
	primaryclass = {astro-ph.CO},
	title = {{A new non-parametric method to infer galaxy cluster masses from
	         weak lensing}},
	volume = {7},
	year = 2024,
}

@article{Driver2011,
	author = {{Driver}, S.~P. and {Hill}, D.~T. and {Kelvin}, L.~S. and {
	          Robotham}, A.~S.~G. and {Liske}, J. and {Norberg}, P. and {Baldry},
	          I.~K. and {Bamford}, S.~P. and {Hopkins}, A.~M. and {Loveday}, J.
	          and {Peacock}, J.~A. and {Andrae}, E. and {Bland-Hawthorn}, J. and
	          {Brough}, S. and {Brown}, M.~J.~I. and {Cameron}, E. and {Ching},
	          J.~H.~Y. and {Colless}, M. and {Conselice}, C.~J. and {Croom}, S.~
	          M. and {Cross}, N.~J.~G. and {de Propris}, R. and {Dye}, S. and {
	          Drinkwater}, M.~J. and {Ellis}, S. and {Graham}, Alister W. and {
	          Grootes}, M.~W. and {Gunawardhana}, M. and {Jones}, D.~H. and {van
	          Kampen}, E. and {Maraston}, C. and {Nichol}, R.~C. and {Parkinson},
	          H.~R. and {Phillipps}, S. and {Pimbblet}, K. and {Popescu}, C.~C.
	          and {Prescott}, M. and {Roseboom}, I.~G. and {Sadler}, E.~M. and {
	          Sansom}, A.~E. and {Sharp}, R.~G. and {Smith}, D.~J.~B. and {Taylor
	          }, E. and {Thomas}, D. and {Tuffs}, R.~J. and {Wijesinghe}, D. and
	          {Dunne}, L. and {Frenk}, C.~S. and {Jarvis}, M.~J. and {Madore}, B.
	          ~F. and {Meyer}, M.~J. and {Seibert}, M. and {Staveley-Smith}, L.
	          and {Sutherland}, W.~J. and {Warren}, S.~J.},
	title = "{Galaxy and Mass Assembly (GAMA): survey diagnostics and core data
	         release}",
	journal = {\mnras},
	year = 2011,
	month = may,
	volume = {413},
	number = {2},
	pages = {971-995},
	doi = {10.1111/j.1365-2966.2010.18188.x},
	archivePrefix = {arXiv},
	eprint = {1009.0614},
	primaryClass = {astro-ph.CO},
	adsurl = {https://ui.adsabs.harvard.edu/abs/2011MNRAS.413..971D},
}

@article{Liske2015,
	author = {{Liske}, J. and {Baldry}, I.~K. and {Driver}, S.~P. and {Tuffs},
	          R.~J. and {Alpaslan}, M. and {Andrae}, E. and {Brough}, S. and {
	          Cluver}, M.~E. and {Grootes}, M.~W. and {Gunawardhana}, M.~L.~P.
	          and {Kelvin}, L.~S. and {Loveday}, J. and {Robotham}, A.~S.~G. and
	          {Taylor}, E.~N. and {Bamford}, S.~P. and {Bland-Hawthorn}, J. and {
	          Brown}, M.~J.~I. and {Drinkwater}, M.~J. and {Hopkins}, A.~M. and {
	          Meyer}, M.~J. and {Norberg}, P. and {Peacock}, J.~A. and {Agius},
	          N.~K. and {Andrews}, S.~K. and {Bauer}, A.~E. and {Ching}, J.~H.~Y.
	          and {Colless}, M. and {Conselice}, C.~J. and {Croom}, S.~M. and {
	          Davies}, L.~J.~M. and {De Propris}, R. and {Dunne}, L. and {Eardley
	          }, E.~M. and {Ellis}, S. and {Foster}, C. and {Frenk}, C.~S. and {H
	          {\"a}u{\ss}ler}, B. and {Holwerda}, B.~W. and {Howlett}, C. and {
	          Ibarra}, H. and {Jarvis}, M.~J. and {Jones}, D.~H. and {Kafle}, P.~
	          R. and {Lacey}, C.~G. and {Lange}, R. and {Lara-L{\'o}pez}, M.~A.
	          and {L{\'o}pez-S{\'a}nchez}, {\'A}. R. and {Maddox}, S. and {Madore
	          }, B.~F. and {McNaught-Roberts}, T. and {Moffett}, A.~J. and {
	          Nichol}, R.~C. and {Owers}, M.~S. and {Palamara}, D. and {Penny},
	          S.~J. and {Phillipps}, S. and {Pimbblet}, K.~A. and {Popescu}, C.~
	          C. and {Prescott}, M. and {Proctor}, R. and {Sadler}, E.~M. and {
	          Sansom}, A.~E. and {Seibert}, M. and {Sharp}, R. and {Sutherland},
	          W. and {V{\'a}zquez-Mata}, J.~A. and {van Kampen}, E. and {Wilkins}
	          , S.~M. and {Williams}, R. and {Wright}, A.~H.},
	title = "{Galaxy And Mass Assembly (GAMA): end of survey report and data
	         release 2}",
	journal = {\mnras},
	year = 2015,
	month = sep,
	volume = {452},
	number = {2},
	pages = {2087-2126},
	doi = {10.1093/mnras/stv1436},
	archivePrefix = {arXiv},
	eprint = {1506.08222},
	primaryClass = {astro-ph.GA},
	adsurl = {https://ui.adsabs.harvard.edu/abs/2015MNRAS.452.2087L},
}

@article{Li2024,
	adsurl = {https://ui.adsabs.harvard.edu/abs/2024arXiv241103152L},
	archiveprefix = {arXiv},
	author = {{Li}, Shun-Sheng and {Hoekstra}, Henk and {Kuijken}, Konrad and {
	          Schaller}, Matthieu and {Schaye}, Joop},
	eid = {arXiv:2411.03152},
	eprint = {2411.03152},
	journal = {arXiv e-prints},
	month = nov,
	pages = {arXiv:2411.03152},
	primaryclass = {astro-ph.GA},
	title = {{Weak lensing constraints on the stellar-to-halo mass relation of
	         galaxy groups with simulation-informed scatter}},
	year = 2024,
}

@article{Liu2024,
	adsurl = {https://ui.adsabs.harvard.edu/abs/2024arXiv241107525L},
	archiveprefix = {arXiv},
	author = {{Liu}, Zhenjie and {Xu}, Kun and {Zhang}, Jun and {Wang}, Wenting
	          and {Liu}, Cong},
	eid = {arXiv:2411.07525},
	eprint = {2411.07525},
	journal = {arXiv e-prints},
	month = nov,
	pages = {arXiv:2411.07525},
	primaryclass = {astro-ph.GA},
	title = {{The Dependence of Dark Matter Halo Properties on the Morphology of
	         Their Central Galaxies from Weak Lensing}},
	year = 2024,
}

@article{Viola2015,
	adsurl = {https://ui.adsabs.harvard.edu/abs/2015MNRAS.452.3529V},
	archiveprefix = {arXiv},
	author = {{Viola}, M. and {Cacciato}, M. and {Brouwer}, M. and {Kuijken}, K.
	          and {Hoekstra}, H. and {Norberg}, P. and {Robotham}, A.~S.~G. and {
	          van Uitert}, E. and {Alpaslan}, M. and {Baldry}, I.~K. and {Choi},
	          A. and {de Jong}, J.~T.~A. and {Driver}, S.~P. and {Erben}, T. and
	          {Grado}, A. and {Graham}, Alister W. and {Heymans}, C. and {
	          Hildebrandt}, H. and {Hopkins}, A.~M. and {Irisarri}, N. and {
	          Joachimi}, B. and {Loveday}, J. and {Miller}, L. and {Nakajima}, R.
	          and {Schneider}, P. and {Sif{\'o}n}, C. and {Verdoes Kleijn}, G.},
	doi = {10.1093/mnras/stv1447},
	eprint = {1507.00735},
	journal = {\mnras},
	month = oct,
	number = {4},
	pages = {3529--3550},
	primaryclass = {astro-ph.GA},
	title = {{Dark matter halo properties of GAMA galaxy groups from 100 square
	         degrees of KiDS weak lensing data}},
	volume = {452},
	year = 2015,
}

@article{Kuijken2019,
	adsurl = {https://ui.adsabs.harvard.edu/abs/2019A\&A...625A...2K},
	archiveprefix = {arXiv},
	author = {{Kuijken}, K. and {Heymans}, C. and {Dvornik}, A. and {Hildebrandt
	          }, H. and {de Jong}, J.~T.~A. and {Wright}, A.~H. and {Erben}, T.
	          and {Bilicki}, M. and {Giblin}, B. and {Shan}, H. -Y. and {Getman},
	          F. and {Grado}, A. and {Hoekstra}, H. and {Miller}, L. and {
	          Napolitano}, N. and {Paolilo}, M. and {Radovich}, M. and {Schneider
	          }, P. and {Sutherland}, W. and {Tewes}, M. and {Tortora}, C. and {
	          Valentijn}, E.~A. and {Verdoes Kleijn}, G.~A.},
	doi = {10.1051/0004-6361/201834918},
	eid = {A2},
	eprint = {1902.11265},
	journal = {\aap},
	month = may,
	pages = {A2},
	primaryclass = {astro-ph.GA},
	title = {{The fourth data release of the Kilo-Degree Survey: ugri imaging
	         and nine-band optical-IR photometry over 1000 square degrees}},
	volume = {625},
	year = 2019,
}

@article{Wright2020,
	adsurl = {https://ui.adsabs.harvard.edu/abs/2020A\&A...637A.100W},
	archiveprefix = {arXiv},
	author = {{Wright}, Angus H. and {Hildebrandt}, Hendrik and {van den Busch},
	          Jan Luca and {Heymans}, Catherine},
	doi = {10.1051/0004-6361/201936782},
	eid = {A100},
	eprint = {1909.09632},
	journal = {\aap},
	month = may,
	pages = {A100},
	primaryclass = {astro-ph.CO},
	title = {{Photometric redshift calibration with self-organising maps}},
	volume = {637},
	year = 2020,
}

@article{Giblin2021,
	adsurl = {https://ui.adsabs.harvard.edu/abs/2021A\&A...645A.105G},
	archiveprefix = {arXiv},
	author = {{Giblin}, Benjamin and {Heymans}, Catherine and {Asgari}, Marika
	          and {Hildebrandt}, Hendrik and {Hoekstra}, Henk and {Joachimi},
	          Benjamin and {Kannawadi}, Arun and {Kuijken}, Konrad and {Lin},
	          Chieh-An and {Miller}, Lance and {Tr{\"o}ster}, Tilman and {van den
	          Busch}, Jan Luca and {Wright}, Angus H. and {Bilicki}, Maciej and {
	          Blake}, Chris and {de Jong}, Jelte and {Dvornik}, Andrej and {Erben
	          }, Thomas and {Getman}, Fedor and {Napolitano}, Nicola R. and {
	          Schneider}, Peter and {Shan}, HuanYuan and {Valentijn}, Edwin},
	doi = {10.1051/0004-6361/202038850},
	eid = {A105},
	eprint = {2007.01845},
	journal = {\aap},
	month = jan,
	pages = {A105},
	primaryclass = {astro-ph.CO},
	title = {{KiDS-1000 catalogue: Weak gravitational lensing shear measurements
	         }},
	volume = {645},
	year = 2021,
}

@article{Hildebrandt2021,
	adsurl = {https://ui.adsabs.harvard.edu/abs/2021A\&A...647A.124H},
	archiveprefix = {arXiv},
	author = {{Hildebrandt}, H. and {van den Busch}, J.~L. and {Wright}, A.~H.
	          and {Blake}, C. and {Joachimi}, B. and {Kuijken}, K. and {Tr{\"o}
	          ster}, T. and {Asgari}, M. and {Bilicki}, M. and {de Jong}, J.~T.~
	          A. and {Dvornik}, A. and {Erben}, T. and {Getman}, F. and {Giblin},
	          B. and {Heymans}, C. and {Kannawadi}, A. and {Lin}, C. -A. and {
	          Shan}, H. -Y.},
	doi = {10.1051/0004-6361/202039018},
	eid = {A124},
	eprint = {2007.15635},
	journal = {\aap},
	month = mar,
	pages = {A124},
	primaryclass = {astro-ph.CO},
	title = {{{KiDS-1000 catalogue: Redshift distributions and their calibration
	         } }},
	volume = {647},
	year = 2021,
}

@article{Taylor2011,
	author = {{Taylor}, Edward N. and {Hopkins}, Andrew M. and {Baldry}, Ivan K.
	          and {Brown}, Michael J.~I. and {Driver}, Simon P. and {Kelvin}, Lee
	          S. and {Hill}, David T. and {Robotham}, Aaron S.~G. and {
	          Bland-Hawthorn}, Joss and {Jones}, D.~H. and {Sharp}, R.~G. and {
	          Thomas}, Daniel and {Liske}, Jochen and {Loveday}, Jon and {Norberg
	          }, Peder and {Peacock}, J.~A. and {Bamford}, Steven P. and {Brough}
	          , Sarah and {Colless}, Matthew and {Cameron}, Ewan and {Conselice},
	          Christopher J. and {Croom}, Scott M. and {Frenk}, C.~S. and {
	          Gunawardhana}, Madusha and {Kuijken}, Konrad and {Nichol}, R.~C.
	          and {Parkinson}, H.~R. and {Phillipps}, S. and {Pimbblet}, K.~A.
	          and {Popescu}, C.~C. and {Prescott}, Matthew and {Sutherland}, W.~
	          J. and {Tuffs}, R.~J. and {van Kampen}, Eelco and {Wijesinghe}, D.},
	title = "{Galaxy And Mass Assembly (GAMA): stellar mass estimates}",
	journal = {\mnras},
	year = 2011,
	month = dec,
	volume = {418},
	number = {3},
	pages = {1587-1620},
	doi = {10.1111/j.1365-2966.2011.19536.x},
	archivePrefix = {arXiv},
	eprint = {1108.0635},
	primaryClass = {astro-ph.CO},
	adsurl = {https://ui.adsabs.harvard.edu/abs/2011MNRAS.418.1587T},
}

@article{Brouwer2021,
	adsurl = {https://ui.adsabs.harvard.edu/abs/2021A\&A...650A.113B},
	archiveprefix = {arXiv},
	author = {{Brouwer}, Margot M. and {Oman}, Kyle A. and {Valentijn}, Edwin A.
	          and {Bilicki}, Maciej and {Heymans}, Catherine and {Hoekstra}, Henk
	          and {Napolitano}, Nicola R. and {Roy}, Nivya and {Tortora},
	          Crescenzo and {Wright}, Angus H. and {Asgari}, Marika and {van den
	          Busch}, Jan Luca and {Dvornik}, Andrej and {Erben}, Thomas and {
	          Giblin}, Benjamin and {Graham}, Alister W. and {Hildebrandt},
	          Hendrik and {Hopkins}, Andrew M. and {Kannawadi}, Arun and {Kuijken
	          }, Konrad and {Liske}, Jochen and {Shan}, HuanYuan and {Tr{\"o}ster
	          }, Tilman and {Verlinde}, Erik and {Visser}, Manus},
	doi = {10.1051/0004-6361/202040108},
	eid = {A113},
	eprint = {2106.11677},
	journal = {Astronomy \& Astrophysics},
	month = jun,
	pages = {A113},
	primaryclass = {astro-ph.GA},
	title = {{The weak lensing radial acceleration relation: Constraining
	         modified gravity and cold dark matter theories with KiDS-1000}},
	volume = {650},
	year = 2021,
}

@article{Dvornik2017,
	adsurl = {https://ui.adsabs.harvard.edu/abs/2017MNRAS.468.3251D},
	archiveprefix = {arXiv},
	author = {{Dvornik}, Andrej and {Cacciato}, Marcello and {Kuijken}, Konrad
	          and {Viola}, Massimo and {Hoekstra}, Henk and {Nakajima}, Reiko and
	          {van Uitert}, Edo and {Brouwer}, Margot and {Choi}, Ami and {Erben}
	          , Thomas and {Fenech Conti}, Ian and {Farrow}, Daniel J. and {
	          Herbonnet}, Ricardo and {Heymans}, Catherine and {Hildebrandt},
	          Hendrik and {Hopkins}, Andrew M. and {McFarland}, John and {Norberg
	          }, Peder and {Schneider}, Peter and {Sif{\'o}n}, Crist{\'o}bal and
	          {Valentijn}, Edwin and {Wang}, Lingyu},
	doi = {10.1093/mnras/stx705},
	eprint = {1703.06657},
	journal = {\mnras},
	month = jul,
	number = {3},
	pages = {3251--3265},
	primaryclass = {astro-ph.CO},
	title = {{A KiDS weak lensing analysis of assembly bias in GAMA galaxy
	         groups}},
	volume = {468},
	year = 2017,
}

@article{Guzik2001,
	adsurl = {https://ui.adsabs.harvard.edu/abs/2001MNRAS.321..439G},
	archiveprefix = {arXiv},
	author = {{Guzik}, Jacek and {Seljak}, Uro{\v s}},
	doi = {10.1046/j.1365-8711.2001.04081.x},
	eprint = {astro-ph/0007067},
	journal = {\mnras},
	month = mar,
	number = {3},
	pages = {439--449},
	primaryclass = {astro-ph},
	title = {{Galaxy-dark matter correlations applied to galaxy-galaxy lensing:
	         predictions from the semi-analytic galaxy formation models}},
	volume = {321},
	year = 2001,
}

@article{Oguri2011,
	adsurl = {https://ui.adsabs.harvard.edu/abs/2011MNRAS.414.1851O},
	archiveprefix = {arXiv},
	author = {{Oguri}, Masamune and {Hamana}, Takashi},
	doi = {10.1111/j.1365-2966.2011.18481.x},
	eprint = {1101.0650},
	journal = {\mnras},
	month = jul,
	number = {3},
	pages = {1851--1861},
	primaryclass = {astro-ph.CO},
	title = {{Detailed cluster lensing profiles at large radii and the impact on
	         cluster weak lensing studies}},
	volume = {414},
	year = 2011,
}

@article{Covone2014,
	adsurl = {https://ui.adsabs.harvard.edu/abs/2014ApJ...784L..25C},
	archiveprefix = {arXiv},
	author = {{Covone}, Giovanni and {Sereno}, Mauro and {Kilbinger}, Martin and
	          {Cardone}, Vincenzo F.},
	doi = {10.1088/2041-8205/784/2/L25},
	eid = {L25},
	eprint = {1402.4815},
	journal = {\apjl},
	month = apr,
	number = {2},
	pages = {L25},
	primaryclass = {astro-ph.CO},
	title = {{Measurement of the Halo Bias from Stacked Shear Profiles of Galaxy
	         Clusters}},
	volume = {784},
	year = 2014,
}

@article{Tinker2010,
	adsurl = {https://ui.adsabs.harvard.edu/abs/2010ApJ...724..878T},
	archiveprefix = {arXiv},
	author = {{Tinker}, Jeremy L. and {Robertson}, Brant E. and {Kravtsov},
	          Andrey V. and {Klypin}, Anatoly and {Warren}, Michael S. and {Yepes
	          }, Gustavo and {Gottl{\"o}ber}, Stefan},
	doi = {10.1088/0004-637X/724/2/878},
	eprint = {1001.3162},
	journal = {\apj},
	month = dec,
	number = {2},
	pages = {878--886},
	primaryclass = {astro-ph.CO},
	title = {{The Large-scale Bias of Dark Matter Halos: Numerical Calibration
	         and Model Tests}},
	volume = {724},
	year = 2010,
}

@article{Lelli2016,
	adsurl = {https://ui.adsabs.harvard.edu/abs/2016AJ....152..157L},
	archiveprefix = {arXiv},
	author = {{Lelli}, Federico and {McGaugh}, Stacy S. and {Schombert}, James
	          M.},
	doi = {10.3847/0004-6256/152/6/157},
	eid = {157},
	eprint = {1606.09251},
	journal = {The Astronomical Journal},
	month = dec,
	number = {6},
	pages = {157},
	primaryclass = {astro-ph.GA},
	title = {{SPARC: Mass Models for 175 Disk Galaxies with Spitzer Photometry
	         and Accurate Rotation Curves}},
	volume = {152},
	year = 2016,
}

@article{McGaugh2020b,
	adsurl = {https://ui.adsabs.harvard.edu/abs/2020RNAAS...4...45M},
	author = {{McGaugh}, Stacy S. and {Lelli}, Federico and {Schombert}, James
	          M.},
	doi = {10.3847/2515-5172/ab8471},
	eid = {45},
	journal = {Research Notes of the American Astronomical Society},
	month = mar,
	number = {4},
	pages = {45},
	title = {{Scaling Relations for Molecular Gas and Metallicity: Impact on the
	         Baryonic Tully-Fisher Relation}},
	volume = {4},
	year = 2020,
}

@article{Johnston2007,
	adsurl = {https://ui.adsabs.harvard.edu/abs/2007ApJ...656...27J},
	archiveprefix = {arXiv},
	author = {{Johnston}, David E. and {Sheldon}, Erin S. and {Tasitsiomi}, Argyro and {Frieman}, Joshua A. and {Wechsler}, Risa H. and {McKay}, Timothy A.},
	doi = {10.1086/510060},
	eprint = {astro-ph/0507467},
	journal = {\apj},
	month = feb,
	number = {1},
	pages = {27–41},
	primaryclass = {astro-ph},
	title = {{Cross-Correlation Lensing: Determining Galaxy and Cluster Mass Profiles from Statistical Weak-Lensing Measurements}},
	volume = {656},
	year = 2007
}

@article{Famaey2024,
	adsurl = {https://ui.adsabs.harvard.edu/abs/2025PhRvD.111l3042F},
	archiveprefix = {arXiv},
	author = {{Famaey}, Benoit and {Pizzuti}, Lorenzo and {Saltas}, Ippocratis D.},
	doi = {10.1103/dccw-srks},
	eid = {123042},
	eprint = {2410.02612},
	journal = {\prd},
	month = jun,
	number = {12},
	pages = {123042},
	primaryclass = {astro-ph.CO},
	title = {{Nature of the missing mass of galaxy clusters in MOND: The view from gravitational lensing}},
	volume = {111},
	year = 2025
}

@article{Bilicki2021,
	adsurl = {https://ui.adsabs.harvard.edu/abs/2021A\&A...653A..82B},
	author = {{Bilicki}, M. and {Dvornik}, A. and {Hoekstra}, H. and {Wright}, A.~H. and {Chisari}, N.~E. and {Vakili}, M. and {Asgari}, M. and { Giblin}, B. and {Heymans}, C. and {Hildebrandt}, H. and {Holwerda}, B.~W. and {Hopkins}, A. and {Johnston}, H. and {Kannawadi}, A. and {Kuijken}, K. and {Nakoneczny}, S.~J. and {Shan}, H.~Y. and { Sonnenfeld}, A. and {Valentijn}, E.},
	doi = {10.1051/0004-6361/202140352},
	eid = {A82},
	journal = {\aap},
	month = sep,
	pages = {A82},
	title = {{Bright galaxy sample in the Kilo-Degree Survey Data Release 4. Selection, photometric redshifts, and physical properties}},
	volume = {653},
	year = 2021
}

@ARTICLE{Strateva2001,
       author = {{Strateva}, Iskra and {Ivezi{\'c}}, {\v{Z}}eljko and {Knapp}, Gillian R. and {Narayanan}, Vijay K. and {Strauss}, Michael A. and {Gunn}, James E. and {Lupton}, Robert H. and {Schlegel}, David and {Bahcall}, Neta A. and {Brinkmann}, Jon and {Brunner}, Robert J. and {Budav{\'a}ri}, Tam{\'a}s and {Csabai}, Istv{\'a}n and {Castander}, Francisco Javier and {Doi}, Mamoru and {Fukugita}, Masataka and {Gy{\H{o}}ry}, Zsuzsanna and {Hamabe}, Masaru and {Hennessy}, Greg and {Ichikawa}, Takashi and {Kunszt}, Peter Z. and {Lamb}, Don Q. and {McKay}, Timothy A. and {Okamura}, Sadanori and {Racusin}, Judith and {Sekiguchi}, Maki and {Schneider}, Donald P. and {Shimasaku}, Kazuhiro and {York}, Donald},
        title = "{Color Separation of Galaxy Types in the Sloan Digital Sky Survey Imaging Data}",
      journal = {\aj},
         year = 2001,
        month = oct,
       volume = {122},
       number = {4},
        pages = {1861-1874},
          doi = {10.1086/323301},
archivePrefix = {arXiv},
       eprint = {astro-ph/0107201},
 primaryClass = {astro-ph},
       adsurl = {https://ui.adsabs.harvard.edu/abs/2001AJ....122.1861S}
}

@ARTICLE{Lelli2014feedback,
       author = {{Lelli}, Federico and {Verheijen}, Marc and {Fraternali}, Filippo},
        title = "{Dynamics of starbursting dwarf galaxies. III. A H I study of 18 nearby objects}",
      journal = {\aap},
         year = 2014,
        month = jun,
       volume = {566},
          eid = {A71},
        pages = {A71},
          doi = {10.1051/0004-6361/201322657},
archivePrefix = {arXiv},
       eprint = {1404.6252},
 primaryClass = {astro-ph.GA},
       adsurl = {https://ui.adsabs.harvard.edu/abs/2014A&A...566A..71L}
}

@ARTICLE{Marasco2023,
       author = {{Marasco}, A. and {Belfiore}, F. and {Cresci}, G. and {Lelli}, F. and {Venturi}, G. and {Hunt}, L.~K. and {Concas}, A. and {Marconi}, A. and {Mannucci}, F. and {Mingozzi}, M. and {McLeod}, A.~F. and {Kumari}, N. and {Carniani}, S. and {Vanzi}, L. and {Ginolfi}, M.},
        title = "{Shaken, but not expelled: Gentle baryonic feedback from nearby starburst dwarf galaxies}",
      journal = {\aap},
         year = 2023,
        month = feb,
       volume = {670},
          eid = {A92},
        pages = {A92},
          doi = {10.1051/0004-6361/202244895},
archivePrefix = {arXiv},
       eprint = {2209.02726},
 primaryClass = {astro-ph.GA},
       adsurl = {https://ui.adsabs.harvard.edu/abs/2023A&A...670A..92M}
}

@ARTICLE{Leroy2015,
       author = {{Leroy}, Adam K. and {Walter}, Fabian and {Martini}, Paul and {Roussel}, H{\'e}l{\`e}ne and {Sandstrom}, Karin and {Ott}, J{\"u}rgen and {Weiss}, Axel and {Bolatto}, Alberto D. and {Schuster}, Karl and {Dessauges-Zavadsky}, Miroslava},
        title = "{The Multi-phase Cold Fountain in M82 Revealed by a Wide, Sensitive Map of the Molecular Interstellar Medium}",
      journal = {\apj},
         year = 2015,
        month = dec,
       volume = {814},
       number = {2},
          eid = {83},
        pages = {83},
          doi = {10.1088/0004-637X/814/2/83},
archivePrefix = {arXiv},
       eprint = {1509.02932},
 primaryClass = {astro-ph.GA},
       adsurl = {https://ui.adsabs.harvard.edu/abs/2015ApJ...814...83L}
}

@ARTICLE{Concas2017,
       author = {{Concas}, A. and {Popesso}, P. and {Brusa}, M. and {Mainieri}, V. and {Erfanianfar}, G. and {Morselli}, L.},
        title = "{Light breeze in the local Universe}",
      journal = {\aap},
         year = 2017,
        month = oct,
       volume = {606},
          eid = {A36},
        pages = {A36},
          doi = {10.1051/0004-6361/201629519},
archivePrefix = {arXiv},
       eprint = {1701.06569},
 primaryClass = {astro-ph.GA},
       adsurl = {https://ui.adsabs.harvard.edu/abs/2017A&A...606A..36C}
}

@ARTICLE{Concas2019,
       author = {{Concas}, A. and {Popesso}, P. and {Brusa}, M. and {Mainieri}, V. and {Thomas}, D.},
        title = "{Two-face(s): ionized and neutral gas winds in the local Universe}",
      journal = {\aap},
         year = 2019,
        month = feb,
       volume = {622},
          eid = {A188},
        pages = {A188},
          doi = {10.1051/0004-6361/201732152},
archivePrefix = {arXiv},
       eprint = {1710.08423},
 primaryClass = {astro-ph.GA},
       adsurl = {https://ui.adsabs.harvard.edu/abs/2019A&A...622A.188C}
}

@ARTICLE{McQuinn2019,
       author = {{McQuinn}, Kristen. B.~W. and {van Zee}, Liese and {Skillman}, Evan D.},
        title = "{Galactic Winds in Low-mass Galaxies}",
      journal = {\apj},
         year = 2019,
        month = nov,
       volume = {886},
       number = {1},
          eid = {74},
        pages = {74},
          doi = {10.3847/1538-4357/ab4c37},
archivePrefix = {arXiv},
       eprint = {1910.04167},
 primaryClass = {astro-ph.GA},
       adsurl = {https://ui.adsabs.harvard.edu/abs/2019ApJ...886...74M}
}

@ARTICLE{Concas2022,
       author = {{Concas}, Alice and {Maiolino}, Roberto and {Curti}, Mirko and {Hayden-Pawson}, Connor and {Cirasuolo}, Michele and {Jones}, Gareth C. and {Mercurio}, Amata and {Belfiore}, Francesco and {Cresci}, Giovanni and {Cullen}, Fergus and {Mannucci}, Filippo and {Marconi}, Alessandro and {Cappellari}, Michele and {Cicone}, Claudia and {Peng}, Yingjie and {Troncoso}, Paulina},
        title = "{Being KLEVER at cosmic noon: Ionized gas outflows are inconspicuous in low-mass star-forming galaxies but prominent in massive AGN hosts}",
      journal = {\mnras},
         year = 2022,
        month = jun,
       volume = {513},
       number = {2},
        pages = {2535-2562},
          doi = {10.1093/mnras/stac1026},
archivePrefix = {arXiv},
       eprint = {2203.11958},
 primaryClass = {astro-ph.GA},
       adsurl = {https://ui.adsabs.harvard.edu/abs/2022MNRAS.513.2535C}
}

@ARTICLE{KelleherLelli2024,
       author = {{Kelleher}, R. and {Lelli}, F.},
        title = "{Galaxy clusters in Milgromian dynamics: Missing matter, hydrostatic bias, and the external field effect}",
      journal = {\aap},
         year = 2024,
        month = aug,
       volume = {688},
          eid = {A78},
        pages = {A78},
          doi = {10.1051/0004-6361/202449968},
archivePrefix = {arXiv},
       eprint = {2405.08557},
 primaryClass = {astro-ph.CO},
       adsurl = {https://ui.adsabs.harvard.edu/abs/2024A&A...688A..78K}
}

@ARTICLE{ManceraPina2025,
       author = {{Mancera Pi{\~n}a}, Pavel E. and {Read}, Justin I. and {Kim}, Stacy and {Marasco}, Antonino and {Benavides}, Jos{\'e} A. and {Glowacki}, Marcin and {Pezzulli}, Gabriele and {Lagos}, Claudia del P.},
        title = "{The galaxy-halo connection of disc galaxies over six orders of magnitude in stellar mass}",
      journal = {\aap},
         year = 2025,
        month = jul,
       volume = {699},
          eid = {A311},
        pages = {A311},
          doi = {10.1051/0004-6361/202554381},
archivePrefix = {arXiv},
       eprint = {2505.22727},
 primaryClass = {astro-ph.GA},
       adsurl = {https://ui.adsabs.harvard.edu/abs/2025A&A...699A.311M}
}

@ARTICLE{McQuinn2022,
       author = {{McQuinn}, Kristen. B.~W. and {Adams}, Elizabeth A.~K. and {Cannon}, John M. and {Fuson}, Jackson and {Skillman}, Evan D. and {Brooks}, Alyson and {Rhode}, Katherine L. and {Haynes}, Martha P. and {Inoue}, John L. and {Marine}, Joshua and {Salzer}, John. J. and {Talluri}, Anjana K.},
        title = "{The Turndown of the Baryonic Tully-Fisher Relation and Changing Baryon Fraction at Low Galaxy Masses}",
      journal = {\apj},
         year = 2022,
        month = nov,
       volume = {940},
       number = {1},
          eid = {8},
        pages = {8},
          doi = {10.3847/1538-4357/ac9285},
archivePrefix = {arXiv},
       eprint = {2203.10105},
 primaryClass = {astro-ph.GA},
       adsurl = {https://ui.adsabs.harvard.edu/abs/2022ApJ...940....8M}
}

@ARTICLE{Zhang2024a,
       author = {{Zhang}, Yi and {Comparat}, Johan and {Ponti}, Gabriele and {Merloni}, Andrea and {Nandra}, Kirpal and {Haberl}, Frank and {Locatelli}, Nicola and {Zhang}, Xiaoyuan and {Sanders}, Jeremy and {Zheng}, Xueying and et al.},
        title = "{The hot circumgalactic medium in the eROSITA All-Sky Survey: I. X-ray surface brightness profiles}",
      journal = {\aap},
         year = 2024,
        month = oct,
       volume = {690},
          eid = {A267},
        pages = {A267},
          doi = {10.1051/0004-6361/202449412},
archivePrefix = {arXiv},
       eprint = {2401.17308},
 primaryClass = {astro-ph.GA},
       adsurl = {https://ui.adsabs.harvard.edu/abs/2024A&A...690A.267Z}
}

@ARTICLE{Zhang2025a,
       author = {{Zhang}, Yi and {Comparat}, Johan and {Ponti}, Gabriele and {Merloni}, Andrea and {Nandra}, Kirpal and {Haberl}, Frank and {Truong}, Nhut and {Pillepich}, Annalisa and {Popesso}, Paola and {Locatelli}, Nicola and et al.},
        title = "{The hot circumgalactic medium in the eROSITA All-Sky Survey: III. Star-forming and quiescent galaxies}",
      journal = {\aap},
         year = 2025,
        month = jan,
       volume = {693},
          eid = {A197},
        pages = {A197},
          doi = {10.1051/0004-6361/202452273},
archivePrefix = {arXiv},
       eprint = {2411.19945},
 primaryClass = {astro-ph.GA},
       adsurl = {https://ui.adsabs.harvard.edu/abs/2025A&A...693A.197Z}
}

@ARTICLE{Zhang2024b,
       author = {{Zhang}, Yi and {Comparat}, Johan and {Ponti}, Gabriele and {Merloni}, Andrea and {Nandra}, Kirpal and {Haberl}, Frank and {Truong}, Nhut and {Pillepich}, Annalisa and {Locatelli}, Nicola and {Zhang}, Xiaoyuan and et al.},
        title = "{The hot circumgalactic medium in the eROSITA All-Sky Survey: II. Scaling relations between X-ray luminosity and galaxies' mass}",
      journal = {\aap},
         year = 2024,
        month = oct,
       volume = {690},
          eid = {A268},
        pages = {A268},
          doi = {10.1051/0004-6361/202449413},
archivePrefix = {arXiv},
       eprint = {2401.17309},
 primaryClass = {astro-ph.GA},
       adsurl = {https://ui.adsabs.harvard.edu/abs/2024A&A...690A.268Z}
}

@ARTICLE{Zhang2025b,
       author = {{Zhang}, Yi and {Shreeram}, Soumya and {Ponti}, Gabriele and {Comparat}, Johan and {Merloni}, Andrea and {Qu}, Zhijie and {Li}, Jiangtao and {Bregman}, N. Joel and {Fang}, Taotao},
        title = "{On the baryon budget in the X-ray-emitting circumgalactic medium of Milky Way-mass galaxies}",
      journal = {arXiv e-prints},
         year = 2025,
        month = nov,
          eid = {arXiv:2511.17313},
        pages = {arXiv:2511.17313},
          doi = {10.48550/arXiv.2511.17313},
archivePrefix = {arXiv},
       eprint = {2511.17313},
 primaryClass = {astro-ph.GA},
       adsurl = {https://ui.adsabs.harvard.edu/abs/2025arXiv251117313Z}
}

@ARTICLE{Bregman2022ApJ...928...14B,
       author = {{Bregman}, Joel N. and {Hodges-Kluck}, Edmund and {Qu}, Zhijie and {Pratt}, Cameron and {Li}, Jiang-Tao and {Yun}, Yansong},
        title = "{Hot Extended Galaxy Halos around Local L* Galaxies from Sunyaev-Zeldovich Measurements}",
      journal = {\apj},
         year = 2022,
        month = mar,
       volume = {928},
       number = {1},
          eid = {14},
        pages = {14},
          doi = {10.3847/1538-4357/ac51de},
archivePrefix = {arXiv},
       eprint = {2107.14281},
 primaryClass = {astro-ph.GA},
       adsurl = {https://ui.adsabs.harvard.edu/abs/2022ApJ...928...14B}
}

@ARTICLE{Li2018ApJ...855L..24L,
       author = {{Li}, Jiang-Tao and {Bregman}, Joel N. and {Wang}, Q. Daniel and {Crain}, Robert A. and {Anderson}, Michael E.},
        title = "{Baryon Budget of the Hot Circumgalactic Medium of Massive Spiral Galaxies}",
      journal = {\apjl},
         year = 2018,
        month = mar,
       volume = {855},
       number = {2},
          eid = {L24},
        pages = {L24},
          doi = {10.3847/2041-8213/aab2af},
archivePrefix = {arXiv},
       eprint = {1802.09453},
 primaryClass = {astro-ph.GA},
       adsurl = {https://ui.adsabs.harvard.edu/abs/2018ApJ...855L..24L}
}

@ARTICLE{Pratt2021ApJ...920..104P,
       author = {{Pratt}, Cameron T. and {Qu}, Zhijie and {Bregman}, Joel N.},
        title = "{The Resolved Sunyaev-Zel'dovich Profiles of Nearby Galaxy Groups}",
      journal = {\apj},
         year = 2021,
        month = oct,
       volume = {920},
       number = {2},
          eid = {104},
        pages = {104},
          doi = {10.3847/1538-4357/ac1796},
archivePrefix = {arXiv},
       eprint = {2105.01123},
 primaryClass = {astro-ph.CO},
       adsurl = {https://ui.adsabs.harvard.edu/abs/2021ApJ...920..104P}
}

@BOOK{Draine2011,
       author = {{Draine}, Bruce T.},
        title = "{Physics of the Interstellar and Intergalactic Medium}",
       publisher = {Princeton, NJ : Princeton University Press},
         year = 2011,
       adsurl = {https://ui.adsabs.harvard.edu/abs/2011piim.book.....D}
}

@ARTICLE{Hua2025,
       author = {{Hua}, Zichen and {Lelli}, Federico and {Di Teodoro}, Enrico and {McGaugh}, Stacy and {Schombert}, James},
        title = "{The baryonic mass─size relation of galaxies: I. A dichotomy in star-forming galaxy disks}",
      journal = {\aap},
         year = 2025,
        month = nov,
       volume = {703},
          eid = {A223},
        pages = {A223},
          doi = {10.1051/0004-6361/202555721},
archivePrefix = {arXiv},
       eprint = {2510.17770},
 primaryClass = {astro-ph.GA},
       adsurl = {https://ui.adsabs.harvard.edu/abs/2025A&A...703A.223H}
}

@ARTICLE{Scannapieco2006,
       author = {{Scannapieco}, E. and {Pichon}, C. and {Aracil}, B. and {Petitjean}, P. and {Thacker}, R.~J. and {Pogosyan}, D. and {Bergeron}, J. and {Couchman}, H.~M.~P.},
        title = "{The sources of intergalactic metals}",
      journal = {\mnras},
         year = 2006,
        month = jan,
       volume = {365},
       number = {2},
        pages = {615-637},
          doi = {10.1111/j.1365-2966.2005.09753.x},
archivePrefix = {arXiv},
       eprint = {astro-ph/0503001},
 primaryClass = {astro-ph},
       adsurl = {https://ui.adsabs.harvard.edu/abs/2006MNRAS.365..615S}
}

@ARTICLE{Danforth2006,
       author = {{Danforth}, Charles W. and {Shull}, J. Michael and {Rosenberg}, Jessica L. and {Stocke}, John T.},
        title = "{The Low-z Intergalactic Medium. II. Ly{\ensuremath{\beta}}, O VI, and C III Forest}",
      journal = {\apj},
         year = 2006,
        month = apr,
       volume = {640},
       number = {2},
        pages = {716-740},
          doi = {10.1086/500191},
archivePrefix = {arXiv},
       eprint = {astro-ph/0508656},
 primaryClass = {astro-ph},
       adsurl = {https://ui.adsabs.harvard.edu/abs/2006ApJ...640..716D}
}

@ARTICLE{MitchellSchaye2022,
       author = {{Mitchell}, Peter D. and {Schaye}, Joop},
        title = "{Baryonic mass budgets for haloes in the EAGLE simulation, including ejected and prevented gas}",
      journal = {\mnras},
         year = 2022,
        month = apr,
       volume = {511},
       number = {2},
        pages = {2600-2609},
          doi = {10.1093/mnras/stab3686},
archivePrefix = {arXiv},
       eprint = {2112.08244},
 primaryClass = {astro-ph.GA},
       adsurl = {https://ui.adsabs.harvard.edu/abs/2022MNRAS.511.2600M}
}

@ARTICLE{Silk2003MNRAS.343..249S,
       author = {{Silk}, Joseph},
        title = "{A new prescription for protogalactic feedback and outflows: where have all the baryons gone?}",
      journal = {\mnras},
         year = 2003,
        month = jul,
       volume = {343},
       number = {1},
        pages = {249-254},
          doi = {10.1046/j.1365-8711.2003.06674.x},
archivePrefix = {arXiv},
       eprint = {astro-ph/0212068},
 primaryClass = {astro-ph},
       adsurl = {https://ui.adsabs.harvard.edu/abs/2003MNRAS.343..249S}
}

@ARTICLE{Korsaga2023,
       author = {{Korsaga}, Marie and {Famaey}, Benoit and {Freundlich}, Jonathan and {Posti}, Lorenzo and {Ibata}, Rodrigo and {Boily}, Christian and {Kraljic}, Katarina and {Esparza-Arredondo}, D. and {Ramos Almeida}, C. and {Koulidiati}, Jean},
        title = "{Disk Galaxies Are Self-similar: The Universality of the H I-to-Halo Mass Ratio for Isolated Disks}",
      journal = {\apjl},
         year = 2023,
        month = aug,
       volume = {952},
       number = {2},
          eid = {L41},
        pages = {L41},
          doi = {10.3847/2041-8213/ace364},
archivePrefix = {arXiv},
       eprint = {2307.01035},
 primaryClass = {astro-ph.GA},
       adsurl = {https://ui.adsabs.harvard.edu/abs/2023ApJ...952L..41K}
}

@ARTICLE{Dev2024,
       author = {{Dev}, Ajay and {Driver}, Simon P. and {Meyer}, Martin and {Robotham}, Aaron and {Obreschkow}, Danail and {Popesso}, Paola and {Comparat}, Johan},
        title = "{The baryon census and the mass-density of stars, neutral gas, and hot gas as a function of halo mass}",
      journal = {\mnras},
         year = 2024,
        month = dec,
       volume = {535},
       number = {3},
        pages = {2357-2374},
          doi = {10.1093/mnras/stae2485},
archivePrefix = {arXiv},
       eprint = {2411.00456},
 primaryClass = {astro-ph.CO},
       adsurl = {https://ui.adsabs.harvard.edu/abs/2024MNRAS.535.2357D}
}

@ARTICLE{Marasco2025,
       author = {{Marasco}, A. and {de Blok}, W.~J.~G. and {Maccagni}, F.~M. and {Fraternali}, F. and {Oman}, K.~A. and {Oosterloo}, T. and {Combes}, F. and {McGaugh}, S.~S. and {Kamphuis}, P. and {Spekkens}, K. and {Kleiner}, D. and {Veronese}, S. and {Amram}, P. and {Chemin}, L. and {Brinks}, E.},
        title = "{HI within and around observed and simulated galaxy discs: Comparing MeerKAT observations with mock data from TNG50 and FIRE-2}",
      journal = {\aap},
         year = 2025,
        month = may,
       volume = {697},
          eid = {A86},
        pages = {A86},
          doi = {10.1051/0004-6361/202453172},
archivePrefix = {arXiv},
       eprint = {2503.03818},
 primaryClass = {astro-ph.GA},
       adsurl = {https://ui.adsabs.harvard.edu/abs/2025A&A...697A..86M}
}

@ARTICLE{Schaye2015EAGLE,
       author = {{Schaye}, Joop and {Crain}, Robert A. and {Bower}, Richard G. and {Furlong}, Michelle and {Schaller}, Matthieu and {Theuns}, Tom and {Dalla Vecchia}, Claudio and {Frenk}, Carlos S. and {McCarthy}, I.~G. and {Helly}, John C. and {Jenkins}, Adrian and {Rosas-Guevara}, Y.~M. and {White}, Simon D.~M. and {Baes}, Maarten and {Booth}, C.~M. and {Camps}, Peter and {Navarro}, Julio F. and {Qu}, Yan and {Rahmati}, Alireza and {Sawala}, Till and {Thomas}, Peter A. and {Trayford}, James},
        title = "{The EAGLE project: simulating the evolution and assembly of galaxies and their environments}",
      journal = {\mnras},
         year = 2015,
        month = jan,
       volume = {446},
       number = {1},
        pages = {521-554},
          doi = {10.1093/mnras/stu2058},
archivePrefix = {arXiv},
       eprint = {1407.7040},
 primaryClass = {astro-ph.GA},
       adsurl = {https://ui.adsabs.harvard.edu/abs/2015MNRAS.446..521S}
}

@ARTICLE{Schaye2025,
       author = {{Schaye}, Joop and {Chaikin}, Evgenii and {Schaller}, Matthieu and {Ploeckinger}, Sylvia and {Hu{\v{s}}ko}, Filip and {McGibbon}, Rob and {Trayford}, James W. and {Ben{\'\i}tez-Llambay}, Alejandro and {Correa}, Camila and {Frenk}, Carlos S. and {Richings}, Alexander J. and {Forouhar Moreno}, Victor J. and {Bah{\'e}}, Yannick M. and {Borrow}, Josh and {Durrant}, Anna and {Gebek}, Andrea and {Helly}, John C. and {Jenkins}, Adrian and {Lacey}, Cedric G. and {Ludlow}, Aaron and {Nobels}, Folkert S.~J.},
        title = "{The COLIBRE project: cosmological hydrodynamical simulations of galaxy formation and evolution}",
      journal = {arXiv e-prints},
         year = 2025,
        month = aug,
          eid = {arXiv:2508.21126},
        pages = {arXiv:2508.21126},
          doi = {10.48550/arXiv.2508.21126},
archivePrefix = {arXiv},
       eprint = {2508.21126},
 primaryClass = {astro-ph.GA},
       adsurl = {https://ui.adsabs.harvard.edu/abs/2025arXiv250821126S}
}

@ARTICLE{Dado2025,
       author = {{Dado}, Diego and {Oman}, Kyle A. and {Harborne}, Katherine E. and {Fragkoudi}, Francesca and {Schaye}, Joop and {Schaller}, Matthieu and {Ben{\'\i}tez-Llambay}, Alejandro and {Chaikin}, Evgenii and {Frenk}, Carlos S. and {Hu{\v{s}}ko}, Filip and {Ploeckinger}, Sylvia and {Richings}, Alexander J.},
        title = "{Dynamical disequilibrium in dwarf galaxies: rethinking gas dynamics, rotation curves, and dark matter inference}",
      journal = {arXiv e-prints},
         year = 2025,
        month = dec,
          eid = {arXiv:2512.11033},
        pages = {arXiv:2512.11033},
          doi = {10.48550/arXiv.2512.11033},
archivePrefix = {arXiv},
       eprint = {2512.11033},
 primaryClass = {astro-ph.GA},
       adsurl = {https://ui.adsabs.harvard.edu/abs/2025arXiv251211033D}
}

@ARTICLE{KdN2011,
       author = {{Kuzio de Naray}, Rachel and {Kaufmann}, Tobias},
        title = "{Recovering cores and cusps in dark matter haloes using mock velocity field observations}",
      journal = {\mnras},
         year = 2011,
        month = jul,
       volume = {414},
       number = {4},
        pages = {3617-3626},
          doi = {10.1111/j.1365-2966.2011.18656.x},
archivePrefix = {arXiv},
       eprint = {1012.3471},
 primaryClass = {astro-ph.CO},
       adsurl = {https://ui.adsabs.harvard.edu/abs/2011MNRAS.414.3617K}
}

@ARTICLE{Wright2024MNRAS.532.3417W,
       author = {{Wright}, Ruby J. and {Somerville}, Rachel S. and {Lagos}, Claudia del P. and {Schaller}, Matthieu and {Dav{\'e}}, Romeel and {Angl{\'e}s-Alc{\'a}zar}, Daniel and {Genel}, Shy},
        title = "{The baryon cycle in modern cosmological hydrodynamical simulations}",
      journal = {\mnras},
         year = 2024,
        month = aug,
       volume = {532},
       number = {3},
        pages = {3417-3440},
          doi = {10.1093/mnras/stae1688},
archivePrefix = {arXiv},
       eprint = {2402.08408},
 primaryClass = {astro-ph.GA},
       adsurl = {https://ui.adsabs.harvard.edu/abs/2024MNRAS.532.3417W}
}

@ARTICLE{Medlock2025ApJ...980...61M,
       author = {{Medlock}, Isabel and {Neufeld}, Chloe and {Nagai}, Daisuke and {Angl{\'e}s-Alc{\'a}zar}, Daniel and {Genel}, Shy and {Oppenheimer}, Benjamin D. and {Sims}, Xavier and {Singh}, Priyanka and {Villaescusa-Navarro}, Francisco},
        title = "{Quantifying Baryonic Feedback on the Warm{\textendash}Hot Circumgalactic Medium in CAMELS Simulations}",
      journal = {\apj},
         year = 2025,
        month = feb,
       volume = {980},
       number = {1},
          eid = {61},
        pages = {61},
          doi = {10.3847/1538-4357/ada442},
archivePrefix = {arXiv},
       eprint = {2410.16361},
 primaryClass = {astro-ph.GA},
       adsurl = {https://ui.adsabs.harvard.edu/abs/2025ApJ...980...61M}
}

@misc{McGaugh_2024, 
   author={{McGaugh}, Stacy S},
   title="{Clusters of galaxies ruin everything}", 
   url={https://tritonstation.com/2024/02/06/clusters-of-galaxies-ruin-everything/}, 
   journal = {Triton Station},  
   year = 2024, 
   month = feb, 
          volume = {2024},
       number = {2},
          eid = {6},
   doi = {10.59350/zmemn-5ew89}
}

@ARTICLE{Kent1987,
       author = {{Kent}, Stephen M.},
        title = "{Dark Matter in Spiral Galaxies. II. Galaxies with H I Rotation Curves}",
      journal = {\aj},
         year = 1987,
        month = apr,
       volume = {93},
        pages = {816},
          doi = {10.1086/114366},
       adsurl = {https://ui.adsabs.harvard.edu/abs/1987AJ.....93..816K}
}

@ARTICLE{Milgrom1988,
       author = {{Milgrom}, Mordehai},
        title = "{On the Use of Galaxy Rotation Curves to Test the Modified Dynamics}",
      journal = {\apj},
         year = 1988,
        month = oct,
       volume = {333},
        pages = {689},
          doi = {10.1086/166777},
       adsurl = {https://ui.adsabs.harvard.edu/abs/1988ApJ...333..689M}
}

@ARTICLE{Haubner2025,
       author = {{Haubner}, Konstantin and {Lelli}, Federico and {Di Teodoro}, Enrico and {Duey}, Francis and {McGaugh}, Stacy and {Schombert}, James},
        title = "{A new uncertainty scheme for galaxy distances from flow models}",
      journal = {\aap},
         year = 2025,
        month = apr,
       volume = {696},
          eid = {A185},
        pages = {A185},
          doi = {10.1051/0004-6361/202554164},
archivePrefix = {arXiv},
       eprint = {2503.08491},
 primaryClass = {astro-ph.CO},
       adsurl = {https://ui.adsabs.harvard.edu/abs/2025A&A...696A.185H}
}

@ARTICLE{Andreon2017A&A...606A..24A,
       author = {{Andreon}, S. and {Wang}, J. and {Trinchieri}, G. and {Moretti}, A. and {Serra}, A.~L.},
        title = "{Variegate galaxy cluster gas content: Mean fraction, scatter, selection effects, and covariance with X-ray luminosity}",
      journal = {\aap},
         year = 2017,
        month = oct,
       volume = {606},
          eid = {A24},
        pages = {A24},
          doi = {10.1051/0004-6361/201730722},
archivePrefix = {arXiv},
       eprint = {1706.08356},
 primaryClass = {astro-ph.CO},
       adsurl = {https://ui.adsabs.harvard.edu/abs/2017A&A...606A..24A}
}

@ARTICLE{Besla2007ApJ...668..949B,
       author = {{Besla}, Gurtina and {Kallivayalil}, Nitya and {Hernquist}, Lars and {Robertson}, Brant and {Cox}, T.~J. and {van der Marel}, Roeland P. and {Alcock}, Charles},
        title = "{Are the Magellanic Clouds on Their First Passage about the Milky Way?}",
      journal = {\apj},
         year = 2007,
        month = oct,
       volume = {668},
       number = {2},
        pages = {949-967},
          doi = {10.1086/521385},
archivePrefix = {arXiv},
       eprint = {astro-ph/0703196},
 primaryClass = {astro-ph},
       adsurl = {https://ui.adsabs.harvard.edu/abs/2007ApJ...668..949B}
}

@ARTICLE{Oehm2024Univ...10..143O,
       author = {{Oehm}, Wolfgang and {Kroupa}, Pavel},
        title = "{The Relevance of Dynamical Friction for the MW/LMC/SMC Triple System}",
      journal = {Universe},
         year = 2024,
        month = mar,
       volume = {10},
       number = {3},
          eid = {143},
        pages = {143},
          doi = {10.3390/universe10030143},
archivePrefix = {arXiv},
       eprint = {2403.17999},
 primaryClass = {astro-ph.GA},
       adsurl = {https://ui.adsabs.harvard.edu/abs/2024Univ...10..143O}
}
\bibliographystyle{aasjournal}

\end{document}